\documentclass[australian,english,prl, singlespace, twocolumn]{revtex4-1}
\usepackage[T1]{fontenc}
\usepackage[latin9]{inputenc}
\usepackage{color}
\usepackage{bm}
\usepackage{amsmath}
\usepackage{amssymb}
\usepackage{graphicx}
\usepackage{esint}
\PassOptionsToPackage{normalem}{ulem}
\usepackage{ulem}

\makeatletter
\@ifundefined{definecolor}
 {\usepackage{color}}{}
\makeatother

\IfFileExists{lmodern.sty}{\usepackage{lmodern}}{}

\makeatother

\usepackage{babel}
\begin{document}
\title{Resolving Schr\"{o}dinger's analysis of the Einstein-Podolsky-Rosen
paradox: an incompleteness criterion and weak elements of reality}
\author{C. McGuigan, R.Y. Teh, P.D. Drummond and M.D Reid}
\affiliation{Centre for Quantum Science and Technology Theory, Swinburne University
of Technology, Melbourne 3122, Australia}
\begin{abstract}
The Einstein-Podolsky-Rosen (EPR) paradox was presented as an argument
that quantum mechanics is an incomplete description of physical reality.
However, the premises on which the argument is based are falsifiable
by Bell experiments.  In this paper, we examine the EPR paradox from
the perspective of Schr\"{o}dinger's reply to EPR. Schr\"{o}dinger
pointed out that the correlated states of the paradox enable the simultaneous
measurement of $\hat{x}$ and $\hat{p}$, ``\emph{one by direct,
the other by indirect measurement}''. Schr\"{o}dinger's analysis
takes on a timely importance because the recent experiment of Colciaghi
et al \citep{sch-epr-exp-atom} realizes these correlations for macroscopic
atomic systems. Different to the original argument, Schr\"{o}dinger's
analysis applies to the experiment at the time \emph{when the measurement
settings} \emph{have been fixed}. In this context, a subset of local
realistic assumptions (not negated by Bell's theorem) implies that
$x$ and $p$ are simultaneously precisely defined. Hence, an alternative
argument can be presented that quantum mechanics is incomplete, based
on a set of (arguably) nonfalsifiable premises. We refer to the reduced
premises as \emph{weak local realism} (wLR). As systems are amplified,
macroscopic realism can be invoked, and the assumptions of wLR become
less strict, referred to as \emph{weak macroscopic realism} (wMR).
In this paper, we propose a realization of Schr\"{o}dinger's gedanken
experiment using two-mode squeezed states, where field quadrature
phase amplitudes $\hat{X}$ and $\hat{P}$ replace position and momentum.
Assuming wMR, we derive a practical criterion for the incompleteness
of quantum mechanics, showing that the criterion is feasible for current
experiments. Questions raised by Schr\"{o}dinger about the EPR paradox
are resolved. By performing simulations of forward-backward stochastic
equations based on an objective-field ($Q$-based) model for quantum
mechanics, we illustrate the emergence on amplification of simultaneous
predetermined values for\emph{ }$\hat{X}$ and $\hat{P}$.  The
values can be regarded as \emph{weak elements of reality} (along the
lines of Bell's macroscopic 'beables') that exist for a physical quantity
after measurement settings are fixed.
\end{abstract}
\maketitle

\section{Introduction}

In 1935, Einstein, Podolsky and Rosen (EPR) presented an argument
that quantum mechanics is an incomplete description of physical reality
\foreignlanguage{australian}{\citep{epr-1}}. The argument was based
on two seemingly very reasonable premises: (1) a criterion for reality
and (2) locality. The argument considers two separated particles,
$A$ and $B$, which have correlated positions $x_{A}$ and $x_{B}$,
and anti-correlated momenta $p_{A}$ and $p_{B}$ \citep{epr-1,bell-cs-review}.
It is possible to predict with certainty the outcome of the measurement
of $x_{A}$ of one particle $A$, by measuring $x_{B}$ of the other.
It is also possible to predict with certainty the outcome of the measurement
of $p_{A}$ of particle $A$, by measuring $p_{B}$ of particle $B$.
Based on the premises, EPR concluded that both the position $x_{A}$
and momentum $p_{A}$ of particle $A$ were simultaneously precisely
defined, as \emph{``elements of reality''}, prior to a measurement
being performed. Since there is no such wavefunction description for
the localized particle, EPR concluded that quantum mechanics is incomplete.

Schr\"{o}dinger's response in which he considered a macroscopic object
(a ``cat'') in a superposition of macroscopically distinct states
(``dead and alive'') is well known \citep{s-cat-1}. Less well
known is that at the end of that response, Schr\"{o}dinger further
analyzed EPR's gedanken experiment, by considering the simultaneous
measurement of $\hat{x}$ and $\hat{p}$ of a system $A$, ``one
by direct, the other by indirect measurement''. Schr\"{o}dinger
asked \emph{whether the outcomes for $x$ and $p$ can be simultaneously
precisely predetermined} (prior to a final readout), and, if so, raised
potential inconsistencies with quantum mechanics \citep{s-cat-1}.
In considering the system $A$ after a $p$-measurement on system
$B$, Schr\"{o}dinger remarked: ``the quantum mechanician maintains
that'' the system $A$ ``has a psi-function\emph{'' }in which $p$\emph{
``is fully sharp'',} but $x$\emph{ ``fully indeterminate}''.

Schr\"{o}dinger's analysis has largely been forgotten. The answer
to his question is generally thought to be negative, and to have been
addressed by Bell's work \citep{Bell-2,bell-contextual,fine}. The
spin version of EPR's argument \citep{Bohm} motivated Bell and Greenberger,
Horne and Zeilinger (GHZ) \citep{ghz,mermin-ghz,ghz-1,ghz-clifton-1}
to derive inequalities, which are violated by quantum mechanics. Such
violations falsify all local hidden-variable explanations of Bell
and GHZ experiments, thereby negating the possibility that simultaneous
values for noncommuting spin observables, as implied by the EPR premises,
can exist \citep{bell-cs-review}.

However, in the set-up considered by Schr\"{o}dinger, the measurement
settings \emph{have been specified}, as $\hat{x}$ for $A$ and $\hat{p}$
for $B$ \citep{sch-epr-exp-atom,viewpoint}. This means that simultaneous
predetermined values for $x$ and $p$ can be posited using a set
of \emph{weaker premises} that are not been negated by Bell or GHZ
experiments. These premises apply to the systems as defined at a time
$t_{m}$, \emph{after} the dynamics associated with the fixing of
the measurement settings in the experiment \citep{weak-versus-det,ghz-cat,wigner-friend-macro,manushan-bell-cat-lg}.
In this paper, we explain how the simultaneous precise values for
$x$ and $p$ \emph{can} exist at time $t_{m}$, prior to the irreversible
readout. We address Schr\"{o}dinger's concerns about potential incompatibility
with quantum mechanics, and provide a phase-space model (the objective-field
$Q$-based model \citep{q-contextual,q-frederic,objective-fields-entropy})
 in which the simultaneous values are computed.

Our work is motivated by Bell's concept of macroscopic ``beables''
\citep{beables} and a recent experiment, in which the EPR correlations
proposed by Schr\"{o}dinger have been observed for macroscopic atomic
systems \citep{sch-epr-exp-atom}. Where the systems are macroscopic,
the conclusion that there are predetermined values (at time $t_{m}$)
can be based on the assumption of \emph{macroscopic realism} \citep{legggarg-1,manushan-bell-cat-lg}.

In order to address Schr\"{o}dinger's paradox, we examine the realization
of the EPR argument using field modes \citep{epr-r2}. Here, field
quadrature phase amplitudes $\hat{X}$ and $\hat{P}$ replace position
and momentum in the paradox. The highly correlated EPR state is realized
as the two-mode squeezed state \citep{two-mode-squeezed-state-1,two-mode-squeezed-state-2},
generated from nondegenerate parametric amplification \citep{epr-r2}.
The fields denoted $A$ and $B$ are separated, and then prepared
for a quadrature measurement \citep{epr-rmp,bowen-anu,ou-epr,schnabel-1,schnabel-2,teleportation-sophie-Z}.
The first stage of the measurement involves unitary operations $U_{\theta}^{A}$
and $U_{\phi}^{B}$ that fix the measurement settings. In the proposed
experiment, this constitutes a phase shift followed by an \emph{amplification}
at each site. The choice of setting is to measure $\hat{x}$ at $A$,
and $\hat{p}$ at $B$. The final stage of measurement constitutes
a detection and readout of a meter. In our analysis, we consider the
system at the time $t_{m}$,\emph{ after} the operations $U_{\theta}^{A}$
and $U_{\phi}^{B}$ have been performed, but prior to the final detection
stage of the measurement process. 

For the system at this time $t_{m}$, we apply a set of modified EPR
premises, which we refer to as \emph{weak macroscopic realism} (wMR)
\citep{ghz-cat}. These weaker premises are sufficient to imply the
existence of \emph{simultaneous values} \emph{$x_{A}$ and $p_{A}$}
for the outcomes of measurements $\hat{X}_{A}$ and $\hat{P}_{A}$
respectively. The existence of $x_{A}$ is implied by macroscopic
realism, because the amplitude $\hat{X}_{A}$ of system $A$ has been
\emph{amplified} $-$ eventually, the system is in a superposition
of macroscopically distinct states with definite outcomes for $\hat{X}_{A}$.
The existence of $p_{A}$ is implied by a weak version of EPR's criterion
of reality for a physical quantity. The weak criterion posits that
the value for the measurement of $\hat{P}_{A}$ is predetermined at
time $t_{m}$, because it can predicted with certainty by the measurement
$\hat{P}_{B}$ and, at time $t_{m}$, there is a predetermined value
$p_{B}$ for $\hat{P}_{B}$ as posited by macroscopic realism (since
$\hat{P}_{B}$ has also been amplified). In summary, a modified set
of premises, shown to be \emph{not} falsifiable by Bell experiments
\citep{ghz-cat,wigner-friend-macro,manushan-bell-cat-lg}, imply the
simultaneous values for $x$ and $p$, hence presenting an EPR-type
argument for the incompleteness of quantum mechanics. A similar alternative
EPR argument exists for a two-spin version of Bohm's EPR paradox \citep{ghz-cat},
but this involves extra assumptions and is more difficult experimentally.

In a realistic experiment, it is not possible to predict $\hat{P}_{A}$
with absolute certainty, based on measurement $\hat{P}_{B}$. We hence
derive an ``\emph{incompleteness criterion}'' that can be applied
to an experiment, to conclude the incompleteness of (standard) quantum
mechanics, based on the weak premises. The premises of wMR apply when
the systems at time $t_{m}$ have been amplified. It is also possible
to conclude the incompleteness of quantum mechanics based on the premises
of \emph{weak local realism} (wLR), which apply regardless of amplification.
The incompleteness criterion is the basis for an argument that quantum
mechanics be completed by variables (``\emph{beables}'') that are
not contradicted by Bell's theorem.

The simultaneous values $x_{A}$ and $p_{A}$ (if they exist) pose
apparent inconsistencies with quantum mechanics. Schr\"{o}dinger
pointed out that the value of $\hat{X}_{A}^{2}+\hat{P}_{A}^{2}=1+2\hat{n}$
(where $\hat{n}$ is the number operator) must always be odd, which
seems inconsistent with $x_{A}$ and $p_{A}$ being continuous outcomes
for $\hat{X}_{A}$ and $\hat{P}_{A}$ \citep{s-cat-1}. In this paper,
we show how there is no inconsistency, once the measurement process
is properly considered.

This leaves the question of whether the simultaneous values $x_{A}$
and $p_{A}$ can be explained by quantum theory. To this end, we perform
simulations based on the \emph{objective field ($Q$-based) model}
for quantum mechanics \citep{q-contextual,q-measurement,q-measurement-wMR,objective-fields-entropy}.
The measurement of $\hat{X}_{A}$ is modeled by a direct amplification
of $\hat{X}_{A}$, realized by an interaction $H_{A}$ based on degenerate
parametric amplification. Measurements of $\hat{P}_{A}$, $\hat{X}_{B}$
and $\hat{P}_{B}$ are modeled similarly. The amplification is simulated
by solving stochastic forward-backward equations derived from $H_{A}$,
obtaining solutions for amplitudes $x_{A}(t)$, $p_{A}(t)$, $x_{B}(t)$
and $p(t)$, as defined by the $Q$ function $Q(x_{A},p_{A},x_{B},p_{B})$.
We find that \emph{bands} of $x_{A}(t)$ exist for the state of the
system at the time $t_{m}$, the values of which give a predetermination
of the measurement outcome for the amplified $\hat{X}_{A}$.  Similarly,
we find \emph{bands} for $p_{B}(t)$, which give the value of the
outcome of $\hat{P}_{B}$ (and hence, according to the theory, $\hat{P}_{A}$).
We thus verify the existence of the simultaneous $x_{A}$ and $p_{A}$
by simulation. Similar values could be obtained using the Wigner function
\citep{wigner-dn,wigner}, which is positive for two-mode squeezed
states. However, the Wigner function becomes negative (and hence cannot
produce a similar simulation) for Bell states, which are also predicted
to satisfy the incompleteness criterion \citep{ghz-cat}.

There is no conflict of the simulation with Bell's theorem: the values
$x_{A}$ and $p_{A}$ are not defined for the system as it exists
prior to the unitary interactions $U$ that fix the measurement settings.
Also, there is no violation of the uncertainty relation for system
$A$. The quantum state $|\psi\rangle$ defined for the systems at
time $t_{m}$ is a superposition of states with different outcomes
$x_{A}$ of $\hat{X}_{A}$, and hence satisfies the uncertainty relation.
The values $x_{A}$ and $p_{A}$ of the simulation, if they are to
address Schr\"{o}dinger's questions, would represent a more complete
quantum description.

\emph{Layout of paper:} In Section II, we summarize the EPR set-up
using two-mode squeezed states and present Schr\"{o}dinger's analysis,
explaining inconsistencies raised by Schr\"{o}dinger. In Section
III, we define the weak EPR premises, and derive criteria by which
to establish the incompleteness paradox. In Sections IV and V, we
present the quantum predictions for the criteria, showing that an
experiment is feasible. In Section VI, we present the simulation based
on the objective-field model. A conclusion is given in Section VII.

\section{Realization of Schr\"{o}dinger's EPR set-up \label{sec:An-EPR-realisation}}

\subsection{The EPR solutions}

\begin{figure}[t]
\begin{centering}
\includegraphics[width=1\columnwidth]{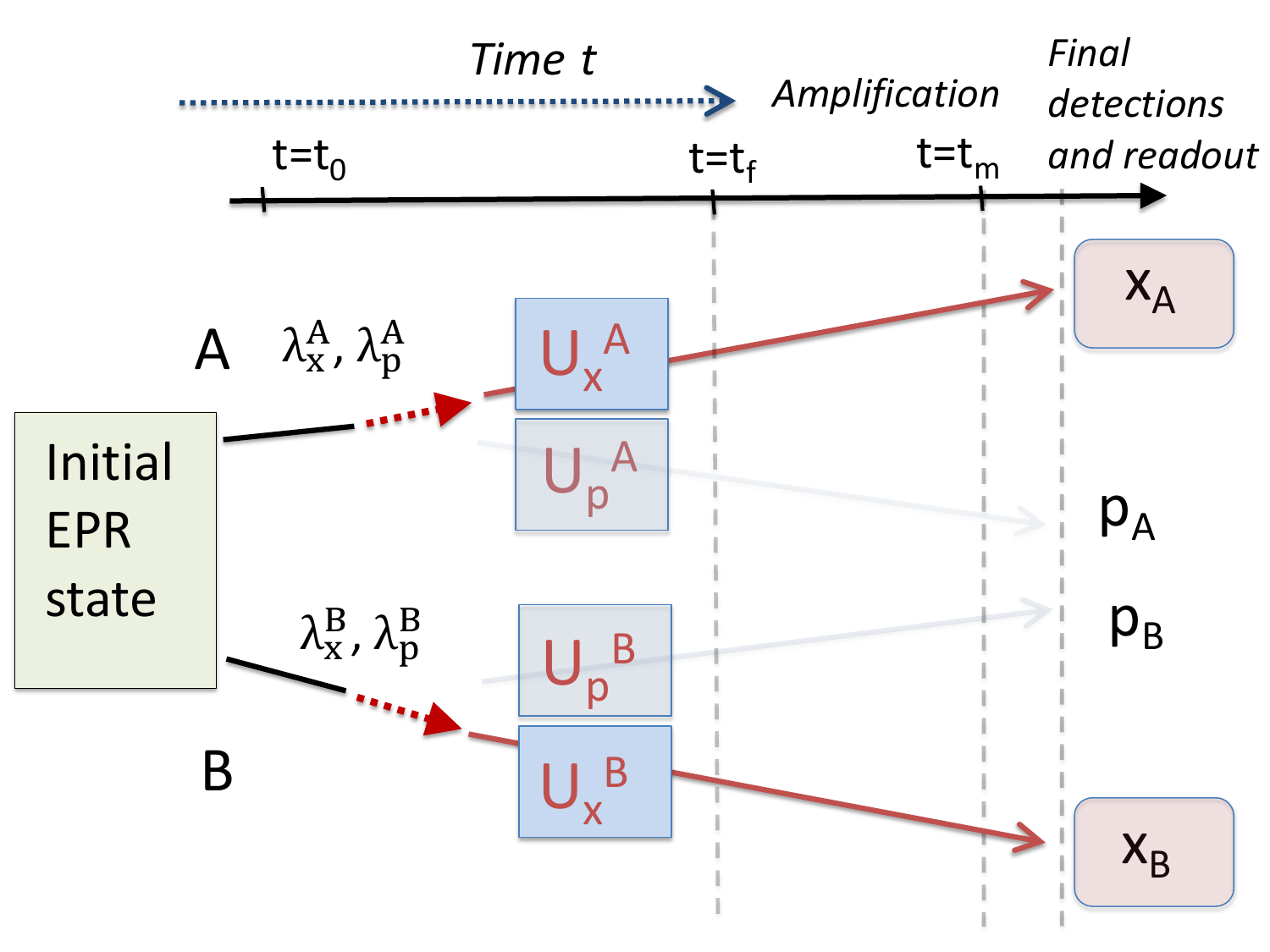}
\par\end{centering}
\centering{}\caption{A diagram of a standard EPR set-up. The two outputs $A$ and $B$
of the EPR state are spatially separated. For each system, $\hat{x}$
or $\hat{p}$ can be measured, indicated by the red dashed switch.
In the first stage of measurement, for each system, the experimentalist
selects the measurement setting ($x$ or $p$) by interacting the
system with a measurement-setting device. The effect is unitary operations
$U^{A}$ and $U^{B}$ on the systems $A$ and $B$. The measurement
is completed by a coupling to a meter and readout. For EPR states,
it is always the case that if both $\hat{x}_{A}$ and $\hat{x}_{B}$
are measured (as in the diagram), the outcomes are the correlated
so that $x_{A}=x_{B}$. If both $\hat{p}_{A}$ and $\hat{p}_{B}$
are measured, the outcomes are anti-correlated, so that $p_{B}=-p_{A}$.
The EPR premises then that imply predetermined values (``elements
of reality'') exist, at time $t_{0}$, for the outcomes of both measurements
$\hat{x}$ and $\hat{p}$, for each system. These are $\lambda_{x}^{A}$,
$\lambda_{p}^{A}$, $\lambda_{x}^{B}$, $\lambda_{p}^{B}$ in the
diagram, so that $x_{A}=\lambda_{x}^{A}$, $p_{A}=\lambda_{p}^{A}$,
$x_{B}=\lambda_{x}^{B}$, $p_{B}=\lambda_{p}^{B}$ where $\lambda_{x}^{A}=\lambda_{x}$
and $\lambda_{p}^{A}=-\lambda_{p}^{B}$. Perfect EPR correlation is
achieved for the solutions (\ref{eq:arraysolns-2}), in the limit
of large amplification $r$. \label{fig:schematic-epr}}
\end{figure}

EPR correlations have been realized in optics using the two-mode squeezed
state, $|\psi(r)\rangle_{ss}$ \citep{epr-r2,epr-rmp} (Fig. \ref{fig:schematic-epr}).
This state is generated by parametric down conversion, modeled by
the Hamiltonian (in a rotating frame)
\begin{equation}
H_{AB}=i\kappa E(\hat{a}^{\dagger}\hat{b}^{\dagger}-\hat{a}\hat{b})\label{eq:ham-tmss}
\end{equation}
where here $\hat{a}$ and $\hat{b}$ are boson destruction operators
for two distinct field modes, denoted $A$ and $B$. Under $H_{AB}$,
a system initially (at time $t=0$) in the two-mode vacuum state $|0\rangle|0\rangle$
evolves into the two-mode squeezed state
\begin{equation}
|\psi(r)\rangle_{ss}=e^{\kappa Et(\hat{a}^{\dagger}\hat{b}^{\dagger}-\hat{a}\hat{b})/\hbar}|0\rangle0\rangle\thinspace.\label{eq:tmss-def}
\end{equation}
We define field quadrature phase amplitudes $\hat{X}_{A}$, $\hat{P}_{A}$,
$\hat{X}_{B}$ and $\hat{P}_{B}$ for each mode as $\hat{X}_{A}=(\hat{a}+\hat{a}^{\dagger})/\sqrt{2},\hat{P}_{A}=(\hat{a}-\hat{a}^{\dagger})/i\sqrt{2}$
and $\hat{X}_{B}=(\hat{b}+\hat{b}^{\dagger})/\sqrt{2},\hat{P}_{B}=(\hat{b}-\hat{b}^{\dagger})/i\sqrt{2}$.
The uncertainty relations
\begin{equation}
\Delta\hat{X}_{A}\Delta\hat{P}_{A}\geq1/2\label{eq:hup}
\end{equation}
and $\Delta\hat{X}_{B}\Delta\hat{P}_{B}\geq1/2$ follow. \textcolor{red}{}In
this paper, we omit the operator ``hats'' where the meaning is clear,
and use the symbols $\hat{X}$and $\hat{P}$ interchangeably with
$\hat{x}$ and $\hat{p}$. The evolution of a system under $H_{AB}$
can also be treated by solving the operator equations. We find\textcolor{red}{\emph{}}
\begin{eqnarray}
\hat{a}(t) & = & \hat{a}(0)\cosh r+\hat{b}^{\dagger}(0)\sinh r\nonumber \\
\hat{b}(t) & = & \hat{b}(0)\cosh r+\hat{a}^{\dagger}(0)\sinh r\label{eq:solns}
\end{eqnarray}
where $r=\kappa Et/\hbar$, leading to solutions
\begin{eqnarray}
\hat{X}_{A}(t) & = & \cosh r\hat{X}_{A}\left(0\right)+\sinh r\hat{X}_{B}\left(0\right)\nonumber \\
\hat{P}_{A}(t) & = & \cosh r\hat{P}_{A}\left(0\right)-\sinh r\hat{P}_{B}\left(0\right)\nonumber \\
\hat{X}_{B}(t) & = & \cosh r\hat{X}_{B}\left(0\right)+\sinh r\hat{X}_{A}\left(0\right)\nonumber \\
\hat{P}_{B}(t) & = & \cosh r\hat{P}_{B}\left(0\right)-\sinh r\hat{P}_{A}\left(0\right)\thinspace.\label{eq:arraysolns-2}
\end{eqnarray}
The amplification of the quadrature field amplitudes is apparent.
Assuming the system is initially in the vacuum state $|0\rangle|0\rangle$,
we find $\langle\hat{a}^{\dagger}\hat{a}\rangle=\sinh^{2}r$, $\langle\hat{a}\hat{b}\rangle=\cosh r\sinh r$
and $\langle\hat{a}\rangle=0$. The moments $\langle\hat{X}_{A}\rangle$,
$\langle\hat{P}_{A}\rangle$, $\langle\hat{X}_{B}\rangle$, $\langle\hat{P}_{B}\rangle$
remain zero on evolution, but the variances $\sigma_{X_{A}}^{2}\equiv\langle\hat{X}_{A}^{2}\rangle-\langle\hat{X}_{A}\rangle$
and $\sigma_{P_{A}}^{2}\equiv\langle\hat{P}_{A}^{2}\rangle-\langle\hat{P}_{A}\rangle^{2}$
amplify:
\begin{eqnarray}
\sigma_{X_{A}}^{2} & = & \cosh^{2}r\langle\hat{X}_{A}(0)^{2}\rangle+\sinh^{2}r\langle\hat{X}_{B}(0)^{2}\rangle\nonumber \\
 & = & \frac{1}{2}\cosh2r\label{eq:varx}
\end{eqnarray}
and
\begin{eqnarray}
\sigma_{P_{A}}^{2} & = & \cosh^{2}r\langle\hat{P}_{A}(0)^{2}\rangle+\sinh^{2}r\langle\hat{P}_{B}(0)^{2}\rangle\nonumber \\
 & = & \frac{1}{2}\cosh2r\thinspace.\label{eq:varp}
\end{eqnarray}

As $r$ increases, it is clear from (\ref{eq:arraysolns-2}) that
ideal EPR correlations are obtained. One can infer the outcome $X_{A}$
of $\hat{X}_{A}$ by measuring $\hat{X}_{B}$, and one can infer the
outcome $P_{A}$ of $\hat{P}_{A}$ by measuring $\hat{P}_{B}$. Following
Refs. \citep{epr-r2,epr-rmp}, we estimate $P_{A}$ as $-g_{p}P_{B}$
and $X_{A}$ by $g_{x}X_{B}$, where $g_{x}$ and $g_{p}$ are real
constants. The variances of $X_{A}-g_{x}X_{B}$ and $P_{A}+g_{p}P_{B}$
are
\begin{eqnarray}
(\Delta(X_{A}-g_{x}X_{B}))^{2} & = & \frac{1}{2}[\cosh2r\left(1+g_{x}^{2}\right)-2g_{x}\sinh2r]\nonumber \\
(\Delta(P_{A}+g_{p}P_{B}))^{2} & = & \frac{1}{2}[\cosh2r\left(1+g_{p}^{2}\right)-2g_{p}\sinh2r]\nonumber \\
\label{eq:var-g}
\end{eqnarray}
where we use the notation $\Delta O\equiv\sqrt{\langle O^{2}\rangle-\langle O\rangle^{2}}$
for the standard deviation, $O$ being the measured quantity. The
minimum values $(\Delta(X_{A}-g_{x}X_{B}))_{min}^{2}$ and $(\Delta(P_{A}+g_{p}P_{B}))_{min}^{2}$
for the variances on optimizing $g_{x}$ and $g_{p}$ are denoted
$\Delta_{inf}^{2}P_{A}$ and $\Delta_{inf}^{2}X_{A}$. We find 
\begin{eqnarray}
\Delta_{inf}^{2}X_{A} & \equiv & (\Delta(X_{A}-g_{x}X_{B}))_{min}^{2}\nonumber \\
 & = & \dfrac{1}{2\cosh2r}\label{eq:soln-min}
\end{eqnarray}
\textcolor{blue}{}and 
\begin{eqnarray}
\Delta_{inf}^{2}P_{A} & \equiv & (\Delta(P_{A}+g_{p}P_{B}))_{min}^{2}\nonumber \\
 & = & \dfrac{1}{2\cosh2r}\thinspace.\label{eq:soln-min-1}
\end{eqnarray}
The optimal values of $g_{x}$ and $g_{p}$ are equal, given by $g_{0}$,
where 
\begin{equation}
g_{0}=\tanh2r\thinspace.\label{eq:g0}
\end{equation}
 We refer to $\Delta_{inf}^{2}X_{A}$ and $\Delta_{inf}^{2}P_{A}$
as the variances in the inferences (``inference variances'') of
$X_{A}$ and $P_{A}$ respectively. The inference variances $\Delta_{inf}^{2}P_{A}$
and $\Delta_{inf}^{2}X_{A}$ correspond exactly to the variances of
the conditional distributions $P(X_{A}|X_{B})$ and $P(P_{A}|P_{B})$,
which are Gaussians, with means $g_{0}X_{B}$\textcolor{blue}{\uline{}}
and $-g_{0}P_{B}$ respectively.\textcolor{blue}{\emph{}}\textcolor{red}{}

The Wigner function $W(x_{A},p_{A},x_{B},p_{B})$ of the two-mode
squeezed state is positive and gives an underlying probability distribution
for the outcomes of the measurements of the quadrature phase amplitudes
(refer Appendix A) \citep{epr-rmp,wigner-dn,bellcv}.\textcolor{red}{{}
}Integrating the Wigner function over $p_{A}$ and $p_{B}$, we
obtain the probability distribution $P(x_{A},x_{B})$ for outcomes
$x_{A}$ and $x_{B}$ of measurements $\hat{X}_{A}$ and $\hat{X}_{B}$.
In this paper, we use $x_{A}$, $x_{B}$, $p_{A}$ and $p_{B}$ interchangeably
with $X_{A}$, $X_{B}$, $P_{A}$ and $P_{B}$ to denote the outcomes
of $\hat{X}_{A}$, $\hat{X}_{B}$, $\hat{P}_{A}$ and $\hat{P}_{B}$.
The distribution $P(x_{A},x_{B})$ is Gaussian,
\begin{eqnarray}
P(x_{A},x_{B}) & = & \dfrac{1}{\pi}e^{-\cosh2r\left(x_{A}^{2}+x_{B}^{2}\right)}e^{2\sinh(2r)x_{A}x_{B}}\nonumber \\
 & = & \dfrac{1}{\pi}e^{-(e^{2r})(x_{A}-x_{B})^{2}/2}e^{-(e^{-2r})(x_{A}+x_{B})^{2}/2}\thinspace.\nonumber \\
\label{eq:joint-x}
\end{eqnarray}
We see that
\begin{eqnarray}
(\Delta(x_{A}\mp x_{B}))^{2} & = & e^{\mp2r}\label{eq:var-sumdiffx}
\end{eqnarray}
and similarly, by deriving\textcolor{red}{}
\begin{eqnarray}
P(p_{A},p_{B}) & = & \dfrac{1}{\pi}e^{-\cosh2r\left(p_{A}^{2}+p_{B}^{2}\right)}e^{-2\sinh(2r)p_{A}p_{B}}\nonumber \\
 & = & \dfrac{1}{\pi}e^{-(e^{2r})(p_{A}+p_{B})^{2}/2}e^{-(e^{-2r})(p_{A}-p_{B})^{2}/2}\thinspace,\nonumber \\
\label{eq:joint-p}
\end{eqnarray}
we find
\begin{eqnarray}
(\Delta(p_{A}\pm p_{B}))^{2} & = & e^{\mp2r}\thinspace.\label{eq:var-sum-diff-p}
\end{eqnarray}
in agreement with the solutions (\ref{eq:var-g}) above (putting $g_{x}=g_{p}=\pm1$).The
conditional distribution 
\begin{eqnarray}
P(x_{A}|x_{B}) & = & \dfrac{\sqrt{\cosh2r}}{\sqrt{\pi}}\frac{e^{-\cosh2r\left(x_{A}^{2}+x_{B}^{2}\right)}e^{2\sinh(2r)x_{A}x_{B}}}{e^{-x_{B}^{2}/\cosh2r}}\nonumber \\
\label{eq:cond-dis-x}
\end{eqnarray}
is also Gaussian, with mean $\mu_{x_{A}|x_{B}}=g_{0}x_{B}$ and variance
\begin{equation}
\sigma_{x_{A}|x_{B}}^{2}=\dfrac{1}{2\cosh2r}\thinspace,\label{eq:var-2}
\end{equation}
in agreement with the estimate $g_{0}X_{B}$ and inference variance
$\Delta_{inf}^{2}X_{A}$ above. We see that the inference variance
$\Delta_{inf}^{2}X_{A}$ can be defined by $\sigma_{x_{A}|x_{B}}^{2}$,
and is independent of $x_{B}$. The solutions for $P(p_{A},p_{B})$
follow in a similar fashion, with 
\begin{eqnarray}
P(p_{A}|p_{B}) & = & \dfrac{\sqrt{\cosh2r}}{\sqrt{\pi}}\frac{e^{-\cosh2r\left(p_{A}^{2}+p_{B}^{2}\right)}e^{-2\sinh(2r)p_{A}p_{B}}}{e^{-p_{B}^{2}/\cosh2r}}\thinspace.\nonumber \\
\label{eq:cond-dis-p}
\end{eqnarray}
 The mean and variance of $P(p_{A}|p_{B})$ are \textcolor{red}{}
$\mu_{p_{A}|p_{B}}=-g_{0}p_{B}$ and 
\begin{equation}
\sigma_{p_{A}|p_{B}}^{2}=\dfrac{1}{2\cosh2r},\label{eq:var-3}
\end{equation}
where we note that we can choose to define $\Delta_{inf}^{2}P_{A}$
by $\sigma_{p_{A}|p_{B}}^{2}$, which is independent of $p_{B}$ for
the two-mode squeezed state.

\subsection{The EPR argument and assumptions}

EPR's argument is based on two assumptions. We summarize below.

\paragraph*{EPR Assertion I: Locality}

It is assumed that measurements made on one system cannot disturb
another system sufficiently spatially separated from it, an assumption
referred to in Bell's work as Locality.

\paragraph*{EPR Assertion II: Criterion for reality}

EPR defined in their paper a criterion sufficient to imply an ``element
of reality'' for a physical quantity. ``If, without in any way disturbing
a system, we can predict with certainty the value of a physical quantity,
then there exists an element of reality corresponding to this physical
quantity'' \citep{epr-1}.

The two assertions are implied by all local realistic theories \citep{bell-cs-review}.
In the context of the EPR correlations, the assertions imply simultaneous
``elements of reality'' for $\hat{X}$ and $\hat{P}$ of system
$A$ (and $B$), if the modes are sufficiently spatially separated
(refer Fig. \ref{fig:schematic-epr} for the experimental arrangement)
\citep{mermin-ghz}. The ``elements of reality'' imply precise values
for the outcomes of measurements $\hat{X}$ and $\hat{P}$, a specification
that cannot be achieved by a quantum state $|\psi\rangle$, because
of the uncertainty principle $\Delta\hat{X}_{A}\Delta\hat{P}_{A}\geq1/2$.
Hence, the conclusion by EPR that quantum mechanics is an incomplete
description of physical reality \citep{epr-1}.

In the more general context where the two systems are not perfectly
correlated, a sufficient condition to realize the EPR paradox is given
by \citep{epr-r2}
\begin{equation}
\Delta_{inf}\hat{X}_{A}\Delta_{inf}\hat{P}_{A}<1/2\thinspace.\label{eq:epr-crit}
\end{equation}
 where $\Delta_{inf}\hat{X}_{A}$ and $\Delta_{inf}\hat{P}_{A}$
are the average errors in the inference of $\hat{X}_{A}$ and $\hat{P}_{A}$
respectively, given measurements made on system $B$, in accordance
with the definitions (\ref{eq:soln-min}) and (\ref{eq:soln-min-1}),
and (\ref{eq:var-2}) and (\ref{eq:var-3}). The condition (\ref{eq:epr-crit})
can be derived as a condition for EPR-steering \citep{epr-steer,steer-prl,steer-pra-2}.
The two-mode squeezed state satisfies the EPR criterion for all $r$,
and the criterion has been experimentally verified \citep{epr-rmp,bec-epr-exp,ou-epr,schnabel-1,schnabel-2,bowen-anu,bec-steer-2,sch-epr-exp-atom}
(see also Ref. \citep{wu-epr}).

As pointed out by Bell \citep{bellcv}, the EPR premises are not falsified
in a standard EPR experiment based on continuous-variable measurements,
such as quadrature phase amplitudes, or position and momentum. This
is evident by the set-up given for the two-mode squeezed state, where
the Wigner function $W(x_{A},p_{A},x_{B},p_{B})$ provides a local
hidden variable model, in which the variables $x_{A}$, $p_{A}$,
$x_{B}$ and $p_{B}$ of the distribution can represent the ``elements
of reality'' that predetermine the outcomes for the measurements.
However, EPR correlations exist for spin measurements, as in Bohm's
version of the paradox \citep{Bohm,bohm-aharonov,aspect-Bohm,wu-epr}.
For spin measurements, the existence of the predetermined spin values
implied by EPR's premises is falsified by the violation of Bell inequalities
\citep{bell-cs-review,Bell-2,bell-1971,bell-brunner-rmp,chsh,ghz-pan,ghz,mermin-ghz,Bouwmeester-ghz-experiment,fine}.

\subsection{Schr\"{o}dinger's analysis of EPR's set-up}

\begin{figure}[t]
\begin{centering}
\includegraphics[width=1\columnwidth]{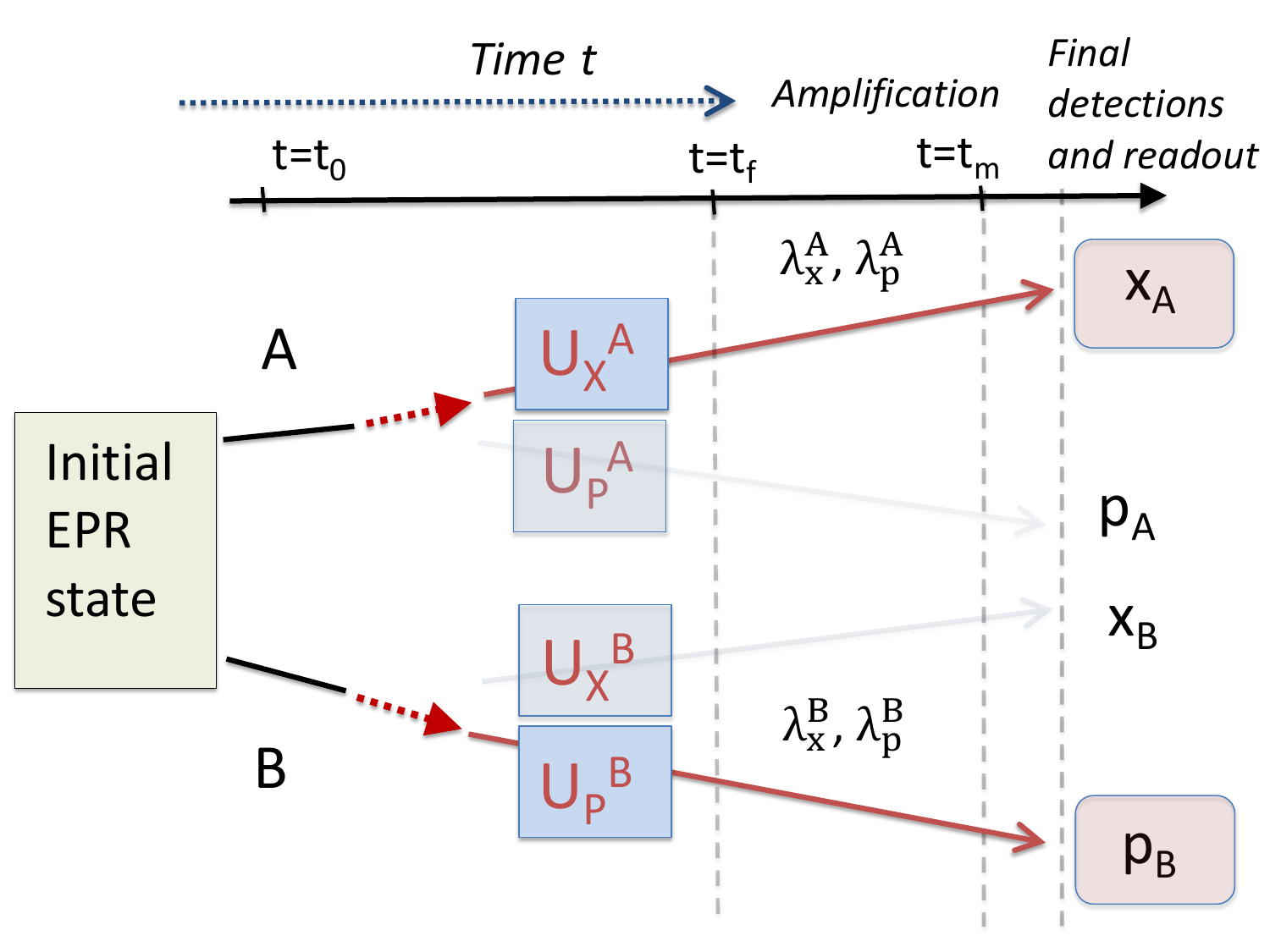}
\par\end{centering}
\centering{}\caption{A diagram of the EPR set-up considered by Schr\"{o}dinger. In the
original EPR argument, either the position $x$ or momentum $p$ can
be measured, indicated by the red dashed switch. The unitary operations
$U^{A}$ and $U^{B}$ are performed on each system, $A$ and $B$,
respectively. The measurement involves further amplification, and
then a final coupling to a meter. In Schr\"{o}dinger's proposed set-up,
the choice is made to measure $x$ at $A$ and $p$ at $B$ (red arrows).
This constitutes simultaneous measurement of $x$ and $p$, ``one
by direct, the other by indirect measurement''. Here, we consider
the validity of the existence of predetermined values $\lambda_{x}^{A}$
and $\lambda_{p}^{A}$ at the time $t_{m}$, \emph{after} the unitary
interactions and amplification. We show that the values can be posited,
based on the premises of weak macroscopic realism (refer Section \ref{sec:Incompleteness-criterion-based}).
\label{fig:schematic-sch}}
\end{figure}

Schr\"{o}dinger's analysis of EPR's set-up refers to a realization
where the measurement settings have \emph{already been established
}for both systems, $A$ and $B$, so as to realize the simultaneous
measurement of $\hat{x}_{A}$ and $\hat{p}_{B}$ \citep{s-cat-1,sch-epr-exp-atom}
(Fig. \ref{fig:schematic-sch}). For EPR states, the measurement of
$\hat{p}_{B}$ is precisely correlated with the value of $\hat{p}_{A}$.
Hence, it seems that one can measure $\hat{x}_{A}$ and $\hat{p}_{A}$
simultaneously, ``one by direct, the other by indirect measurement''
\citep{s-cat-1}. Schr\"{o}dinger's question is \emph{whether there
can be considered predetermined values for the outcomes of the measurements},
$\hat{x}_{A}$ and $\hat{p}_{A}$.  This question is closely related
to questions about measurement in quantum mechanics (the ``measurement
problem'') \citep{born,brukner,bell-against,hossenfelder-toy-collapse,wigner-friend,proietti}:
\emph{When is the value for the outcome of the measurement determined?
}The question is also about nonlocality \citep{wood,price,Bell-2}.\emph{
Does the measurement at $B$ cause the outcome for $\hat{p}_{A}$
at $A$, or was that outcome determined prior?}

According to the EPR premises, the values of the outcomes for measurements
$\hat{x}_{A}$ and $\hat{p}_{A}$ are indeed both simultaneously predetermined$-$
but these premises are falsifiable by the violations of Bell inequalities.
The concept of simultaneously predetermined values for the outcomes
of noncommuting spin measurements (as are implied by the EPR premises
in Bohm's version of the paradox) is falsified by Bell violations
\citep{bell-cs-review,Bell-2,fine,hall}.\emph{ }However, these predetermined
values are defined for the system at the time $t_{0}$, prior to the
unitary interactions that fix the settings (Fig. \ref{fig:schematic-sch}).
Schr\"{o}dinger's question can be applied to the system at the later
time ($t_{f}$ or $t_{m}$) when the measurement settings are fixed.
Hence, as explained in Refs. \citep{ghz-cat,wigner-friend-macro,weak-versus-det},
Schr\"{o}dinger's question is not addressed by the violations of
Bell inequalities.

\begin{figure}[t]
\centering{}\includegraphics[width=1\columnwidth]{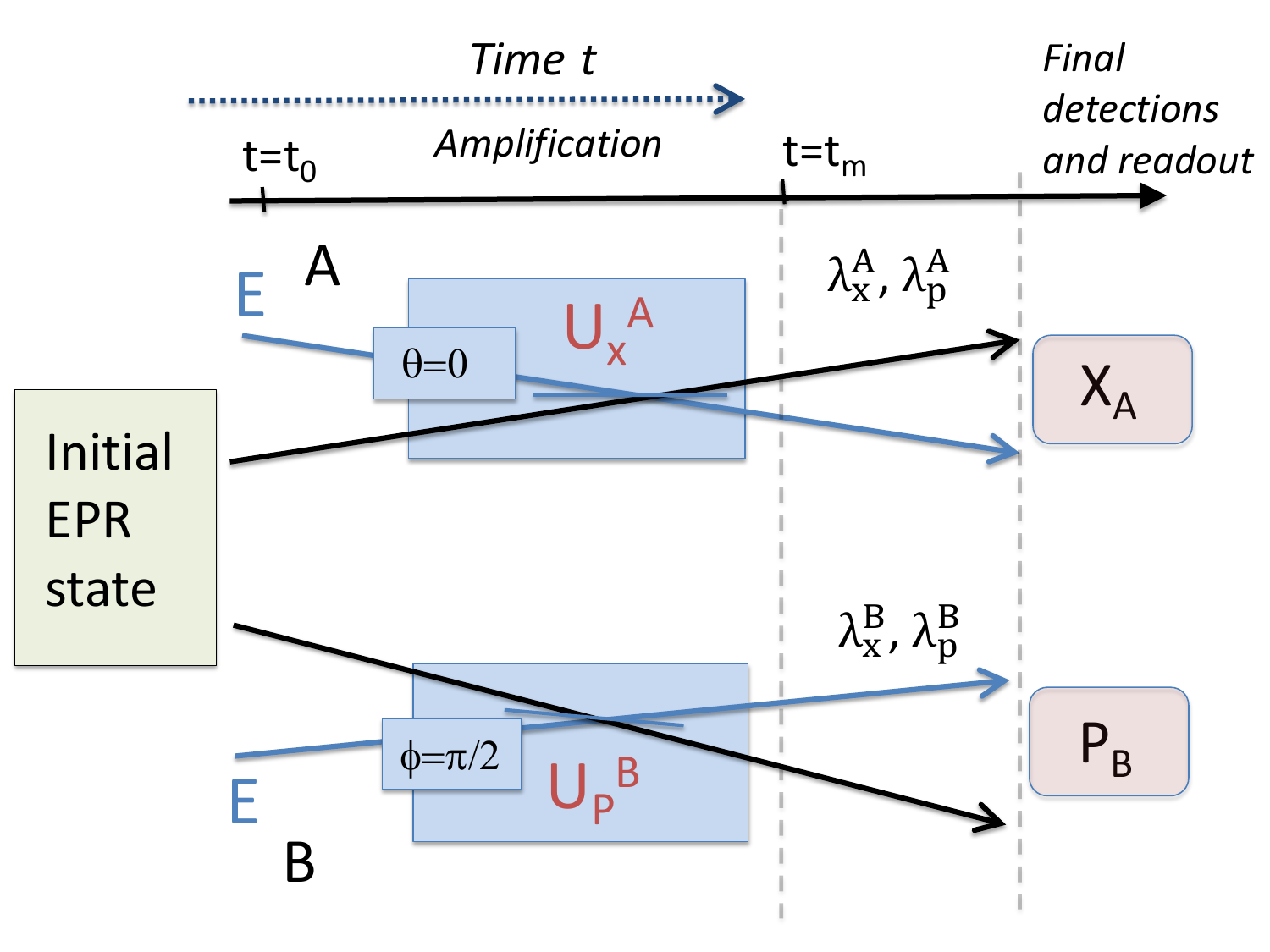}\caption{A diagram of Schr\"{o}dinger's proposed set-up using field quadrature
phase amplitudes $\hat{X}$ and $\hat{P}$ measured by homodyne detection.
The output fields $A$ and $B$ are combined at each site with an
intense coherent field, the local oscillator, $E$. The measurement
setting is determined by the choice of phase shift $\theta$ and $\phi$
of the field relative to $E$ at each site. Here, the choice $\theta=0$
gives measurement of $\hat{X}_{A}$ at $A$; the choice $\phi=\pi/2$
gives measurement of $\hat{P}_{B}$ at $B$. The fields are combined
with $E$ over a beam splitter, leading to the reversible unitary
operations $U_{X}^{A}$ and $U_{P}^{B}$. The final part of the measurement
involves detection of each field and subtraction of resulting photocurrents,
to yield a macroscopic scaled readout of the quadrature amplitudes
at each meter. \label{fig:schematic-lo}}
\end{figure}

Schr\"{o}dinger in his paper gives arguments both for and against
the hypothesis, that there are predetermined value for the outcomes
of the measurement $\hat{x}_{A}$ and $\hat{p}_{A}$, coming to no
definite conclusion. In this section, we analyse the argument put
forward by him that the hypothesis might be negated by consideration
of $\hat{x}_{A}^{2}+\hat{p}_{A}^{2}$. Following Schr\"{o}dinger,
we consider in our formulation $\hat{X}_{A}^{2}+\hat{P}_{A}^{2}$
(Fig. \ref{fig:schematic-lo}). We express this quantity in terms
of the number operator $\hat{n}_{A}=\hat{a}^{\dagger}\hat{a}$, by
expanding out the boson operators. \textcolor{red}{}Manipulation
gives
\begin{eqnarray}
\hat{X}_{A}^{2}+\hat{P}_{A}^{2} & = & (\hat{a}(t)+\hat{a}^{\dagger}(t))^{2}/2-(\hat{a}(t)-\hat{a}^{\dagger}(t))^{2}/2\nonumber \\
 & = & \hat{a}(t)\hat{a}^{\dagger}(t)+\hat{a}^{\dagger}(t)\hat{a}(t)\nonumber \\
 & = & 1+2\hat{a}^{\dagger}(t)\hat{a}(t)\label{eq:sch-sum}
\end{eqnarray}
for which the outcome must always corresponds to an odd number.

\textcolor{red}{}Schr\"{o}dinger's argument is along the following
lines. Assuming the values for $\hat{X}_{A}$ and $\hat{P}_{B}$ are
predetermined, given by $x_{A}$ and $p_{B}$ respectively, then the
outcome for a measurement of $\hat{X}_{A}^{2}$ would be given by
$x_{A}^{2}$, and the outcome for $\hat{P}_{B}^{2}$ given by $p_{B}^{2}$.
Schr\"{o}dinger's set-up constitutes an indirect measurement of $\hat{P}_{A}$.
Assuming the predetermined values, the outcome for $\hat{P}_{A}$
is $-p_{B}$. The outcomes of the measurements of $\hat{X}_{A}$ and
$\hat{P}_{A}$ are continuous-valued. Schr\"{o}dinger's question
was how can this be compatible with $\hat{X}_{A}^{2}+\hat{P}_{A}^{2}$
being restricted to an odd number?

The measurement of $\hat{X}$ or $\hat{P}$ as depicted in Fig. \ref{fig:schematic-lo}
employs a different experimental configuration to that of direct detection,
as used to measure the number $\hat{n}$. The predetermination of
values for $\hat{X}$ and $\hat{P}$ is not necessarily relevant to
that for $\hat{n}$ where a different experimental set-up is used
\citep{bell-contextual,rebuttal-von-neumann-proof-1,rebuttal-von-neumann-proof-2}.
As quoted from Ref. \citep{bell-cs-review}, ``The actual procedure
for measuring $A+B$, when $A$ and $B$ do not commute, is different
from the procedures for measuring $A$ and $B$ separately and does
not presuppose any information about the value of either A or B.''
Bell commented in Ref. \citep{bell-contextual} that ``The result
of an observation may reasonably depend not only on the state of the
system (including hidden variables) but also on the complete description
of the apparatus'' \citep{bell-cs-review}.

However, the set-up of Schr\"{o}dinger offers a different perspective,
because here the observable $\hat{X}_{A}^{2}+\hat{P}_{A}^{2}$ (which
implies measurement of $\hat{n})$ is measured using the \emph{same}
experimental apparatus as for $\hat{X}_{A}$ and $\hat{P}_{A}$, by
the simultaneous measurement of $\hat{X}_{A}$ and $\hat{P}_{B}$.
We will see that while the measurement of $\hat{P}_{B}$ allows a
perfect inference of $\hat{P}_{A}$ for sufficiently large $r$, the
measurement of $P_{B}^{2}$ does not allow a sufficiently accurate
estimation of $\hat{P}_{A}^{2}$ (and hence of $1+2\hat{n}_{A}$)
in absolute terms.

First, we note there is no inconsistency with the prediction from
(\ref{eq:sch-sum}) that, in the limit $r\rightarrow\infty$ where
$g_{0}P_{B}$ accurately gives the outcome for $\hat{P}_{A}$, we
would measure for the expectation values
\begin{equation}
\langle X_{A}^{2}+P_{A}^{2}\rangle\rightarrow\langle X_{A}^{2}+g_{0}^{2}P_{B}^{2}\rangle=\langle1+2\hat{n}_{A}\rangle\thinspace.\label{eq:meancheck}
\end{equation}
The outcome of $\hat{X}_{A}$ is measured by homodyne detection, being
amplified by the local oscillator amplitude $E$, so that the outcome
at the photodetector is $\sim EX_{A}$ (Fig. \ref{fig:schematic-lo})
\citep{bowen-anu,slusher}. The value $X_{A}^{2}$ becomes $E^{2}X_{A}^{2}$
and is the measure of the variance, which (after a renormalisation
which divides by $E^{2}$) is given by $\langle X_{A}^{2}\rangle=\sigma_{X_{A}}^{2}=\frac{1}{2}\cosh2r$.
The value $P_{B}$ at the second detector is similarly amplified by
a second local oscillator field, which we also take to be magnitude
$E$. The variance is $\langle P_{B}^{2}\rangle=\sigma_{P_{B}}^{2}=\frac{1}{2}\cosh2r$.
 Using $g_{0}=\tanh2r$ (Eq. (\ref{eq:g0})), we see that
\begin{eqnarray}
E^{2}\langle X_{A}^{2}+g_{0}^{2}P_{B}^{2}\rangle & = & \frac{E^{2}}{2}(\cosh2r+\tanh^{2}2r\cosh2r)\nonumber \\
 & = & E^{2}\{\cosh2r-\frac{1}{2\cosh2r}\}\thinspace.\label{eq:check}
\end{eqnarray}
The mean number is $\langle\hat{n}_{A}\rangle=\langle\hat{a}^{\dagger}\hat{a}\rangle=\sinh^{2}r$.
We satisfy (\ref{eq:meancheck}) in the limit of large $r$, since
\begin{eqnarray}
E^{2}(1+2\langle\hat{n}_{A}\rangle) & = & E^{2}(1+2\sinh^{2}r)\nonumber \\
 & = & E^{2}\cosh2r\thinspace.
\end{eqnarray}
The ``number $1$'' appearing in the expression (\ref{eq:meancheck})
is measurable in the amplified fields as twice the vacuum quantum
noise, proportional to $E^{2}$ in the detector. This noise level
is calibrated in the experiments, and it is this level which the EPR
correlations are measured against, as the noise in the right-side
of the EPR inequality (\ref{eq:epr-crit}).

To address Schr\"{o}dinger's question, we carefully examine the quantities
measured in the EPR experiment (Fig. \ref{fig:schematic-sch}). The
outputs are $\hat{X}_{A}(t)$, $\hat{P}_{A}(t)$, $\hat{X}_{B}(t)$
and $\hat{P}_{B}(t)$, as given by (\ref{eq:arraysolns-2}). The estimate
of $\hat{P}_{A}(t)$ is $g_{0}\hat{P}_{B}(t)$, where $g_{0}=\tanh2r$.
The error in inferring $P_{A}$ from $g_{0}P_{B}$ is
\begin{eqnarray}
\hat{P}_{A}-g_{0}\hat{P}_{B} & = & (\cosh r+g_{0}\sinh r)\hat{P}_{A}(0)\nonumber \\
 &  & -(\sinh r+g_{0}\cosh r)\hat{P}_{B}(0)\label{eq:2}
\end{eqnarray}
which goes to zero as $r\rightarrow\infty$ when $\cosh\rightarrow\sinh r$.
However, the error in inferring $\hat{P}_{A}^{2}$ from $g_{0}^{2}\hat{P}_{B}^{2}$
is precisely
\begin{eqnarray}
\hat{P}_{A}^{2}-g_{0}^{2}\hat{P}_{B}^{2} & = & (\cosh^{2}r-g_{0}^{2}\sinh^{2}r)\hat{P}_{A}^{2}(0)\nonumber \\
 &  & -(\sinh^{2}r-g_{0}^{2}\cosh^{2}r)\hat{P}_{B}^{2}(0)\label{eq:3}
\end{eqnarray}
which becomes $\hat{P}_{A}^{2}(0)-\hat{P}_{B}^{2}(0)$ for large $r$.
The absolute error in inferring $P_{A}^{2}$ from $P_{B}^{2}$ depends
on the input vacuum fluctuations $P_{A}(0)$ and $P_{B}(0)$ which
are independent, and which do not\emph{ }vanish for $r\rightarrow\infty$.
Mathematically, the paradox arises because $\cosh r-\sinh r\rightarrow0$
as $r\rightarrow\infty$, but always $\cosh^{2}r-\sinh^{2}r=1$. We
expand \emph{$\hat{X}_{A}^{2}+\hat{P}_{A}^{2}$ }as\textcolor{blue}{\emph{
}}\textcolor{red}{}
\begin{eqnarray}
\hat{X}_{A}^{2}(t)+g_{0}^{2}\hat{P}_{B}^{2}(t) & = & \hat{X}_{A}^{2}-\frac{g_{0}^{2}}{2}(\hat{b}(t)-\hat{b}^{\dagger}(t))^{2}\nonumber \\
 & = & \hat{X}_{A}^{2}(t)+g_{0}^{2}\left(\hat{P}_{A}^{2}(t)+\hat{P}_{B}^{2}(0)-\hat{P}_{A}^{2}(0)\right)\nonumber \\
\label{eq:sch-sum-est-1}
\end{eqnarray}
where we substitute for $\hat{P}_{B}$ in terms of $\hat{P}_{A}$.
For large $r$, $\hat{P}_{B}=-\hat{P}_{A},$ since from (\ref{eq:arraysolns-2}),
$\hat{P}_{A}(t)=-\hat{P}_{B}(t)=\hat{P}_{A}(0)-\hat{P}_{B}(0)$, with
$g_{0}\rightarrow1$.  However, for large $r$, there is \emph{infinite
amplification}, so that $\hat{n}_{A}$ is large. It is clear from
(\ref{eq:sch-sum-est-1}) that additional fluctuations are present
at the vacuum level, arising from the initial independent vacuum inputs
of the field modes, which are masked by the amplification which ensures
readouts for $\hat{P}_{A}(t)$ and $\hat{P}_{B}(t)$ \textcolor{blue}{}are
large.

\begin{figure}[t]
\begin{centering}
\includegraphics[width=0.7\columnwidth]{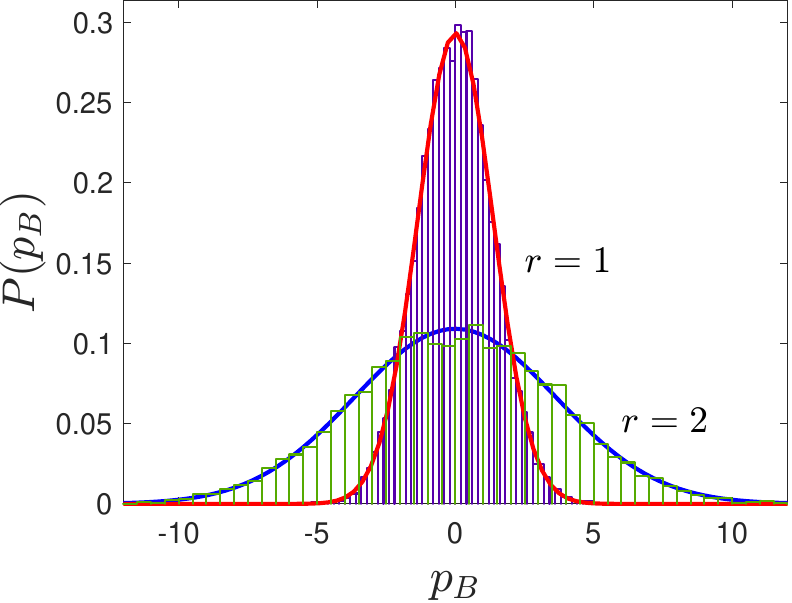}
\par\end{centering}
\centering{}\caption{The probability distribution for the outcome of $p_{B}$ of $\hat{P}_{B}$
is a Gaussian with mean zero and variance $\sigma_{P}$. Here, different
values of the two-mode squeezing parameter $r$ are given. The inference
of $\hat{P}_{A}$ given the measurement of $\hat{P}_{B}$ improves
as $r$ becomes larger.\label{fig:The-probability-distribution}\textcolor{red}{}\textcolor{blue}{}}
\end{figure}

For any finite $r$, there will be an error in estimating $\hat{P}_{A}$
by $\hat{P}_{B}$, as we have seen above. We quantify the error, and
show how this accounts for the discrepancy that concerned Schr\"{o}dinger.
\emph{Any apparent paradox that the value of $\hat{X}_{A}^{2}+\hat{P}_{A}^{2}$
is constrained to be an odd integer, when outcomes for $\hat{X}_{A}$
and $\hat{P}_{A}$ are continuous, can be accounted for by the absolute
error in the measured value of $\hat{X}_{A}^{2}+\hat{P}_{A}^{2}$,
when $\hat{P}_{A}$ is inferred from $\hat{P}_{B}$. The error is
of order $\sim1$, even where the absolute error in the estimate of
$\hat{P}_{A}$ by $\hat{P}_{B}$ becomes negligible.}
\begin{figure}[t]
\begin{centering}
\includegraphics[width=0.7\columnwidth]{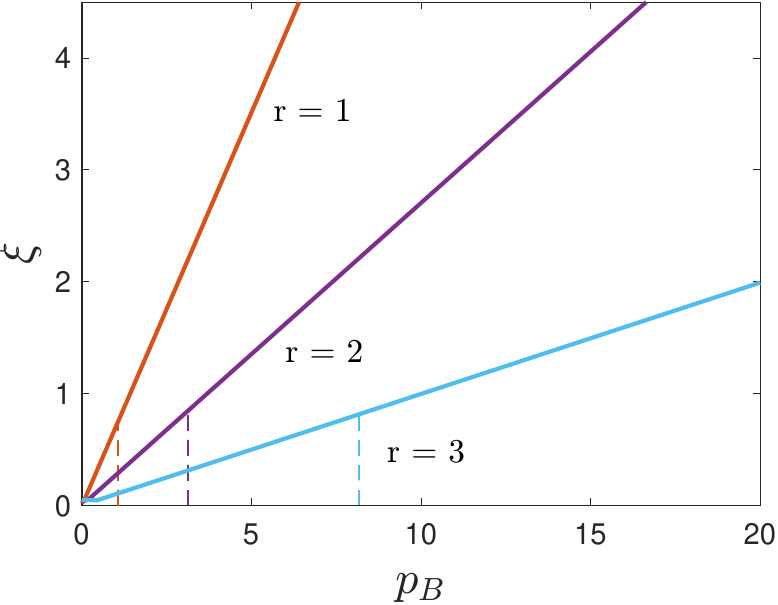}
\par\end{centering}
\bigskip{}

\begin{centering}
\includegraphics[width=0.7\columnwidth]{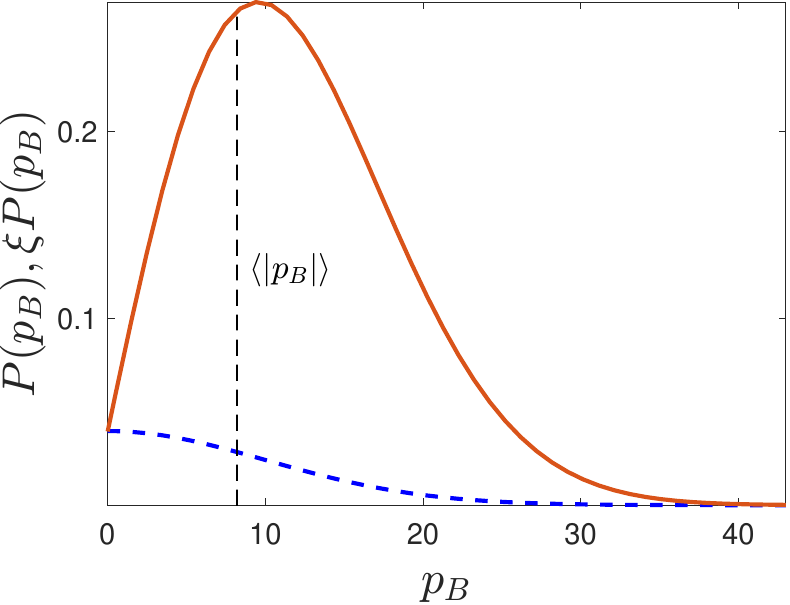}
\par\end{centering}
\centering{}\caption{The top figure plots the absolute error $\xi$ in the inference of
$\hat{P}_{A}^{2}$ given that the outcome of measurement $\hat{P}_{B}$
at $B$ is $p_{B}$, for various values of $r$.\textcolor{red}{}
The lower figure shows the Gaussian distribution $P(p_{B})$ where
$p_{B}\protect\geq0$ (dashed blue curve), for $r=3$. This defines
the half-Gaussian, where $P_{1/2}(p_{B})=2P(p_{B})$, $p_{B}\protect\geq0$.
The product $\xi P(p_{B})$ is also plotted (brown solid curve). The
mean $\langle|p_{B}|\rangle$ of the half-Gaussian is given by the
vertical dashed lines on both figures. \label{fig:estimates-of-error}\textcolor{blue}{}\textcolor{red}{}\textcolor{blue}{}}
\end{figure}

The error in estimating $\hat{P}_{A}$ from measurement of $\hat{P}_{B}$
can be converted to a relative error $e$, and the absolute error
in $\hat{X}_{A}^{2}+\hat{P}_{B}^{2}$, for the given $r$, calculated.
We find the magnitude of the relative error $e$ in the estimate
$p_{A,est}$ of the outcome $p_{A}$ of $\hat{P}_{A}$ is 
\begin{equation}
e=\Delta_{inf}p_{A}/|p_{A,est}|\label{eq:error}
\end{equation}
where here $\Delta_{inf}p_{A}=\frac{1}{\sqrt{2\cosh2r}}$. The relative
error in $p_{A,est}^{2}$ is $2e$. Hence the magnitude $\xi$ of
the absolute error in the estimate of $\hat{P}_{A}^{2}$ is\textcolor{red}{{}
}
\begin{eqnarray}
\xi=2ep_{A,est}^{2} & = & 2(\Delta_{inf}p_{A})|p_{A,est}|\nonumber \\
 & = & \frac{\sqrt{2}}{\sqrt{\cosh2r}}g_{0}|p_{B}|\nonumber \\
 & = & \frac{\sqrt{2}}{\sqrt{\cosh2r}}(\tanh2r)|p_{B}|.\label{eq:final-error}
\end{eqnarray}
Here, $p_{B}$ is the outcome of the measurement of $\hat{P}_{B}$.
We note from the solution (\ref{eq:joint-x}) that the distribution
$P(p_{B})$ for the outcome $p_{B}$ is a Gaussian with mean zero
and variance \textcolor{black}{$\sigma_{P}^{2}=\frac{1}{2}\cosh2r$
(Fig. \ref{fig:The-probability-distribution}):}\textcolor{red}{}\textcolor{red}{\emph{}}\emph{
\begin{equation}
P(p_{B})=\frac{1}{\sqrt{2\pi}\sigma_{p}}e^{-p_{B}^{2}/2\sigma_{p}^{2}}.\label{eq:gausp(x)}
\end{equation}
}We examine the limit of interest, where $r$ is large. In the last
line of (\ref{eq:final-error}), the mean value of $|p_{B}|$ can
be evaluated from the half-Gaussian $P_{1/2}(p_{B})$ defined for
$p_{B}\geq0$, \textcolor{blue}{\emph{}}
\begin{equation}
\langle|p_{B}|\rangle=\frac{\sigma_{P}\sqrt{2}}{\sqrt{\pi}}\rightarrow\frac{e^{r}}{2}\sqrt{\frac{2}{\pi}}=\frac{e^{r}}{\sqrt{2\pi}}.\label{eq:mean-mod}
\end{equation}
We find this corresponds to an error of\textcolor{red}{{} }(as $r\rightarrow\infty$)
\begin{eqnarray}
\xi=2ep_{A,est}^{2} & = & \frac{\sqrt{2}}{\sqrt{\cosh2r}}(\tanh2r)|p_{B}|\nonumber \\
 & \rightarrow & \frac{2}{e^{r}}|p_{B}|=\frac{\sqrt{2}}{\sqrt{\pi}}\sim0.8.\label{eq:large-r-error}
\end{eqnarray}
We see that the error in the mean $p_{A,est}^{2}$ will be of order
$1$, even in the limit where the measurement of $\hat{P}_{A}$ by
$\hat{P}_{B}$ becomes accurate, as $r$ increases. The relative error
$e$ becomes negligible as $r\rightarrow\infty$: $e\rightarrow1/e^{r}|p_{B}|\sim\sqrt{2\pi}e^{-2r}$.
 The Fig.\ref{fig:estimates-of-error} shows the predictions $\xi$
given by (\ref{eq:final-error}) for certain realizable values of
$r$ and for a given outcome $p_{B}$, illustrating consistency with
the results given here in the large $r$ limit.

\section{Incompleteness criterion: weak macroscopic realism and weak local
realism\label{sec:Incompleteness-criterion-based}}

Schr\"{o}dinger's arguments were motivated by EPR's paper and the
EPR premises. However, the EPR argument can be applied to two anti-correlated
spatially-separated spin-$1/2$ systems, prepared in a singlet (Bell)
state \citep{Bohm,bohm-aharonov}. In this case, the EPR premises
imply simultaneous predetermined outcomes for the different components
of the spin of each system, a conclusion that was then negated by
Bell's theorem, which proved that such local hidden variables could
not be compatible with the predictions of quantum mechanics \citep{Bell-2,ghz-1,mermin-ghz,hall,fine}.
Thus, at first glance, Schr\"{o}dinger's arguments appear irrelevant,
since the argument for assuming the predetermined values $x_{A}$,
$p_{A}$, $x_{B}$ and $p_{B}$ is undermined.

\begin{figure}[t]
\begin{centering}
\includegraphics[width=1\columnwidth]{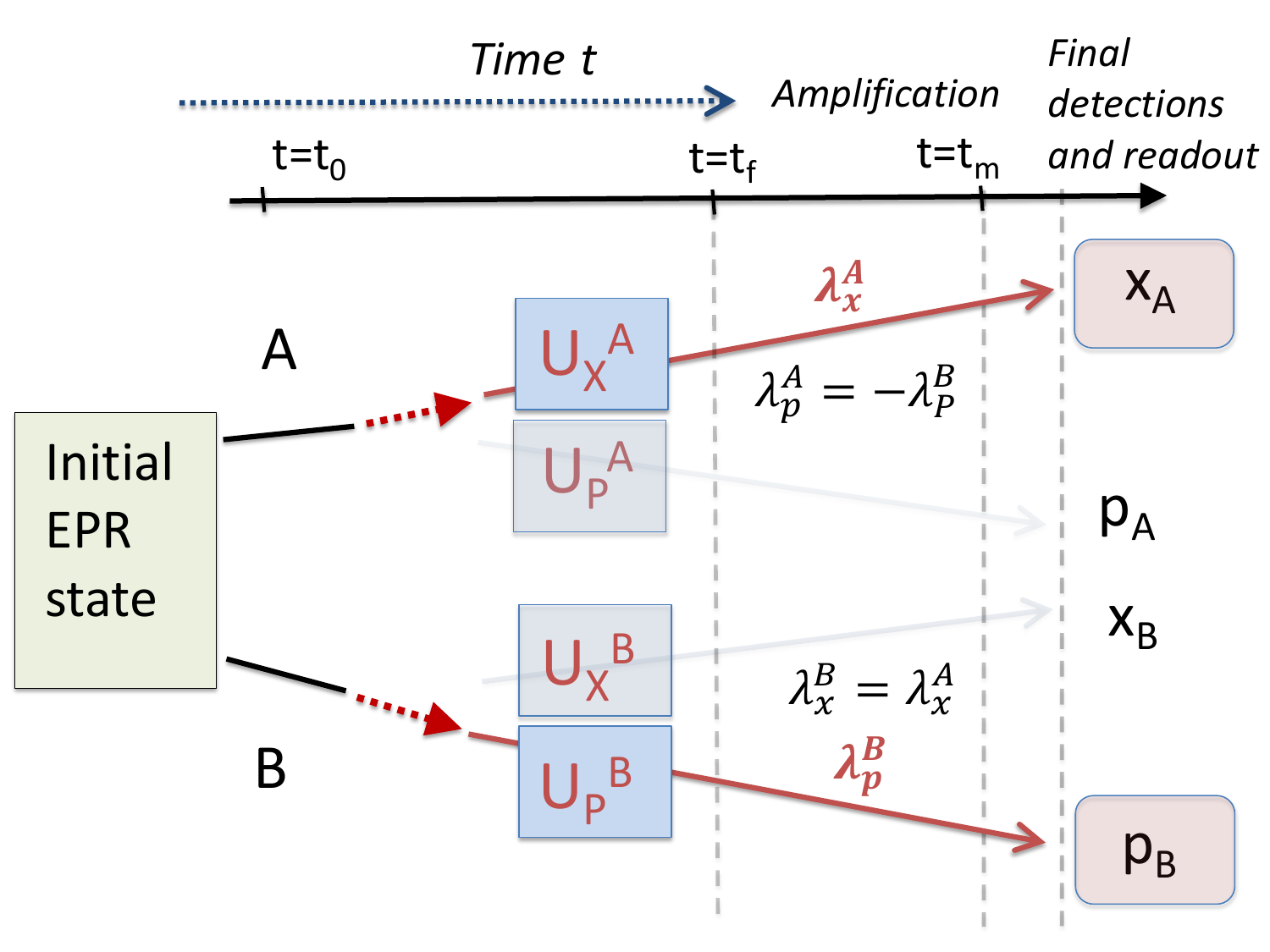}
\par\end{centering}
\centering{}\caption{A diagram showing how the set-up considered by Schr\"{o}dinger implies
an incompleteness of quantum mechanics based on premises not falsifiable
by Bell's theorem. The set-up is as in Fig. \ref{fig:schematic-sch}.
 In Schr\"{o}dinger's proposed experiment, the choice is made to
measure $\hat{x}$ at $A$ and $\hat{p}$ at $B$ (red arrows). This
constitutes simultaneous measurement of $\hat{x}$ and $\hat{p}$,
``one by direct, the other by indirect measurement''. The premises
of weak local realism (wLR)  imply simultaneous values exist at time
$t_{f}$ for the outcomes of the measurements $x_{A}$ and $p_{A}$
(and also of $x_{B}$ and $p_{B}$). The values $\lambda_{x}^{A}$
and $\lambda_{p}^{B}$ (in bold red) are implied by Premise wLR(1);
the values $\lambda_{p}^{A}$ and $\lambda_{x}^{B}$ (in black text)
are implied by Premise wLR(3). The premises of weak macroscopic realism
(wMR) imply similar hidden variables, but at the time $t_{m}$ when
the system is macroscopic. \label{fig:schematic-sch-2}}
\end{figure}

In this paper, we are interested in an alternative EPR-type argument,
which was presented in Ref. \citep{ghz-cat} for macroscopic superposition
states. This argument is based on premises that are \emph{not} falsified
by violation of Bell inequalities, yet which imply the simultaneous
values, $x$ and $p$ (Fig. \ref{fig:schematic-sch-2}). We refer
to these premises as \emph{weak local realism }(wLR). In the stronger
situation where the systems are macroscopic and macroscopic realism
(MR) can be applied, we refer to the premises as \emph{weak macroscopic
realism} (wMR). That the premises of wLR and wMR can also be consistent
with delayed-choice experiments, the three-box-paradox, and Wigner-friend
experiments has been explained in Refs. \citep{wigner-friend-macro,three-box-macro,delayed-choice-cats}.

Schr\"{o}dinger considered the measurement of $\hat{p}_{B}$ at $B$
and $\hat{x}_{A}$ at $A$, which means that the system can be considered
at the times $t_{f}$ (or $t_{m}$), depicted in Fig. \ref{fig:schematic-sch},
\emph{after} the unitary operations that fix the measurement settings.
While the argument that the outcomes for both $\hat{x}_{A}$ and $\hat{p}_{B}$
are simultaneously predetermined was originally based on the EPR premises,
in considering the specific set-up of measurements of $\hat{x}_{A}$
and $\hat{p}_{B}$, Schr\"{o}dinger's questions pertain to the existence
of ``elements of reality'' for the system as it is described \emph{after}
the settings are fixed in the experiment. It is for this set-up that
the weaker premises (wLR and wMR) become relevant.

\subsection{Weak local realism (wLR)}

It is possible to consider a modified set of the EPR premises that
apply to the system considered by Schr\"{o}dinger (Fig. \ref{fig:schematic-sch-2}).
These premises imply ``weak elements of reality''. We follow Ref.
\citep{ghz-cat}, and define three premises, referred to as ``\emph{weak
local realism (wLR)}''. We consider the Bell-EPR experiment, where
there is a measurement-setting device (such as a Stern-Gerlach analyzer
or a polarizing beam splitter) which interacts with the system, thereby
fixing the measurement setting, $\theta$. In the Bell-EPR experiment,
the setting $\theta$ refers to the choice to measure a given spin
component $\hat{S}_{\theta}$. In the original EPR experiment, the
setting determines the choice to measure either $x$ or $p$ (as in
Fig. \ref{fig:schematic-sch-2}). The motivation for the premises
has been discussed in Refs. \citep{ghz-cat,weak-versus-det,wigner-friend-macro}.
The premises are similar to those of EPR and Bell, except that they
refer to systems defined \emph{after} the interactions that fix the
settings, as at time $t_{f}$ in Fig. \ref{fig:schematic-sch-2}.

The terminology ``weak'' is used, since the premises are not sufficient
to imply a Bell inequality. This was shown in Refs. \citep{wigner-friend-macro,weak-versus-det,manushan-bell-cat-lg}
and is illustrated in Figure \ref{fig:schematic-meter}. The premises
are summarized as follows:

\paragraph*{Premise wLR(1): A weak form of realism}

The system $A$ defined at the time $t_{f}$ after the interaction
with the measurement-setting device that fixes the measurement to
be $\hat{S}_{\theta}^{A}$ has a predetermined value $\lambda_{\theta}^{A}$
for the outcome of the measurement, $\hat{S}_{\theta}^{A}$. In Fig.
\ref{fig:schematic-sch-2}, the setting $\theta$ determines whether
$\hat{x}_{A}$ or $\hat{p}_{A}$ will be measured.

\paragraph*{Premise wLR(2): A weak form of locality}

This value is not affected by any interactions or events subsequently
occurring at the spatially separated site.

\paragraph*{Premise wLR(3): A weak EPR criterion for reality}

Suppose the measurement setting has been specified at $B$ i.e. we
consider the system at $B$ after it has interacted with the local
measurement setting device. Suppose the setting is such that the observable
$\hat{S}_{\phi}^{B}$ is to be measured at $B$. In this paper, we
take that $\hat{S}_{\phi}^{B}=\hat{p}_{B}$. If it is possible at
this time (call it $t_{B}$) to predict the value of an observable
$\hat{S}_{\zeta}^{A}$ at $A$ with certainty, based on the final
outcome of $\hat{S}_{\phi}^{B}$ at $B$, then there is an EPR ``element
of reality'' specified for the outcome of $\hat{S}_{\zeta}^{A}$
for system $A$. The value for $\hat{S}_{\zeta}^{A}$ is predetermined,
at the time $t_{B}$, regardless of whether the system $A$ has actually
interacted with the local measurement-setting device at $A$ (in order
to carry out the local measurement of $\hat{S}_{\zeta}^{A}$). In
Schr\"{o}dinger's gedanken experiment, we see that $\hat{S}_{\zeta}^{A}\equiv\hat{p}_{A}$.

\subsection{Weak macroscopic realism}

The Premises can be further strengthened if the system is macroscopic
\citep{ghz-cat,weak-versus-det,wigner-friend-macro,three-box-macro,manushan-bell-cat-lg,delayed-choice-cats}.
There is no strong reason to assume Premise wLR(1), that the outcome
of the measurement is determined at the time $t_{f}$, after the measurement
setting interaction has been completed (although this weak form of
realism is not negated by violation of Bell inequalities \citep{wigner-friend-macro,ghz-cat}).
However, if we consider that the system at $t_{m}$ (defined after
interaction with the measurement-setting device) is in a superposition
(or mixture) of macroscopically distinc\emph{t} states that give a
definite outcome for the measurement $\hat{S}_{\theta}$, then we
can apply the premise of \emph{macroscopic realism}.

Suppose a system $A$ is in a superposition of macroscopically distinct
states, 
\begin{equation}
|\psi\rangle=c_{1}|s_{1}\rangle+c_{2}|s_{2}\rangle,\label{eq:sup1}
\end{equation}
where here $|s_{i}\rangle$ is a macroscopic state giving a definite
outcome $s_{i}$ for $\hat{S}_{\theta}^{A}$, and $|s_{1}-s_{2}|$
is large. For example, it may be assumed that the system has already
interacted with the measurement-setting device, with the setting set
to $\theta$, and that a further amplification, or coupling to a macroscopic
meter, will provide the final readout. Macroscopic realism (MR) asserts
that a system ``with two or more macroscopically distinct states
available to it will at all times \emph{be} in one or other of these
states'' \citep{legggarg-1,emary-review}. This implies that the
outcome of $\hat{S}_{\theta}^{A}$ is predetermined to be one of the
$s_{i}$. We define \emph{weak macroscopic realism (wMR) }according
to the premises below. 

\paragraph*{Premise wMR(1):}

We suppose that a system has at the time $t_{m}$ available to it
two or more macroscopically distinct states $|s_{i}\rangle$, which
have a definite value $s_{i}$ for the outcome of the measurement
$\hat{S}_{\theta}$. This means that the system can be regarded as
being in a mixture or superposition of these states. By considering
the time $t_{m}$ as in Fig. \ref{fig:schematic-sch-2}, we assume
that the system has already been prepared with respect to the measurement
setting $\theta$, so that the values of $s_{i}$ can be directly
measured by an amplification and readout. The premise of (weak) macroscopic
realism asserts that the system can be ascribed a predetermined value
$s_{j}\in\{s_{i}\}$ for the outcome of the measurement $\hat{S}_{\theta}$.

\paragraph*{Premise wMR(2):}

The predetermined value $s_{i}$ defined in premise wMR(1) is not
subsequently affected by any interactions or events at the second
site.

\paragraph*{Premise wMR(3):}

Suppose the measurement setting has been specified at $B$ i.e. we
consider the system at $B$ after it has interacted with the local
measurement setting device. Suppose the setting is such that the observable
$\hat{S}_{\phi}^{B}$ is to be measured and that the system $B$ can
be considered to be in a superposition or a mixture of macroscopically
distinct eigenstates of $\hat{S}_{\phi}^{B}$, so that wMR as defined
in Premise wMR(1) can be applied. If it is possible at this time $T_{B}$
to predict the value of an observable $\hat{S}_{\zeta}^{A}$ at $A$
with certainty, based on the final outcome of $\hat{S}_{\phi}^{B}$
at $B$, then there is an element of reality specified for the outcome
of $\hat{S}_{\zeta}^{A}$ for system $A$, at this time. The value
for $\hat{S}_{\zeta}^{A}$ is predetermined, at the time $T_{B}$,
regardless of whether the system $A$ has interacted with the local
measurement-setting device at $A$ or has been amplified.

\begin{figure}[t]
\begin{centering}
\includegraphics[width=1\columnwidth]{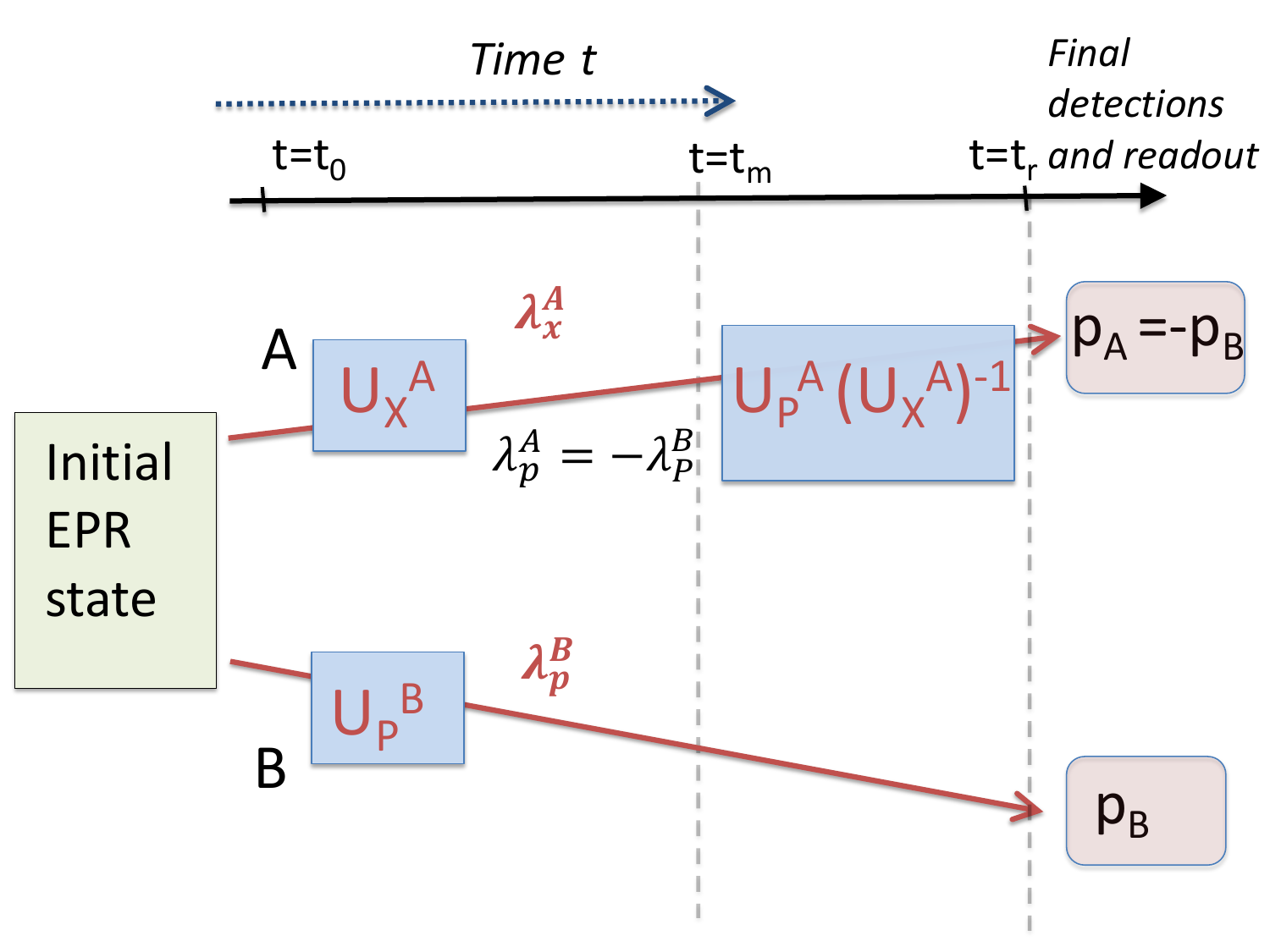}
\par\end{centering}
\centering{}\caption{A diagram of Schr\"{o}dinger's EPR set-up as in Figs. \ref{fig:schematic-sch}
and \ref{fig:schematic-sch-2}, showing how at the time $t_{m}$,
system $B$ acts as a \emph{meter} for $\hat{p}_{A}$ at $A$. According
to Premise wLR(3) (and, if the system is macroscopic, Premise wMR(3)),
the value for the outcome of $\hat{p}_{A}$ is predetermined at this
time, given by $-\lambda_{p}^{A}$. The value of $\lambda_{p}^{A}$
can be revealed at $A$ by the reversal of $U_{x}^{A}$ followed by
$U_{p}^{A}$ and a readout. The outcome will always be $-p_{B}$,
where $p_{B}$ is the outcome of $\hat{p}$ at $B$. However, if a
further unitary operation is performed at $B$ prior to the time $t_{r}$,
to change the setting at $B$, then the system $B$ is no longer a
meter. In that case, the premises of wLR and wMR no longer imply that
the outcome for $\hat{p}_{A}$ is given as $-\lambda_{p}^{B}$. This
contrasts with the conclusions of the original EPR assumptions, which
are stronger, implying simultaneous values $\lambda_{x}^{A}$ and
$\lambda_{p}^{A}$ (defined at time $t_{0}$) that are not changed
by any operations at $B$. Hence, the premises wLR and wMR are not
inconsistent with nonlocality (as in the violation of a Bell inequality).\label{fig:schematic-meter}}
\end{figure}

The premise of macroscopic realism (MR) has been studied previously
in many contexts. Leggett and Garg proposed to test MR by deriving
inequalities that are violated by certain types of macroscopic superposition
states \citep{legggarg-1}. However, the inequalities assumed not
only MR, but also macroscopic noninvasive measurability. Hence, reported
violations of the inequalities (e.g. \citep{macro-cat-ibm,macro-exp-knee,emary-review,macro-lg-cat-states,macro-ali,Maroney})
do not negate wMR.

Similarly, negations of macroscopic realism in the context of violations
of macroscopic Bell inequalities (where measurements distinguish between
macroscopically distinct states) \citep{macro-fuzzy-bell-Jeong,macro-bell-first,macro-lg-cat-states,macro-bell-arora-ali-asadian}
are based on the assumption that the ``elements of reality'' are
defined for the system as it exists \emph{prior} to the interaction
with the measurement-setting devices (e.g. polarizing beam splitters).
This distinguishes a more general definition of macroscopic realism
\citep{manushan-bell-cat-lg}. Hence, the violations of macroscopic
Bell inequalities do not negate wMR (Figure \ref{fig:schematic-meter}).

We also comment that wMR does not posit that the system is in one
of the \emph{states} $|s_{1}\rangle$ or $|s_{2}\rangle$ prior to
the measurement $\hat{S}_{\theta}^{A}$. It is also not assumed that
the ``state'' of the system (at the time $t_{m}$) associated with
a definite value $s_{i}$ need be a \emph{quantum} state. Indeed,
it cannot be, and MR constrained in this way can be negated \citep{manushan-bell-cat-lg,Maroney}.

The premises of wMR can be illustrated by the consideration of a system
$S$ entangled with its measuring device (a meter, $M$), in the state
\citep{ghz-cat}
\begin{eqnarray}
|\psi_{M-S}\rangle & = & \frac{1}{\sqrt{2}}\{|+1\rangle_{M}|\uparrow\rangle_{S}+|-1\rangle_{M}|\downarrow\rangle_{S}\}.\nonumber \\
\label{eq:meter}
\end{eqnarray}
The states $|\pm1\rangle_{M}$ of the meter are macroscopically distinct
and are correlated with the spin-$1/2$ states, $|\uparrow\rangle$
and $|\downarrow\rangle$, of the system. The entangled state is created
at a time $t_{m}$, and any interactions that fix the measurement
settings have already taken place. The value of the meter observable
(whether $+1$ or $-1$) gives the outcome for the spin of the system,
by correlation. Premise wMR(1) implies that the outcome of the meter
is determined at time $t_{m}$, and Premise wMR(3) implies that the
outcome of the spin of system $S$ is also determined at the time
$t_{m}$ (Fig. \ref{fig:schematic-meter}).

\subsection{Application of the wLR and wMR premises to Schr\"{o}dinger's set-up:
an incompleteness argument}

Consider Schr\"{o}dinger's set-up where the setting has been determined
to be $\hat{x}$ at $A$, and $\hat{p}$ at $B$, as in Fig. \ref{fig:schematic-sch-2}.
We consider the time $t_{f}$ when the system at $A$ has interacted
with the local measurement-setting device to prepare for the final
stage of measurement of $\hat{x}_{A}$, and system $B$ has interacted
with a local measurement-setting device to prepare for the final stage
of measurement of $\hat{p}_{B}$. If we assume weak local realism
(wLR), then the system $A$ defined at the time $t_{f}$ has a predetermined
value $x_{A}=\lambda_{x}^{A}$ for the outcome of $\hat{x}$ at $A$,
by Premise wLR(1). Similarly, the system $B$ defined at time $t_{f}$
has a predetermined value $p_{B}=\lambda_{p}^{B}$ for the outcome
of $\hat{p}$ at $B$. In the limit of large $r$, one can infer with
certainty the value of $\hat{p}_{A}$ from the outcome of $\hat{p}_{B}$.
By Premise wLR(3), this means that the outcome of $\hat{p}$ at $A$
is also predetermined, given by $p_{A}=-\lambda_{p}^{B}=-p_{B}$.
Hence, following along the lines of EPR's argument,\emph{ the premises
are sufficient to imply that the values of both $\hat{x}_{A}$ and
$\hat{p}_{A}$ are predetermined }(Fig. \ref{fig:schematic-sch-2}).
The quantum state of system $A$ at time $t_{f}$ would be considered
incomplete, since both of two noncommuting observables $x$ and $p$
are determined at the time $t_{f}$.

The above argument is based on wLR and is strengthened if the experimental
realization involves macroscopically distinct states, which means
we can justify the argument based on wMR. It is usual that further
\emph{amplification of the system will take place}, as part of the
measurement procedure e.g. the signal is combined with the intense
local oscillator field $E$ (Figure \ref{fig:schematic-lo}). Following
Ref. \citep{ghz-cat}, we can define $t_{f}$ (as $t_{m}$) accordingly.

In this section, we establish a \emph{criterion for the incompleteness
argument} that applies in a more general case relevant to an experiment.
In a realistic experiment, the correlation between the systems need
not be maximum, as we see from the solutions (\ref{eq:arraysolns-2})
for finite $r$. Moreover, where amplification occurs, the states
considered for the system may be macroscopic, but not always macroscopically
distinct. We hence require to generalize the premises. There are two
quantities to measure.

\subsubsection{Measuring the uncertainty in the inference of $\hat{P}_{A}$}

First, we consider the error in inferring $\hat{P}_{A}$ via measurement
of $\hat{P}_{B}$. The third premise wMR(3) needs modification for
the case of nonideal correlation. Based on the original modification
of the EPR argument as applied to the two-mode squeezed state, the
premise wLR(3) is extended: The level of predetermination $\sigma_{inf,p_{A}}$
of $\hat{p}_{A}$ is determined by the uncertainty in the inference
of $\hat{p}_{A}$, given the measurement at $B$ i.e. $\sigma_{inf,p_{A}}=\Delta_{inf}p_{A}.$
 The inference variance is measured as \citep{epr-rmp}
\begin{equation}
\sigma_{inf,p_{A}}^{2}=\sum_{J}P_{J}\sigma_{p_{A}|J}^{2}\label{eq:vari-infcond}
\end{equation}
where $P_{J}$ is the probability for an outcome $p_{B,J}$ on measurement
of $\hat{p}_{B}$ at $B$ and $\sigma_{p_{A}|J}^{2}$ is the variance
of the conditional distribution $P(p_{A}|p_{B,J})$.

However, the field at $B$ is amplified after the choice of setting
to measure $P_{B}$ and we prefer to apply the stronger premise of
wMR(3).  In the relevant experiments, the value of $\sigma_{inf,p_{A}}^{2}$
is indeed measured after amplification, which is part of the usual
measurement process, as in homodyne detection (Fig. \ref{fig:schematic-lo}).
It is necessary however to quantify the level at which wMR is applied,
by that meaning to quantify the degree of separation of the states
that are specified to be macroscopically distinct by the premise.
This depends on the nature of the amplification process that is used
and specific models are given in Sections \ref{sec:incompleteness-criterion:-quantu}
and \ref{sec:Amplification-model-for}.

\subsubsection{Measuring the uncertainty associated with ``element of reality''
for $\hat{X}_{A}$}

We also require to quantify the uncertainty $\sigma_{real,x_{A}}$
in the inferred ``element of reality'' for $x$ at $A$, given the
measurements made at $A$ and based on the premise wMR(1). The $\sigma_{real,x_{A}}$
is evaluated by measurement of the distribution $P(x_{A})$, which
is measurable at $A$. We consider three different cases.

\textbf{\emph{Case I: Superpositions of macroscopically distinct eigenstates:}}
We first examine where the system at time $t_{m}$ is in a superposition
of two macroscopically distinct states,
\begin{equation}
|\psi\rangle=\frac{1}{\sqrt{2}}(|x_{1}\rangle+|-x_{1}\rangle)\label{eq:sup2}
\end{equation}
the $|\pm x_{1}\rangle$ being eigenstates of $\hat{x}_{A}$, with
eigenvalues $\pm x_{1}$. The measurement of the distribution $P(x_{A})$
for the outcomes of $\hat{x}_{A}$ gives two distinct separated peaks
each with a variance of $0$. The premise of wMR(1) implies that the
system prior to measurement was in one or other ``state'', meaning
a state that gives a predetermined value for the outcome of $\hat{x}_{A}$
i.e. the system has a well defined outcome $x_{A}$ with value either
$x_{1}$ or $-x_{1}$. Hence, for this case, the premise wMR(1) allows
us to conclude that $\sigma_{real,x_{A}}^{2}=0$.

\textbf{\emph{Case II. Superpositions of macroscopically distinct
states: }}We next consider the system in a superposition of type
\begin{equation}
|\psi\rangle=\frac{1}{\sqrt{2}}(|\psi_{1}\rangle+|\psi_{2}\rangle)\label{eq:sup3}
\end{equation}
where the states $|\psi_{1}\rangle$ and $|\psi_{2}\rangle$ are macroscopically
distinct, but the outcomes for $\hat{x}_{A}$ correspond to a range
of values. Here, we need to generalize the extrapolate the definition
of wMR(1). Suppose the outcomes for $x_{A}$ if the system in the
state $|\psi_{1}\rangle$ are given by the range $R_{1}=[-\infty,x_{1}]$,
and by $R_{2}=[x_{2},\infty]$ if the system is in state $|\psi_{2}\rangle$.
The states $|\psi_{1}\rangle$ and $|\psi_{2}\rangle$ are considered
macroscopically distinct if $|x_{2}-x_{1}|$ is large. In this case,
we may consider the observable $\hat{S}$ defined as $-1$ if the
outcomes of $\hat{S}$ are in the first range $R_{1}$, and $+1$
if in the second range $R_{2}$. The Premise wMR(1) posits that the
system is \emph{either} such that the outcome for $\hat{x}_{A}$ is
predetermined to be in the range $R_{1}$, \emph{or} \emph{else} such
that the outcome is predetermined to be in the range $R_{2}$. The
distributions for each range $R_{I}$ ($I=1,2$) can be determined,
based on the measurement of $P(x_{A})$ for the state $|\psi\rangle$,
and the variance $\sigma_{x_{A}|I}^{2}$ associated with each part
of the distribution (given as $P(x_{A}|S=\pm1)$) evaluated. The variance
$\sigma_{real,x_{A}}^{2}$ is determined as the average 
\begin{equation}
\sigma_{real,x_{A}}^{2}=\sum_{I}P_{I}\sigma_{x_{A}|I}^{2}\label{eq:real-inf}
\end{equation}
where $P_{I}$ is the probability for obtaining an outcome in range
$R_{I}$. The value of $\sigma_{real,x_{A}}^{2}$ is based on the
premise wMR (1) and the \emph{measurable} distribution $P(x_{A})$,
and not assumptions about the specific form of $|\psi_{I}\rangle$.

\begin{figure}[t]
\textcolor{blue}{\includegraphics[width=0.8\columnwidth]{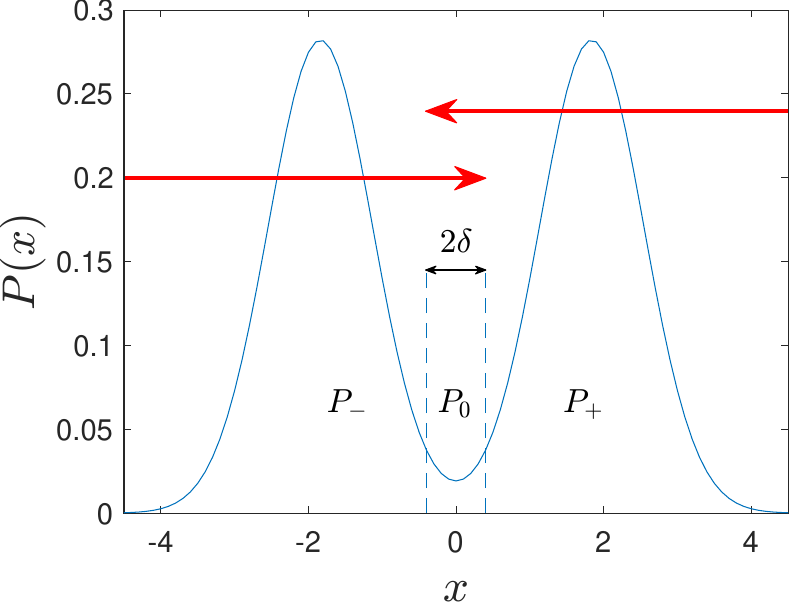}}

\caption{Probability distribution $P(x)$, where $P_{+}$, $P_{-}$ and $P_{0}$
correspond to the probabilities for ranges of outcome given by $x>x_{1}$,
$x<-x_{1}$ and $-x_{1}\protect\leq x\protect\leq x_{1}$ respectively,
where $x_{1}>0$ (so that $\delta=x_{1}$). The two solid red arrows
indicate the overlapping ranges $R_{1}=[-\infty,x_{1}]$ and $R_{2}=[-x_{1},\infty]$.
Here, we depict the distribution $P(x)$ for a system in a cat state,
defined as a superposition $(|\alpha\rangle+i|-\alpha\rangle)/\sqrt{2}$
of two coherent states, where $\alpha=1$. \textcolor{red}{}\textcolor{blue}{}\textcolor{red}{{}
\label{fig:diagram-bins-cat}}}
\end{figure}

\textbf{\emph{Case III: Applying weak macroscopic realism for states
with a macroscopic range:}} A realistic situation often involves a
system in a superposition of states with a wide range of outcomes
for the relevant observable $\hat{X}$, as in Fig. \ref{fig:diagram-bins-cat}
\citep{scat-rmp-frowis,eric-cats-crit}. The outcomes can have a macroscopic
range, but cannot necessarily be categorized into just two macroscopically
distinct parts. Nonetheless, following \citep{eric-cats-crit}, we
can apply wMR in these cases.

We consider a superposition of the form
\begin{equation}
|\psi\rangle\sim|\psi_{-}\rangle+|\psi_{0}\rangle+|\psi_{+}\rangle\label{eq:sup4}
\end{equation}
where the outcomes $x$ of $\hat{x}_{A}$ for $|\psi_{-}\rangle$
and $|\psi_{+}\rangle$ are macroscopically distinct, but those of
$|\psi_{-}\rangle$ and $|\psi_{0}\rangle$ (and of $|\psi_{+}\rangle$
and $|\psi_{0}\rangle$) need not be. We assume that the operations
and interactions that fix the measurement setting to be, for example,
$\hat{x}$ (rather than $\hat{p}$) in the experiment have been carried
out, as depicted by the system at time $t_{m}$ in Fig. \ref{fig:schematic-lo}.
We suppose that the domain of outcomes for $|\psi_{-}\rangle$, $|\psi_{0}\rangle$
and $|\psi_{+}\rangle$ are $[-\infty,x_{1}]$, $[-x_{1},x_{1}]$
and $[-x_{1},\infty]$ respectively ($x_{1}>0$). As an example, Fig.
\ref{fig:diagram-bins-cat} depicts the distribution $P(x)$ obtained
for a system in a superposition of two coherent states $|\alpha\rangle$
and $|-\alpha\rangle$ \citep{yurke-stoler-1,cat-milburn-holmes-1-1,cat-revivals}
(referred to as ``cat states'' \citep{cat-wang-cat,cat-leghtas-cat,cat-states-nobel,cat-states-nobel-2,omran-cats,collapse-revival-bec-2,wrigth-walls-gar-1}).
Although the two states are often considered macroscopically distinct
for $\alpha$ large, there is always an overlap region.

The premise of wMR posits that a system will be \emph{in one or other
of two macroscopically distinct states}. This implies that it is ruled
out that the system can be simultaneously in the two states $|\psi_{-}\rangle$
and $|\psi_{+}\rangle$. Hence, wMR posits that the system is \emph{either}
in a ``state'' $I=1$ with a predetermined outcome in range $\mathcal{R}_{1}=[-\infty,x_{1}]$\emph{,
or else} in a ``state'' $I=2$ with a predetermined outcome in range
$\mathcal{R}_{2}=[-x_{1},\infty]$. Hence, as $x_{1}$ becomes large,
wMR rules out macroscopic superposition states of type $|x_{1}\rangle+|-x_{1}\rangle$,
but not superposition states in general. All superposition states
of type $|x_{2}\rangle+|x_{3}\rangle$ where $|x_{2}-x_{3}|<2|x_{1}|$
are permitted by the description, since they are described by either
$|\psi_{-}\rangle+|\psi_{0}\rangle$ or $|\psi_{0}\rangle+|\psi_{+}\rangle$.
Hence, the value of $2\delta=2|x_{1}|$ determines the ``level''
at which realism is applied. In other words, a loss of ``realism''
at a microscopic level is allowed by the wMR description, but a failure
of macroscopic realism (as in superpositions of $|\psi_{-}\rangle$
and $|\psi_{+}\rangle$) is not allowed. Hence, a system satisfying
wMR (at the level of $2\delta$) can always be described as a mixture
of the states $|\psi_{-}\rangle+|\psi_{0}\rangle$ and $|\psi_{0}\rangle+|\psi_{+}\rangle$.

In this case, bounds on the variances $\sigma_{x_{A}|I}^{2}$ associated
with each range $R_{1}\equiv\mathcal{R}_{-0}$ and $R_{2}\equiv\mathcal{R}_{0+}$
can be determined, based on the probability distribution $P(x_{A})$
for $\hat{x}_{A}$. The $\sigma_{real,x_{A}}^{2}$ is evaluated as
the weighted average 
\begin{equation}
\sigma_{real,x_{A}}^{2}=\sum_{I}P_{I}\sigma_{x_{A}|I}^{2}\label{eq:real-inf-av}
\end{equation}
where $P_{I}$ is the probability the system is in a ``state'' $I$,
so that it is predetermined to give outcomes in range $R_{I}$. Note
that we are unable to determine exact values for $P_{I}$ or $\sigma_{x_{A}|I}^{2}$
from the measured probability distribution $P(x_{A})$, but it becomes
possible to determine bounds on these quantities. As above, the inferred
value of $\sigma_{real,x_{A}}^{2}$ is to be based on the premise
wMR (1) and the measurable distribution $P(x_{A})$, and does not
require knowledge of the underlying state $|\psi\rangle$. The derivation
of bounds on (\ref{eq:real-inf-av}) due to wMR can be presented without
the assumption that the underlying states giving the predetermination
of outcomes need to be quantum states, as we shall see in Section
\ref{sec:incompleteness-criterion:-quantu}.

\subsection{Incompleteness criterion}

We will consider the wMR premises with $\hat{S}_{\theta}^{A}\equiv\hat{x}_{A}$,
$\hat{S}_{\phi}^{B}\equiv\hat{p}_{B}$ and $\hat{S}_{\zeta}^{A}=\hat{p}_{A}$,
which is the situation considered by Schr\"{o}dinger, as in Figures
\ref{fig:schematic-sch} and \ref{fig:schematic-sch-2}. If the variances
$\sigma_{real,x_{A}}^{2}$ and $\sigma_{inf,p_{A}}$ associated with
the range of predetermination implied by wMR for the outcomes of $\hat{S}_{\theta}^{A}\equiv\hat{x}_{A}$
and $\hat{S}_{\zeta}^{A}=\hat{p}_{A}$, respectively, is sufficiently
small, so that
\begin{equation}
\sigma_{real,x_{A}}\sigma_{inf,p_{A}}<1/2,\label{eq:real-inf-crit}
\end{equation}
then we can arrive at Schr\"{o}dinger's conclusion that there are
simultaneously precisely defined values for $x$ and $p$ at $A$,
in a way that is inconsistent with a local representation of a quantum
state for $A$. We refer to this as the ``incompleteness criterion''.
The conclusion is based on the validity of the wMR premises.

\emph{Proof:} We can perform simultaneous measurements of $\hat{x}_{A}$
and $\hat{p}_{B}$, as in Figs. \ref{fig:schematic-sch} and \ref{fig:schematic-sch-2}.
The measurement of $\hat{x}_{A}$ allows us to infer (according to
wMR(1)) that the system $A$ was in a ``state'' $I$ for $x_{A}$,
with associated variance $\sigma_{x_{A}|I}^{2}$ for $x_{A}$ (refer
Eq. (\ref{eq:real-inf-av})). The measurement of $\hat{p}_{B}$ allows
us to infer (according to wMR(3)) that system $A$ was in a ``state''
$J$ for $p_{A}$, with the associated variance $\sigma_{p_{A}|J}^{2}$
for $\hat{p}_{A}$ (refer Eq. (\ref{eq:vari-infcond})). That is,
according to wMR, there is a joint probability that the system $A$
was in a ``state'' with a given $I$ and $J$: For each such state,
the predetermination of $x_{A}$ and $p_{A}$ is given as $\sigma_{x_{A}|I}$
and $\sigma_{p_{A}|J}$. We define the joint ``simultaneous states''
as indexed over $K$, where each $K$ implies a given $I$ and $J$.
We write $K\equiv(I,J)$. A probability $P_{K}$ is defined for each
state, so that $P_{I}=\sum_{J}P_{(I,J)}$ and $P_{J}=\sum_{I}P_{(I,J)}$.
Each state $K$ has a given $I$ and $J$ and hence a well-defined
$\sigma_{x_{A}|I}$ and $\sigma_{p_{A}|J}$. If for each combination
$I$ and $J$, the value $\sigma_{x_{A}|I}\sigma_{p_{A}|J}\geq1/2$,
then it would follow that $\sigma_{real,x_{A}}\sigma_{inf,p_{A}}\geq1/2$.
This follows on considering the Cauchy-Schwarz inequality: 
\begin{eqnarray}
[\sum_{K}P_{K}\sigma_{x_{A}|K}^{2}][\sum_{K}P_{K}\sigma_{p_{A}|K}^{2}] & \geq & [\sum_{K}P_{K}\sigma_{x_{A}|K}\sigma_{p_{A}|K}]^{2}\nonumber \\
 & \geq & [\sum_{K}P_{K}]^{2}/4=1/4.\label{eq:proof1}
\end{eqnarray}
Yet, 
\begin{eqnarray}
\sum_{K}P_{K}\sigma_{x_{A}|K}^{2} & = & \sum_{I,J}P_{(I,J)}\sigma_{x_{A}|(I,J)}^{2}\nonumber \\
 & = & \sum_{I,J}P_{(I,J)}\sigma_{x_{A}|I}^{2}=\sum_{I}P_{I}\sigma_{x_{A}|I}^{2}\nonumber \\
 & = & \sigma_{real,x_{A}}^{2}\label{eq:proof2}
\end{eqnarray}
where we apply (\ref{eq:real-inf-av}) (or \ref{eq:real-inf}) and
have used that $\sigma_{x_{A}|K}^{2}$ depends only on $I$ (not $J$).
Applying similarly for $\sigma_{p_{A}|K}^{2}$, we find 
\begin{eqnarray}
\sum_{K}P_{K}\sigma_{p_{A}|K}^{2} & = & \sum_{I,J}P_{(I,J)}\sigma_{p_{A}|(I,J)}^{2}\nonumber \\
 & = & \sum_{I,J}P_{(I,J)}\sigma_{p_{A}|J}^{2}=\sum_{J}P_{J}\sigma_{p_{A}|J}^{2}\nonumber \\
 & = & \sigma_{inf,p_{A}}^{2}.\label{eq:proof2-1}
\end{eqnarray}
which from (\ref{eq:vari-infcond}) gives the required result. $\square$

\section{incompleteness criterion: quantum predictions\label{sec:incompleteness-criterion:-quantu}}

\textcolor{red}{}We now consider how the predictions of quantum
mechanics can satisfy the incompleteness criterion (\ref{eq:real-inf-crit}).
In this paper, we examine the predictions of the two-mode squeezed
state. We select to measure $X_{A}$ and $P_{B}$ as in Figs. \ref{fig:schematic-sch},
\ref{fig:schematic-lo} and \ref{fig:schematic-sch-2}. We ask how
to justify the incompleteness criterion based on wMR.

\subsection{Amplification by homodyne detection: estimating $\sigma_{inf,P_{A}}^{2}$}

The quadrature phase amplitudes are amplified by the homodyne detection
process, depicted in Fig. \ref{fig:schematic-lo}. Suppose we select
to measure $\hat{X}_{A}$ and $\hat{P}_{B}$, with the choice of phases
$\theta=0$ and $\phi=\pi/2$ (relative to the pump phase determined
by (\ref{eq:ham-tmss})). The outputs of the balanced homodyne detectors,
where the intensities at the two output ports are subtracted, are
proportional to $E\hat{X}_{A}$ and $E\hat{P}_{B}$ respectively,
where $E^{2}$ is the intensity of the local oscillator at each site.
The factor $E$ hence represents an amplification factor, which we
generally call $G$. The amplification means that a continuous-variable
value $EX_{A}$ (or $EP_{B}$) is detected.

Consider the measurement and detection of the quadrature amplitude
$X$, denoted $X(t_{m})$ after amplification. The distribution $P(X)$
for outcomes $X(t)$ can be measured experimentally. The outcomes
$X(t_{m})=EX$ are binned into regions. The outcome $X_{I}$ corresponds
to the band of outcomes $X(t_{m})$ between $X_{I}-\Delta/2$ and
$X_{I}+\Delta/2$. With $E$ sufficiently large, $\Delta$ can be
arbitrarily large in absolute terms. We consider an outcome $X_{K}(t_{m})$
adjacent to $X_{I}(t_{m})$ in the binning process. Following the
procedure of Fig. \ref{fig:diagram-bins-cat}, we will justify that
the outcomes $X_{K}(t_{m})$ and $X_{I}(t_{m})$ are macroscopically
distinct with sufficient amplification, by defining overlap regions
that extend beyond the bins defined by $X_{K}$ and $X_{I}$, by an
amount $\delta$. The overlap region of width $2\delta$ is small
compared to $\Delta$, and defines the fuzzy ``edge'' of the bins.
As $E$ increases, the probability of obtaining an outcome in the
overlap region is negligible, justifying that the outcomes in the
different bins are macroscopically distinct (and can be considered
to correspond to macroscopically distinct states) at this time $t_{m}$.

The premise of weak macroscopic realism (wMR) can then be applied,
to posit predetermined outcomes for the measured quantities. Suppose
$\hat{P}_{B}$ is measured at a site $B$. The amplified quantity
$\hat{P}_{B}(t_{m})$ is detected directly to give a readout of $EP_{B}$
at $B$. The outcome $P_{B}$ is inferred directly from the detected
value by dividing by $E$. According to the premise of wMR(1) as applied
to the system $B$ at time $t_{m}$, the value of $\hat{P}_{B}(t_{m})$
is predetermined at that time, given by the variable $P_{B}(t_{m})$,
say. This implies that the outcome for $\hat{P}_{B}$ is given by
$P_{B}(t_{m})/E$, and is hence \emph{also} predetermined at that
time. Similarly, the premise of wMR(1) justifies a predetermination
of the outcome of $X_{A}$ at the time $t_{m}$, after amplification.

It is possible to measure $\hat{P}_{B}$ and $\hat{P}_{A}$. The measurement
takes place, selecting $\theta=\phi=\pi/2$ (Fig. \ref{fig:schematic-lo}).
At the time $t_{m}$, both fields are amplified by $E$ and detected,
so that $\sigma_{inf,P_{A}}^{2}$ can be measured. The measurement
is of the amplified quantities, so that
\begin{equation}
\sigma_{inf,P_{A}}^{2}=\langle[P_{A}(t_{m})-g_{0}P_{B}(t_{m})]^{2}\rangle/E^{2}.\label{eq:inf-amp-P}
\end{equation}
The measurement is predicted to yield $\sigma_{inf,P_{A}}^{2}=1/(2\cosh2r)$.
According to wMR(3), the value for $P_{A}$ is predetermined to the
level of $\sigma_{inf,P_{A}}^{2}$ at the time $t_{m}$, when $P_{B}$
is amplified, regardless of whether $P_{A}$ or $X_{A}$ is measured
at $A$. At this time $t_{m}$, an ``element of reality'' exists
for $P_{A}$.

We suppose that an experiment yields the value $\sigma_{inf,P_{A}}^{2}$
on measurement with amplification, as given by (\ref{eq:inf-amp-P}).
Then wMR(3) justifies the assumption that the value of $P_{A}$ is
predetermined to this level, and the value $\sigma_{inf,P_{A}}^{2}$
can be used to test the incompleteness criterion (\ref{eq:real-inf-crit}).
It is important to quantify the level $2\delta$, of ``macroscopic
distinctness'', that is assumed in the application of the premise
wMR. This can be estimated by the value of $E$, and will be explained
in Section \ref{sec:Amplification-model-for}.

\subsection{Estimating $\sigma_{real,X_{A}}^{2}$}

To test the incompleteness criterion, we also require to estimate
$\sigma_{real,X_{A}}^{2}$. The same approach as in Section \ref{sec:incompleteness-criterion:-quantu}.A
can be used. However, it is useful to see that larger bin widths $\Delta$
are possible.

We first examine a simple example where we divide the outcome domain
for the field quadrature phase amplitude $\hat{X}$ at $A$ into two
bins, giving positive and negative outcomes. Surprisingly, this restriction
on the knowledge of $X$ is enough to satisfy the incompleteness criterion,
as we will show. The distribution for $X$ is already amplified, as
given by Eq. (\ref{eq:arraysolns-2}). The distribution $P(X)$ is
Gaussian with variance $\sigma_{X_{A}}^{2}=\frac{1}{2}\cosh2r$. 

\begin{figure}[t]
\textcolor{blue}{\includegraphics[width=0.8\columnwidth]{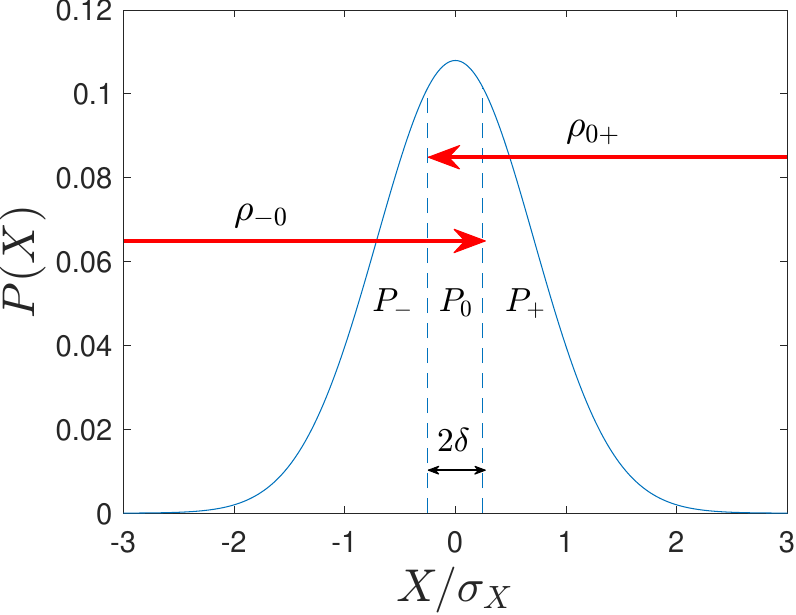}}

\caption{Probability distribution $P(X)$, where $P_{+}$, $P_{-}$ and $P_{0}$
correspond to the probabilities for ranges of outcome given by $X>x_{1}$,
$X<-x_{1}$ and $-x_{1}\protect\leq X\protect\leq x_{1}$ respectively
(here, $x_{1}=\delta>0$). The variance of $P(X)$ is $\sigma_{X}^{2}=\frac{1}{2}\cosh2r$.\textcolor{red}{{}
} Here, we take $x_{1}\equiv\delta=0.25\sigma_{X}$, which as $r$
increases becomes large in absolute terms. The value of $2\delta=2|x_{1}|$
determines the level at which weak macroscopic realism (wMR) is applied.
Where $x_{1}$ is large, the premise of wMR posits that the system
is \emph{either} in a state with range $[-\infty,x_{1}]$ (indicated
by the left red arrow) \emph{or} a state with with range $[-x_{1},\infty]$
(indicated by the right red arrow).  A larger $\delta=|x_{1}|$ corresponds
to a stronger justification of the wMR premise.\textcolor{red}{\label{fig:diagram-bins}}\textcolor{blue}{}\textcolor{red}{}}
\end{figure}

We follow Sec. \ref{sec:Incompleteness-criterion-based}.C.2 and
divide the outcomes $X$ for $\hat{X}_{A}$ into three regions (denoted
$+$, $0$ and $-$) as in the Fig. \ref{fig:diagram-bins}. The probability
for obtaining a value $X$ in each is denoted as $P_{+}$, $P_{-}$
and $P_{0}$, where $P_{+}$ is the probability for outcome $X\geq x_{1}$,
$P_{-}$ is the probability for outcome $X\leq-x_{1}$, and $P_{0}$
is the probability for outcome $-x_{1}\leq X\leq x_{1}$. Here, we
choose $x_{1}$ so that $P_{0}$ is small, but $r$ is large so that
the spread of outcomes enables $\delta=|x_{1}|$ to be large in absolute
terms.\textcolor{red}{}

The premise of wMR posits generally the a system will be \emph{in
one or other of two macroscopically distinct states}. As explained
in Sec. \ref{sec:Incompleteness-criterion-based}.C.2, \textcolor{blue}{}the
motivation for considering the overlap region $0$ is to allow application
of the wMR premise. The premise of wMR posits that the system prior
to measurement is in one of two ``states'', which we designate for
convenience by $\rho_{0+}$ and $\rho_{-0}$, where $\rho_{0+}$ gives
outcomes for $\hat{X}$ greater than $-x_{1}$, and $\rho_{-0}$ gives
outcomes for $\hat{X}$ less than $x_{1}$ (Fig. \ref{fig:diagram-bins}).
Here, we symbolize the ``states'' by $\rho_{-0}$ and $\rho_{0+}$
but it is \emph{not} intended that these ``states'' be restricted
to be quantum states. The states simply reflect a predetermination
of the outcomes for $\hat{X}$, that the outcomes are in one or other
of the two ranges, $R_{1}\equiv\mathcal{R}_{10}=[-\infty,x_{1}]$
and $R_{2}\equiv\mathcal{R}_{02}=[-x_{1},\infty]$. In particular,
the assumption wMR (at the level of $2\delta$) means that the system
is \emph{either} in a ``state'' $\rho_{-0}$ \emph{or} in a ``state''
$\rho_{0+}$. Hence, it is ruled out that the system can be simultaneously
in two ``states'' giving outcomes $X_{0}$ and $-X_{0}$ respectively,
where $X_{0}$ is larger than $2\delta=2x_{1}$ \citep{eric-cats-crit}.
Where $\rho_{0+}$ and $\rho_{-0}$ are quantum states, wMR rules
out superpositions of states $|X_{0}\rangle$ and $|-X_{0}\rangle$,
where $|X\rangle$ is an eigenstate of $\hat{X}$ (with eigenvalue
$X$). As $x_{1}$ becomes large in an absolute sense, wMR rules out
macroscopic superposition states, but \emph{not }superposition states
in general, and is consistent with the system being described as a
mixture of $\rho_{-0}$ and $\rho_{0+}$.

The incompleteness criterion (\ref{eq:real-inf-crit}) requires us
to evaluate $\sigma_{real,X_{A}}^{2}$, given wMR. We use (\ref{eq:real-inf-av})
and take $I=1$ and $2$ for $\rho_{-0}$ and $\rho_{0+}$ respectively.
According to wMR, the system is \emph{either} in $\rho_{0+}$ \emph{or}
$\rho_{-0}$ i.e. the ensemble is described as a mixture of such states.
We denote the variance in $X$ for the system given by $I$ as $\sigma_{X_{A}|I}^{2}$.
Then, using Eq. (\ref{eq:real-inf-av}), we require to evaluate
\begin{equation}
\sigma_{real,X_{A}}^{2}=\sum_{I=1,2}P_{I}\sigma_{X_{A}|I}^{2},\label{eq:real-inf-av-1}
\end{equation}
where $P_{1}$ and $P_{2}=1-P_{1}$ are the probabilities that the
system is in the state designated $\rho_{-0}$ and $\rho_{+0}$ respectively.
We will see that the variances of the distributions given by $\rho_{0+}$
and $\rho_{-0}$ are bounded above, which allows the incompleteness
criterion to be satisfied.

A limit of interest to us is where $x_{1}$ is large, but the ratio
$x_{1}/\sigma_{X}$ is small. Given the $X$ are further amplified
by the factor $E$ (via homodyne detection) as explained in Section
\ref{sec:incompleteness-criterion:-quantu}.A, it can be justified
that the positive and negative bins ($I=1,2$) become macroscopically
distinct for large $E$. Examining $P(X)$ as in Fig. \ref{fig:diagram-bins},
it is clear that as $x_{1}/\sigma_{X}\rightarrow0$, the variances
$\sigma_{X_{A}|I}^{2}$ for each $I=1,2$ correspond to the variance
of the half-Gaussian $P_{1/2}(X)$, given by 
\begin{equation}
Var_{1/2}=\sigma_{X}^{2}(1-\frac{2}{\pi}).\label{eq:var-half}
\end{equation}
Since $\sigma_{X}^{2}=\frac{1}{2}\cosh2r$, $\sigma_{real,X_{A}}^{2}\rightarrow\frac{1}{2}(1-\frac{2}{\pi})\cosh2r$.
If we take $\sigma_{inf,P_{A}}^{2}=1/(2\cosh2r)$, then the product
of the variances becomes
\begin{equation}
\sigma_{real,X_{A}}\sigma_{inf,P_{A}}\rightarrow\frac{1}{2}(1-\frac{2}{\pi})^{1/2}\label{eq:test}
\end{equation}
so that the incompleteness criterion (\ref{eq:real-inf-crit}) is
satisfied.

One can also consider $x_{1}/\sigma_{X}$ to be finite, and give a
quantification of the degree $\delta\equiv|x_{1}|$ at which macroscopic
realism is assumed, in order to bound $\sigma_{real,X_{A}}$, for
the state prior to amplification $E$. We denote the full Gaussian
$P(X)$ by $P_{g}(X)$, and divide into the three regions as in Fig.
\ref{fig:diagram-bins}. In practice, in an experiment, the distribution
$P(X)$ would be measured. Suppose that the distribution for the
underlying state $\rho_{-0}$ is $P_{1}(X)$, and that for $\rho_{0+}$
is $P_{2}(X)$. While we cannot deduce these distributions by measurement,
it is possible to derive an upper bound $U_{B}$ on the associated
variances $\sigma_{X_{A}|I=1}^{2}$ and $\sigma_{X_{A}|I=2}^{2}$.
Derivations are given in the Appendix B.

\section{Amplification model for measurement\label{sec:Amplification-model-for}}

It is possible to model the amplification of the quadrature phase
amplitudes using degenerate parametric amplification, so that the
values of $\delta$ can be more carefully quantified, and the measurement
process simulated. The model can be experimentally realized.

In the model (Fig. \ref{fig:schematic-sch-2}), we consider that after
the two-mode squeezed state has been generated and the two fields
spatially separated, there is a further local amplification of the
quadrature ($X$ or $P$) of the fields according to 
\begin{equation}
H_{A}=i\hbar\frac{g_{A}}{2}(\hat{a}^{\dagger2}-\hat{a}^{2})\label{eq:HampA}
\end{equation}
where $g_{A}>0$ for mode $A$ and 
\begin{equation}
H_{B}=i\hbar\frac{g_{B}}{2}(\hat{b}^{\dagger2}-\hat{b}^{2})\label{eq:HampB}
\end{equation}
where $g_{B}<0$ for mode $B$. The initial system is given by the
two-mode squeezed state, generated at time $t$ in the solutions of
(\ref{eq:arraysolns-2}). The final solutions at a time $t_{m}$,
after amplification for a time $T$, are
\begin{eqnarray}
\hat{X}_{A}\left(t_{m}\right) & = & \hat{X}_{A}\left(t\right)e^{g_{A}T}\nonumber \\
\hat{P}_{A}\left(t_{m}\right) & = & \hat{P}_{A}\left(t\right)e^{-g_{A}T}\nonumber \\
\hat{X}_{B}\left(t_{m}\right) & = & \hat{X}_{B}\left(t\right)e^{g_{B}T}\nonumber \\
\hat{P}_{B}\left(t_{m}\right) & = & \hat{P}_{B}\left(t\right)e^{-g_{B}T}.\label{eq:amp-sol}
\end{eqnarray}
\textcolor{blue}{}The choice to amplify (and hence to measure) either
$X$ or $P$ at each site is determined by the signs of $g_{A}$ and
$g_{B}$. Where $g_{A}>0$, $X_{A}$ is amplified, and hence measured.
If $g_{B}<0$, then $P_{B}$ is amplified, and hence measured. The
interactions $H_{A}$, $H_{B}$ and the solutions (\ref{eq:amp-sol})
are those that model the squeezing of quantum fluctuations for the
single field modes \citep{yuen,wu-squeezing-exp}. The final solutions
are
\begin{eqnarray}
\hat{X}_{A}\left(t_{m}\right) & = & e^{g_{A}T}(\cosh rX_{A}\left(0\right)+\sinh rX_{B}\left(0\right))\nonumber \\
\hat{P}_{A}\left(t_{m}\right) & = & e^{-g_{A}T}(\cosh rP_{A}\left(0\right)-\sinh rP_{B}\left(0\right))\nonumber \\
\hat{X}_{B}\left(t_{m}\right) & = & e^{g_{B}T}(\cosh rX_{B}\left(0\right)+\sinh rX_{A}\left(0\right))\nonumber \\
\hat{P}_{B}\left(t_{m}\right) & = & e^{-g_{B}T}(\cosh rP_{B}\left(0\right)-\sinh rP_{A}\left(0\right)).\nonumber \\
\label{eq:solnsarray-3}
\end{eqnarray}
At the stage denoted by the time $t_{m}$, the operations are fully
reversible. The distributions for the final outcomes of the amplified
quadratures at $A$ and $B$ are Gaussian, with the amplified variances
of $\sigma_{amp}^{2}=\frac{1}{2}e^{2|g_{A}|T}\cosh2r$ and $\frac{1}{2}e^{2|g_{B}|T}\cosh2r$.
\textcolor{red}{}We will take $|g_{A}|=|g_{B}|=|g|$. The amplification
factor is $G=e^{|g|T}$ in the above amplification model. This model
is useful, because of the simulation presented in Section \ref{sec:Simulation}.

Macroscopic realism is applied to the systems at the time $t_{m}$,
to posit predetermined outcomes for the measured quantities. Suppose
$\hat{X}$ is measured at a site $A$. The amplified quantity $\hat{X}_{A}(t_{m})$
is later detected directly to give a readout of $e^{|g|T}X_{A}(t)$
at $A$. The outcome $X_{A}(t)$ is inferred directly from the detected
value by dividing by $G$. According to the premise of weak macroscopic
realism (wMR) as applied to the system $A$ at time $t_{m}$, the
value of $\hat{X}_{A}(t_{m})$ is predetermined at that time, given
by the variable $x_{A}(t_{m})$, say. This implies that the outcome
for $\hat{X}_{A}(t)$ is given by $x_{A}(t_{m})/G$, and is hence
\emph{also} predetermined at that time. \textcolor{black}{}
\begin{figure}[t]
\begin{centering}
\textcolor{black}{}
\par\end{centering}
\centering{}\textcolor{black}{\includegraphics[width=0.8\columnwidth]{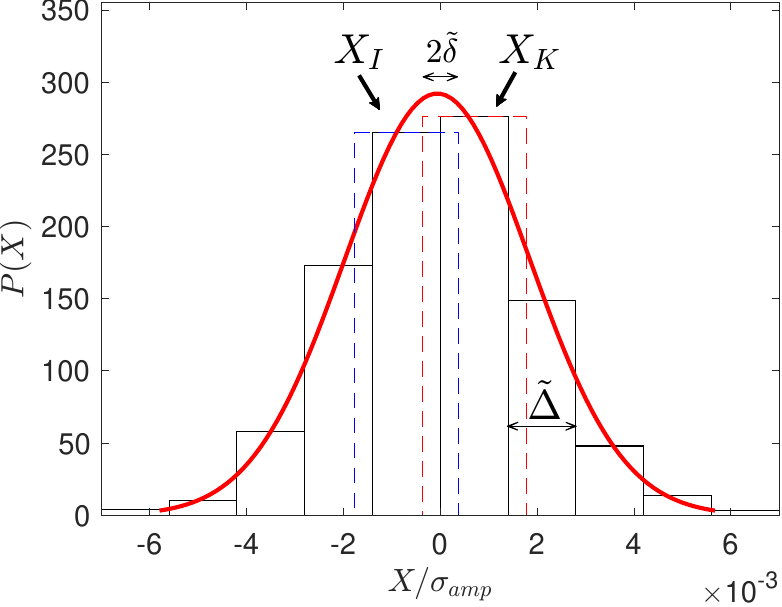}\caption{The distribution for $P(X)$ at the time $t_{m}$, after direct amplification
of quadrature $X$ by a factor $G$. The variance is $\sigma_{amp}^{2}=G^{2}\sigma_{X}^{2}$
where $\sigma_{X}^{2}=\frac{1}{2}\cosh2r$. We divide the domain into
bins of width $\widetilde{\Delta}=\Delta/\sigma_{amp}$. The overlap
regions are of width $2\widetilde{\delta}=2\delta/\sigma_{amp}$ as
indicated by the dashed lines. The outcomes $X_{I}$ and $X_{K}$
are justified to be macroscopically distinct, in the limit of large
amplification. The measure of the level of ``macroscopic distinctness''
is given by $\delta$.  \textcolor{red}{\label{fig:Amplification-multiple-bin}}\textcolor{blue}{{}
}}
}
\end{figure}

To be more precise, we refer to Fig. \ref{fig:Amplification-multiple-bin},
where there has been an amplification of $X$ by a fixed value $G$.\textcolor{blue}{\emph{}}
The distribution can be measured experimentally. The outcomes $X(t_{m})$
are binned into regions. The outcome $X_{I}$ corresponds to the band
of outcomes $X(t_{m})$ between $X_{I}-\Delta/2$ and $X_{I}+\Delta/2$.
With $g$ sufficiently large, $\Delta$ can be arbitrarily large in
absolute terms. We consider an outcome $X_{K}(t_{m})$ adjacent to
$X_{I}(t_{m})$ in the binning process. Following the procedure of
Fig. \ref{fig:diagram-bins}, we will justify that the outcomes $X_{K}(t_{m})$
and $X_{I}(t_{m})$ are macroscopically distinct (as quantified by
the value $2\delta$), by defining overlap regions that extend beyond
the bins defined by $X_{K}$ and $X_{I}$, by an amount $\delta$,
as shown by the blue and red dashed lines in Fig. \ref{fig:Amplification-multiple-bin}.

We consider the domain $-\infty<X<\infty$ divided into an infinite
number of such bins, indexed by the symbol $J$. Assuming wMR (at
a level $2\delta)$, the system can be assumed to be in one of the
set of ``states'', symbolized by $\rho_{J\delta}$, which have a
predetermination of outcomes for $X$ in the region given by $X_{J}-\Delta/2-\delta\leq X\leq X_{J}+\Delta/2+\delta$.
Here, $\delta<\Delta$. As explained in the Sections \ref{sec:Incompleteness-criterion-based}
and \ref{sec:incompleteness-criterion:-quantu}, this assumption,
if $\rho_{J\delta}$ are quantum states, allows \emph{all} superpositions
of the type $|X_{i}\rangle+|X_{j}\rangle$ (where $|X_{i}\rangle$
is an eigenstate of $X$) where $|X_{i}-X_{j}|\leq2\delta$, since
such a superposition can be given by one of the $\rho_{K\delta}$.
The description however does not allow for superpositions across the
bins (e.g. $X_{I}$ and $X_{K}$ in the Fig. \ref{fig:Amplification-multiple-bin})
where $|X_{i}-X_{j}|$ exceeds $2\delta$. The assumption of wMR does
not require to assume that the predetermined ``states'' $\rho_{J\delta}$
are necessarily quantum states: It is simply posited that the system
is always in a state where there is a predetermination on the range
of outcomes, so the system cannot be regarded as simultaneously being
in two ``states'' with a separation of outcomes of more than $2\delta$.

Hence, wMR implies an upper bound on the variance $Var_{X_{J}}$ for
the ``states'' $\rho_{J\delta}$ associated with these predetermined
values: We also note that all superposition of states where $|X_{i}-X_{j}|>\Delta+2\delta$
are excluded by the assumption. We find \citep{exp-variance}
\begin{equation}
Var_{X_{J}}\leq(\Delta+2\delta)^{2}/4\thinspace.\label{eq:var-5}
\end{equation}
Here, the overlap $\delta$ can be large as |$g|\rightarrow\infty$.
The \emph{inferred} value for the outcome $X_{A}$ of $\hat{X}_{A}$
(if the amplified value is detected in the bin given by $X_{J}$)
is $X_{J}/G$. This leads to an upper bound on the inferred variance
$\sigma_{X_{A}|J}^{2}$ given by
\begin{equation}
\sigma_{X_{A}|J}^{2}\leq(\Delta+2\delta)^{2}/4G^{2}\thinspace.\label{eq:var-6}
\end{equation}
The model ensures $G$ is large so that $\sigma_{X_{A}|J}^{2}$ is
small. This is a realistic model for the measurement of $\hat{X}_{A}$.
We see that the binned outcomes $X_{I}$ and $X_{K}$ at the time
$t_{m}$ can be regarded as macroscopically distinct with sufficient
amplification, since the $\delta$ can be large in absolute terms
and yet small relative to $\Delta$, so that the probability of detecting
a result in the overlap regions can be made increasingly negligible
(in the ideal limit where $G\rightarrow\infty$). The binned outcomes
imply the readout values of $X_{I}/G$ and $X_{K}/G$.

We consider Schr\"{o}dinger's set-up, where a measurement of $X$
at $A$ is considered. The premise of macroscopic realism posits that
the outcome for $X$ is determined to the level that the system is
in one or other of the states $\rho_{I\delta}$ at the time $t_{m}$
prior to a final readout. The precision for the actual inferred result
of the measurement based on the value $X_{I}$ in a binned region
is 
\begin{equation}
\sigma_{real,X_{A}}^{2}=\sum_{J}P_{J}\sigma_{X_{A}|J}^{2}\leq(\Delta+2\delta)^{2}/4G^{2}\label{eq:var-x-part}
\end{equation}
(refer Eq. (\ref{eq:real-inf-av})). This gives the level of predetermination
for the outcome $X$, in the measurement.

Now, we require also to consider the measurement of the inferred variance
$\sigma_{inf,P_{A}}^{2}$.  If we measure $P$ at $A$, then both
fields $P_{A}$ and $P_{B}$ are amplified. Let us consider measurement
of $P_{B}$ at $B$. The amplified outcomes $P_{B}(t_{m})$ are binned,
similar to Fig. \ref{fig:Amplification-multiple-bin}, but with a
\emph{small} bin-width of $\Delta_{p}$ . The outcome $P_{I}$ corresponds
to the band of outcomes $P(t_{m})$ between $P_{I}-\Delta_{p}/2$
and $P_{I}+\Delta_{p}/2$. We consider the set of bins, given by $P_{J}$.
Assuming wMR (at a level $\delta$), the system is in a state $\rho_{J\delta}^{(P)}$
where the outcomes are predetermined to lie within the range $P_{J}-\Delta_{p}/2-\delta\leq P\leq P_{J}+\Delta_{p}/2+\delta$.
Any error in the assignment of the value $P_{I}$ on measurement is
not more than $\pm\Delta_{p}$, and similarly for the binned values
$P_{A}(t_{m})$ at $A$. Hence, we see that $P_{A}(t_{m})-g_{p}P_{B}(t_{m})$
has a maximum-error bound of $\sim\pm2\Delta_{p}$, which reduces
to $\pm2\Delta_{p}/G$ in the inferred value, $P_{A}-g_{p}P_{B}$.
The value of $P_{A}-g_{p}P_{B}$ is of order $1/\sqrt{2\cosh2r}\sim e^{-r}$,
so that the relative error is $\frac{2\Delta_{p}}{G}e^{r}$. Let
us assume the high but experimentally feasible squeeze parameter of
$r=2$, for which $\cosh2r\sim e^{2r}/2$ and $e^{r}\sim7.4$\textcolor{blue}{\emph{}}.
Taking $e^{r}\gg e^{-r}$, we place an upper bound on $\sigma_{inf,P_{A}}$:
\begin{equation}
\sigma_{inf,P_{A}}=\langle(P_{A}-g_{p}P_{B})^{2}\rangle^{1/2}\leq(e^{-r}+\frac{2\Delta_{p}}{G})\thinspace.\label{eq:inf-p-amp-1}
\end{equation}

Next, we consider two Cases. First, in Case I, we consider a very
large amplification factor $G$, and small bin-widths so that $2\Delta_{p}/G$
is small, we can ignore the error in (\ref{eq:inf-p-amp-1}) and put
\begin{equation}
\sigma_{inf,P_{A}}^{2}=\langle(P_{A}-g_{p}P_{B})^{2}\rangle=\frac{1}{2\cosh2r}\thinspace.\label{eq:inf-p-amp}
\end{equation}
This is analogous to the case considered for homodyne detection, in
Section \ref{sec:incompleteness-criterion:-quantu}.A. Examples of
possible parameters are: $r=2$, $e^{-r}=0.13$, $G=500$, $\Delta_{p}/G\sim0.01$
and $\delta/G\sim0.004$ so that $\Delta_{p}=5$ and $\delta=2$,
for which $2\Delta_{p}/G=0.02$. Otherwise, we consider Case II, where
the error in estimating the values of $P$ cannot be omitted. Examples
of parameters are: $r=2$, $G=12,$ $\Delta_{p}=3$ and $\delta=2$.

For an incompleteness paradox, we require
\begin{equation}
\sigma_{real,X_{A}}\sigma_{inf,P_{A}}<1/2\label{eq:crit-real2}
\end{equation}
which, for Case I in the limit of very large $G$ and small bin-widths,
becomes $\sigma_{real,X}^{2}<\frac{1}{2}\cosh2r$. Using (\ref{eq:var-x-part}),
the inequality is satisfied  if $\Delta+2\delta<Ge^{r}.$ If we take
$G=500$, we find the inequality reduces to $\Delta+2\delta<3700$.
Consider the original domain in $X$ which has variance $\frac{1}{2}\cosh2r\sim e^{2r}/4$
so that the standard deviation is $\sim e^{r}/2=3.7$. This is to
be suitably divided into bins $X$ to $X+\epsilon$ (so that amplified
domain is $GX$ to $GX+\epsilon G$). We choose $\epsilon=\Delta/G=1.5$
so that $\Delta=750$. Then \textcolor{blue}{}$\Delta+2\delta<3700$
is certainly satisfied with $\delta=2$ (consistent with that used
in Case I for $P_{B}$) . The value $\delta=2$ is well beyond the
level of the original quantum noise value $\sim1/\sqrt{2}\sim0.7$,
hence justifying the application of macroscopic realism at time $t_{m}$.

Now we examine Case II, which is less optimal but applicable to an
experimental optical realization of amplification with $H_{A}$ and
$H_{B}$. We take $r=2$, $e^{r}\sim7.4$ and $G=12$, which is a
similar amplification factor to $r=2$ ($g_{B}\sim2.5$). The original
domain in $X$ with standard deviation $\sim3.7$ is divided into
bins $X$ to $X+\epsilon$. We choose $\epsilon=\Delta/G=1.5$ so
that $\Delta=18$. We take $\delta/G=0.17$ so that $\delta=2$. We
note that for measurement of $\sigma_{inf,P_{A}}$, we can take $\Delta_{p}=\delta$.
Consider $r=2$, $G=12,$ $\Delta_{p}=2$ and $\delta=2$, and taking
$e^{r}\gg e^{-r}$, we use (\ref{eq:inf-p-amp-1}). Then we require
for the incompleteness criterion
\begin{equation}
(\Delta+2\delta)(e^{-r}+\frac{2\Delta_{p}}{G})<G\label{eq:test-1}
\end{equation}
which is satisfied for $\delta=2$.\textcolor{red}{}

\section{Simulations based on the $Q$-function: weak elements of reality\label{sec:Simulation}}

We next address the question raised by Schr\"{o}dinger: Are the
values for the outcomes of the measurement of $\hat{x}_{A}$ and $\hat{p}_{A}$
both predetermined, at the time $t_{m}$, and does this conflict with
quantum mechanics? In this section, we provide a realization of such
a predetermination, showing how the values $x_{A}$ and $p_{A}$ posited
by the wMR premises are the outcomes of the measurements, in a way
that is consistent with quantum theory.

\subsection{Stochastic equations modeling measurements made on the EPR state}

The Hamiltonian $H_{AB}$ of Eq. (\ref{eq:ham-tmss}) generates the
two-mode squeezed state 
\begin{equation}
|\psi_{epr}\rangle=(1-\eta^{2})^{1/2}\sum_{n=0}^{\infty}\tanh^{n}r\thinspace|n\rangle_{A}|n\rangle_{B}\label{eq:tmss}
\end{equation}
which possesses EPR correlations \citep{epr-r2,epr-rmp}. Here $\eta=\tanh r$
and $|n\rangle_{A/B}$ are number states. The two-mode EPR state
$|\psi\rangle$ can be represented uniquely by the $Q$ function,
\begin{eqnarray}
Q_{epr} & = & \frac{1}{\pi^{2}}\langle\alpha|\langle\beta|\psi\rangle\langle\psi|\alpha\rangle|\beta\rangle\nonumber \\
 & = & \dfrac{1}{\pi^{2}}(1-\eta^{2})e^{\left(\alpha^{*}\beta^{*}+\alpha\beta\right)\tanh r}e^{-(|\alpha|^{2}+|\beta|^{2})}\thinspace.\label{eq:q}
\end{eqnarray}
Letting $\alpha=(x_{A}+ip_{A})/\sqrt{2}$ and $\beta=(x_{B}+ip_{B})/\sqrt{2}$,
the $Q$ function becomes\textcolor{blue}{}
\begin{eqnarray}
Q_{epr}(\bm{\lambda},t_{0}) & = & \dfrac{(1-\eta^{2})}{4\pi^{2}}e^{-\frac{1}{4}(x_{A}-x_{B})^{2}(1+\eta)}e^{-\frac{1}{4}(p_{A}+p_{B})^{2}(1+\eta)}\nonumber \\
 &  & \times e^{-\frac{1}{4}(x_{A}+x_{B})^{2}(1-\eta)}e^{-\frac{1}{4}(p_{A}-p_{B})^{2}(1-\eta)}\label{eq:epr-q}
\end{eqnarray}
where $\bm{\lambda}=(x_{A},x_{B},p_{A},p_{B})$. As $r\rightarrow\infty$,
$|\psi_{epr}\rangle$ is an eigenstate of $\hat{X}_{A}-\hat{X}_{B}$
and $\hat{P}_{A}+\hat{P}_{B}$ \citep{epr-1}. Note that the $x_{A}$,
$p_{A}$, $x_{B}$ and $p_{B}$ are defined differently to the same
symbols used in Sec. \ref{sec:Incompleteness-criterion-based}. Here,
they are defined as phase-space variables, not as the outcomes of
measurements $\hat{X}_{A}$, $\hat{P}_{A}$, $\hat{X}_{B}$ and $\hat{P}_{B}$.

\begin{figure}[t]
\centering{}\includegraphics[width=1\columnwidth]{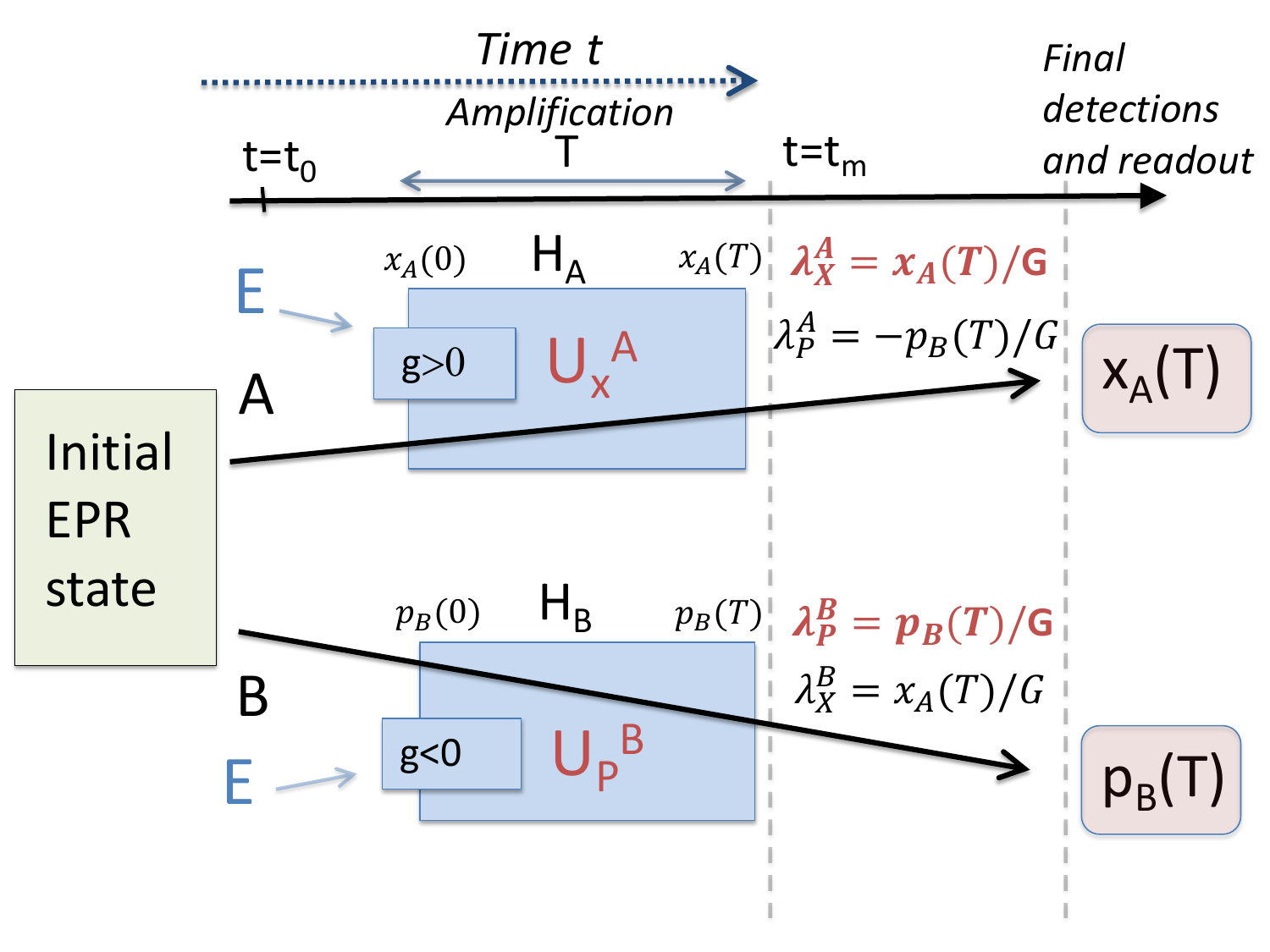}\caption{A diagram of Schr\"{o}dinger's proposed set-up as modeled by the
objective-field $Q$-based simulation. The two outputs $A$ and $B$
of the two-mode squeezed state are spatially separated.  In the first
stage of measurement, for each system, the experimentalist selects
the measurement setting ($X$ or $P$) by interacting the field with
a phase-shifting measurement device that determines sign of $g$.
The field then interacts with a medium to undergo degenerate parametric
amplification, modeled as $H_{A}$ and $H_{B}$. The effect is reversible
unitary operations $U^{A}$ and $U^{B}$ on the fields $A$ and $B$,
respectively. In Schr\"{o}dinger's proposed set-up, the choice is
to measure $X$ at $A$ and $P$ at $B$. The measurement is finalized
by a direct detection, given as the value of $x(T)$ and $p(T)$ in
the simulation of $H_{A}$ and $H_{B}$, where $T$ is the interaction
time. The amplification is $G=e^{|g|T}$. Hence, the inferred results
of the measurements are given as $x_{A}(T)/G$ and $p_{B}(T)/G$.
The simulation realizes values for the outcomes describing the system
at time $t_{m}$, prior to the irreversible readout.\label{fig:schematic-simulation}}
\end{figure}

The measurement of the quadratures $\hat{X}_{A}$ and $\hat{X}_{B}$
(or $\hat{P}_{A}$ and $\hat{P}_{B}$) is modeled as direct amplification
according to the Hamiltonians $H_{A}$ and $H_{B}$, given by Eqs.
(\ref{eq:HampA}) and (\ref{eq:HampB}), where $g_{A}$ and $g_{B}$
can be either positive or negative.\textcolor{blue}{\emph{}} The
final outcome of the measurement is given as the amplified detected
value divided by $G=e^{|g|T}$, where $T$ is the time of amplification.

Following Refs. \citep{q-contextual,q-measurement,q-measurement-wMR},
the interactions $H_{A}$ and $H_{B}$ are solved by converting to
an equation of motion for $Q$ and then transformed to equivalent
forward-backward stochastic equations for the amplitudes $x_{A}$,
$p_{A}$, $x_{B}$ and $p_{B}$ of the $Q$ function \citep{q-husimi,drummond-time,q-contextual}.
The dynamics for the measurement of $\hat{X}_{A}$ and $\hat{X}_{B}$
is given by ($K=A,B$) are\textcolor{blue}{}
\begin{eqnarray}
\frac{dx_{K}}{dt_{-}} & = & -g_{K}x_{K}+\xi_{1K}\left(t\right)\nonumber \\
\frac{dp_{K}}{dt} & = & -g_{K}p_{K}+\xi_{2K}\left(t\right)\label{eq:stoch-eqns}
\end{eqnarray}
where $g_{K}>0$. Details are given in \citep{q-measurement,q-measurement-wMR,drummond-time}.
The noises satisfy $\left\langle \xi_{\mu K}\left(t\right)\xi_{\nu K}\left(t'\right)\right\rangle =|g_{K}|\delta_{\mu\nu}\delta\left(t-t'\right)$,
with noise terms for $A$ and $B$ being independent. Here, $t_{-}=-t$
and the retrocausal equation is solved stochastically with the boundary
condition at the time $T$, after the amplification. This means we
evaluate the $Q$ function at the time $T$ (i.e. after amplification),
this function defining the probability distribution for the ``initial''
amplitudes in the simulation. Hence, the equation for $x_{K}$ is
solved in the ``backward'' direction, from the future boundary condition
given by the $Q$ function $Q_{epr}(\bm{\lambda},T)$. However, because
the equations are separable with respect to $x$ and $p$, the relevant
boundary condition is given by the marginal obtained by integrating
$Q_{epr}(\bm{\lambda},T)$ over $p_{A}$ and $p_{B}$.

Transforming to variables $x_{\pm}=x_{A}\pm x_{B}$, $p_{\pm}=p_{A}\pm p_{B}$,
the dynamical equations for $x_{+}$ and $x_{-}$ are 
\begin{align}
\frac{dx_{+}}{dt_{-}} & =-gx_{+}+\xi_{1+}\left(t\right)\nonumber \\
\frac{dx_{-}}{dt_{-}} & =-gx_{-}+\xi_{1-}\left(t\right)\label{eq:xmeasure}
\end{align}
with future boundary conditions. Those for $p_{+}$ and $p_{-}$ are
\begin{align}
\frac{dp_{+}}{dt} & =-gp_{+}+\xi_{2+}\left(t\right)\nonumber \\
\frac{dp_{-}}{dt} & =-gp_{-}+\xi_{2-}\left(t\right)\label{eq:pdecay}
\end{align}
with boundary conditions at the the initial time. Here $\left\langle \xi_{\mu+}\left(t\right)\xi_{\nu+}\left(t'\right)\right\rangle =2g\delta_{\mu\nu}\delta\left(t-t'\right)$
and $\left\langle \xi_{\mu-}\left(t\right)\xi_{\nu-}\left(t'\right)\right\rangle =2g\delta_{\mu\nu}\delta\left(t-t'\right)$.\textcolor{black}{{}
}The boundary condition for the backward trajectories $x_{\pm}$ is
determined by the marginal
\begin{eqnarray}
Q_{epr}(x_{+},x_{-},T) & = & \frac{e^{-x_{-}^{2}/2\sigma_{-}^{2}(T)}e^{-x_{+}^{2}/2\sigma_{+}^{2}(T)}}{2\pi\sigma_{+}(T)\sigma_{-}(T)}\label{eq:final-q}
\end{eqnarray}
\textcolor{black}{where $\sigma_{\pm}^{2}\left(t\right)=1+e^{2gT}e^{\pm2r}$.}
The correlation between the $x_{A}$ and $x_{B}$ is evident. The
equation for $p$ is solved in the forward direction, the boundary
condition hence given by the initial $Q$ function.

The simulation can be carried out for measurements of $P_{A}$ and
$P_{B}$ in which case $g_{K}<0$ and the equation for each $p_{K}$
is solved in the backward-time direction.\textcolor{red}{{} }We find
\begin{eqnarray}
\frac{dp_{K}}{dt_{-}} & = & -gp_{K}+\xi_{1K}\left(t\right)\nonumber \\
\frac{dx_{K}}{dt} & = & -gx_{K}+\xi_{2K}\left(t\right).\label{eq:pmeasure}
\end{eqnarray}
The solutions are identical as those for $x_{\pm}$, but with $p_{\pm}$
replacing $x_{\mp}$, the marginal used for the backward equations
being $Q_{epr}(p_{-},p_{+},T)$.

\subsection{Simulation of a measurement made on a superposition state}

Before interpreting the results of the EPR simulation, we review the
results for the simulation of a measurement of $\hat{X}$ on a single
mode prepared in a superposition 
\begin{equation}
|\psi\rangle=\frac{1}{\sqrt{2}}(|x_{1}\rangle+i|x_{2}\rangle\label{eq:sup}
\end{equation}
where $|x_{j}\rangle$ is an eigenstate of $\hat{X}=(\hat{a}^{\dagger}+\hat{a})/\sqrt{2}$,
with eigenvalue $x_{j}$. This simulation has been presented previously
\citep{q-measurement}. The measurement of $\hat{X}$ is modeled as
a direct amplification, according to the Hamiltonian $H_{A}$ \citep{q-measurement}.
We note that the eigenstate $|x_{j}\rangle$ has a zero variance,
$(\Delta\hat{X})^{2}\equiv\langle\hat{X}^{2}\rangle-\langle\hat{X}\rangle^{2}=0$,
whereas a coherent state has a variance of $(\Delta\hat{X}_{A})^{2}=1/2$.
However, the $Q$ function of the eigenstate $|x_{1}\rangle$ is a
Gaussian with mean $x_{1}$ and variance $\sigma_{vac}^{2}=1/2$,
which is at the level of the quantum vacuum. The noise level $\sigma_{vac}^{2}$
associated with the $Q$ function of the eigenstate is hence ``hidden''
i.e. undetected.
\begin{figure}[t]
\begin{centering}
\includegraphics[width=0.7\columnwidth]{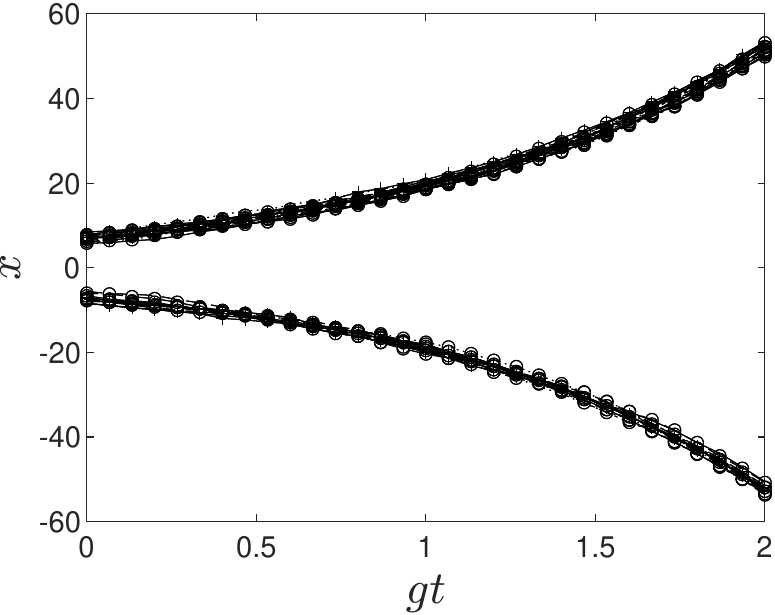}
\par\end{centering}
\bigskip{}

\begin{centering}
\includegraphics[width=0.7\columnwidth]{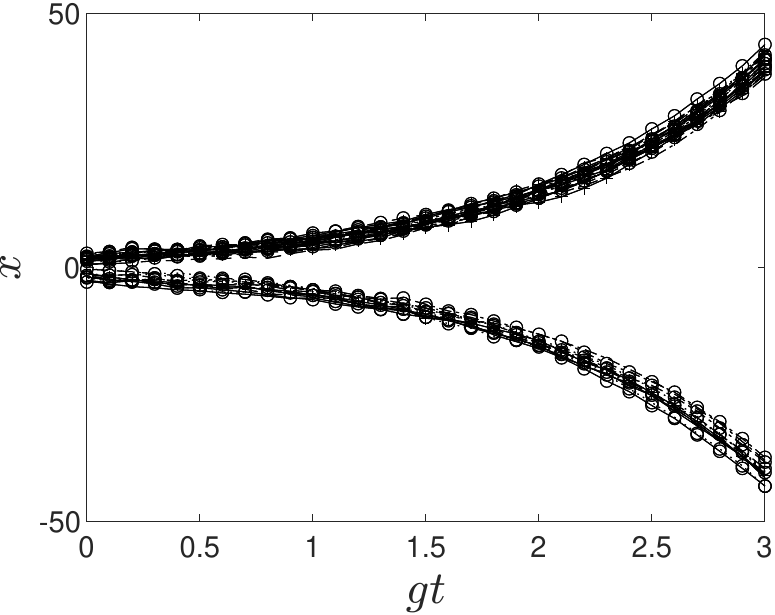}
\par\end{centering}
\caption{Simulation solutions for $x$ where there is amplification of $\hat{X}$.
The initial system is in a superposition of two eigenstates where
$x_{1}=5$, $x_{2}=-5$ (top) and $x_{1}=1$ and $x_{2}=-1$ (lower).\textcolor{red}{{}
}The eigenstates $|x_{1}\rangle$ and $|-x_{1}\rangle$ are modeled
as states that are highly squeezed in the quadrature phase amplitude
$\hat{X}$. Here $gT=2$ (top) and $gT=3$ (lower).\label{fig:sup-dynamics-2}\textcolor{red}{}\textcolor{blue}{}}
\end{figure}

The simulation of the measurement interaction $H_{A}$ is given by
the forward-backward equations\textcolor{blue}{}

\begin{eqnarray}
\frac{dx}{dt_{-}} & = & -gx+\xi_{1}\left(t\right)\nonumber \\
\frac{dp}{dt} & = & -gp+\xi_{2}\left(t\right)\label{eq:stoch-eqns-1}
\end{eqnarray}
which are defined similarly to (\ref{eq:stoch-eqns}), but restricted
to the single mode $A$. The results of the simulation are given in
Figure \ref{fig:sup-dynamics-2}, which presents the solutions for
$x$, the amplified quadrature $\hat{X}$.  After sufficient amplification
of $\hat{X}$, the system in the state (\ref{eq:sup}) evolves into
a macroscopic cat-like state which is a superposition of the two macroscopically-distinct
amplified states.

Following an objective-field ($Q$-based) model for quantum mechanics
\citep{q-contextual,objective-fields-entropy,q-frederic}, we model
the final stage of the measurement as a \emph{direct detection of
the amplitude $x(t)$ once the fields are macroscopic}, which is after
evolution under $H_{A}$ for a time $T$, the amplitude becoming $x(T)$.
The measured amplitude inferred by the detection is $x(T)/G$. Importantly,
we see from Figure \ref{fig:sup-dynamics-2} that the measured amplitudes
always correspond to \emph{one or other of the eigenvalues} $x_{1}$
or $x_{2}$, associated with the eigenstates $|x_{1}\rangle$ or $|x_{2}\rangle$.
This leads to Born's rule \citep{q-measurement}.

After sufficient amplification, at time $t_{m}$, the trajectories
$x(t)$ form \emph{bands }with noise at the hidden vacuum level $\sigma_{vac}^{2}$,
around each eigenvalue value (Fig. \ref{fig:sup-dynamics-2}, top).
Hence, at the time $t_{m}$, after sufficient amplification, the amplitude
$x(t_{m})$ \emph{is} the value of the detected amplified quantity,
the inferred result $x(t_{m})/G$ of the measurement giving either
$x_{1}$ or $x_{2}$ (depending on which band the $x(t_{m}$) belongs
to). The Fig. \ref{fig:sup-dynamics-2} (lower) shows the system initially
in a superposition of states that are not macroscopically distinct.
However, after sufficient amplification at the time e.g. $t_{m}=1.5/g$,
the superpositions become macroscopic and the amplitudes belong to
one or other of two bands, the two outcomes being distinct.

The important conclusion of the simulation is the existence of the
band of amplitudes $x(t_{m})$ (Fig. \ref{fig:sup-dynamics-2}). We
see that after sufficient amplification, at time $t_{m}$, there is
a one-to-one correspondence between the value $x(t_{m})$ and any
final amplitude $x(t_{f})$ ($t_{f}>t_{m}$) if there is further amplification,
after $t_{m}$ (refer \ref{fig:sup-dynamics-2}, top). There is consistency
with the assumption that the value of the amplitude $x(t_{m})$ determines
the branch for the outcome of the measurement (in this case whether
positive or negative), and hence the outcome of the measurement. In
other words, in this model, the \emph{outcome} of the measurement
$X$ is determined by $x(t_{m})$ at this time $t_{m}$.

\subsection{Simulation of the measurements made on the EPR system}

\begin{figure}[t]
\begin{centering}
\includegraphics[width=0.7\columnwidth]{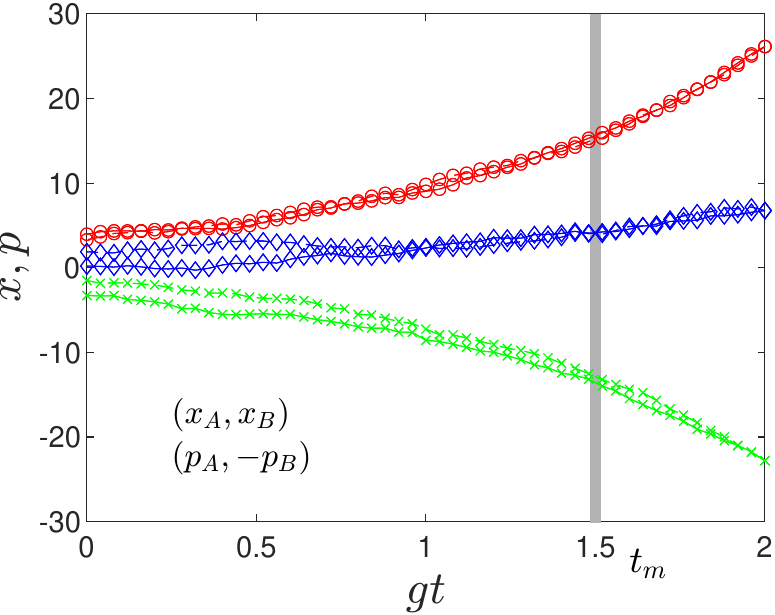}
\par\end{centering}
\bigskip{}

\begin{centering}
\includegraphics[width=0.7\columnwidth]{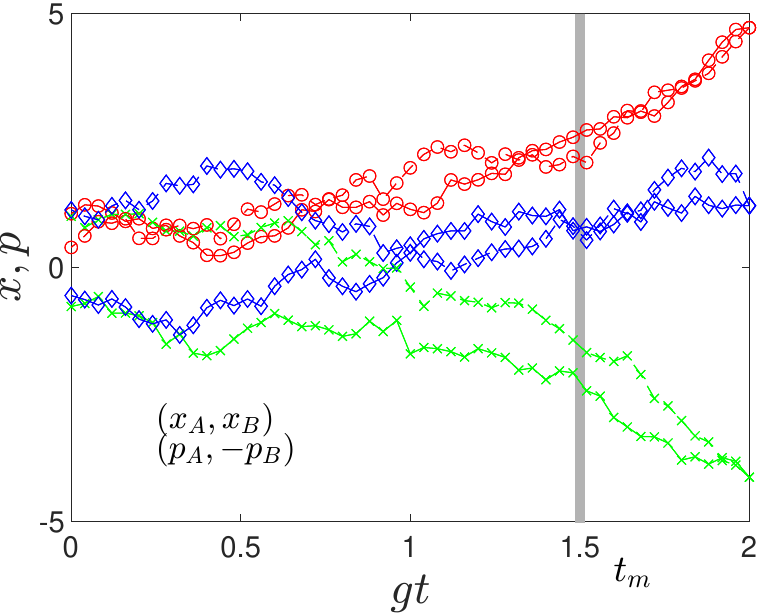}
\par\end{centering}
\caption{Simulation solutions for the EPR system. Example of trajectories for
$x_{A}$ and $x_{B}$ where there is amplification of $\hat{X}_{A}$
and $\hat{X}_{B}$. The solutions of the same color depict the paired
trajectories for the same run in the simulation, thus revealing the
correlated outcomes for $\hat{X}_{A}$ and $\hat{X}_{B}$. The same
solutions give trajectories for $p_{A}$ and $-p_{B}$, showing the
simulation of the measurement of $\hat{P}_{A}$ and $-\hat{P}_{B}$,
revealing anticorrelated outcomes. Here, $r=2$ (top) and $r=0.5$
(lower) and $gT=2$. At $r=0.5$, the EPR correlation is relatively
poor, and is masked by quantum noise at the initial time, but is measurable
as satisfying the EPR criterion (\ref{eq:epr-crit}).\label{fig:epr1}\textcolor{red}{}\textcolor{blue}{}}
\end{figure}

Solutions of the EPR equations (\ref{eq:stoch-eqns}) are shown Figs.
\ref{fig:epr1} and \ref{fig:epr2}. Details of the simulation method
are given in Refs. \citep{q-measurement-wMR,q-measurement}.

First, in Fig. \ref{fig:epr1}, we consider amplification of $\hat{X}$
at both sites ($g_{A}$, $g_{B}>0$). After the amplification, at
time $t_{m}$, there are bands of the amplified amplitudes, $x_{A}(t_{m})$
and $x_{B}(t_{m})$. As above, we find that the values $x_{A}(t_{m})$
and $x_{B}(t_{m})$ define the outcomes for $\hat{X}_{A}$ and $\hat{X}_{B}$
at the time $t_{m}$, as any noise input from the future boundary
condition has negligible impact on the final scaled values $x_{A}(t_{m})/G$
and $x_{B}(t_{m})/G$. Hence, the lines drawn in the figures \emph{give}
the values $x_{A}$ and $x_{B}$ for the outcomes of $\hat{X}_{A}$
and $\hat{X}_{B}$: i.e. in the model, the outcomes are predetermined
at the time $t_{m}$.

However, we also see that for a given run of the simulation, the values
$x_{A}(t_{m})$ and $x_{B}(t_{m})$ are correlated. This is evident
in Fig. \ref{fig:epr1} where the correlated trajectories from the
same run (i.e. with the same noise inputs) are in the same color.
The top figure shows trajectories for the highly correlated EPR state
where $r=2$. The measured variance of the difference $\hat{X}_{A}-\hat{X}_{B}$
becomes zero as $r\rightarrow\infty$. A similar result is obtained
for measurement of $\hat{P}_{A}$ and $\hat{P}_{B}$, where $g_{A}$,
$g_{B}<0$. The variance of the sum $\hat{P}_{A}+\hat{P}_{B}$ becomes
zero with increasing $r$. This confirms that the EPR correlations
are predicted by the individual trajectories.

\subsection{Schr\"{o}dinger's set-up}

\begin{figure}[t]
\begin{centering}
\par\end{centering}
\begin{centering}
\includegraphics[width=0.7\columnwidth]{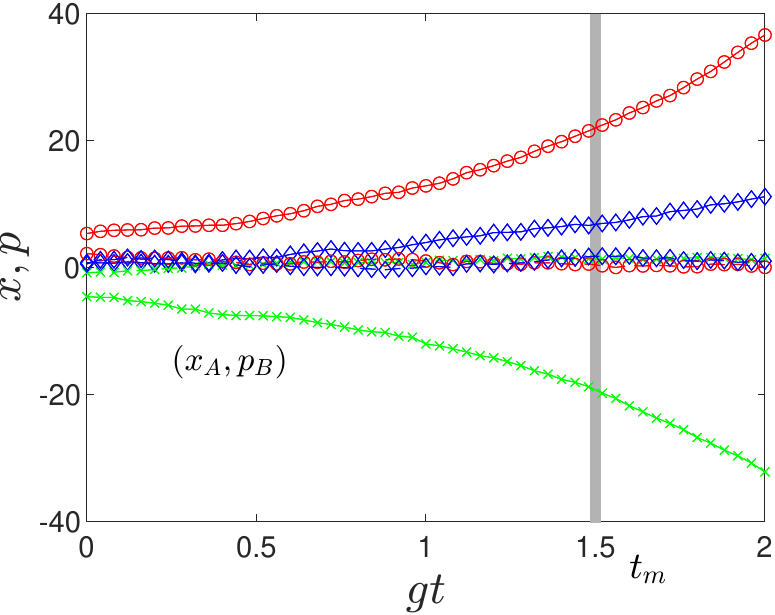}\bigskip{}
\includegraphics[width=0.7\columnwidth]{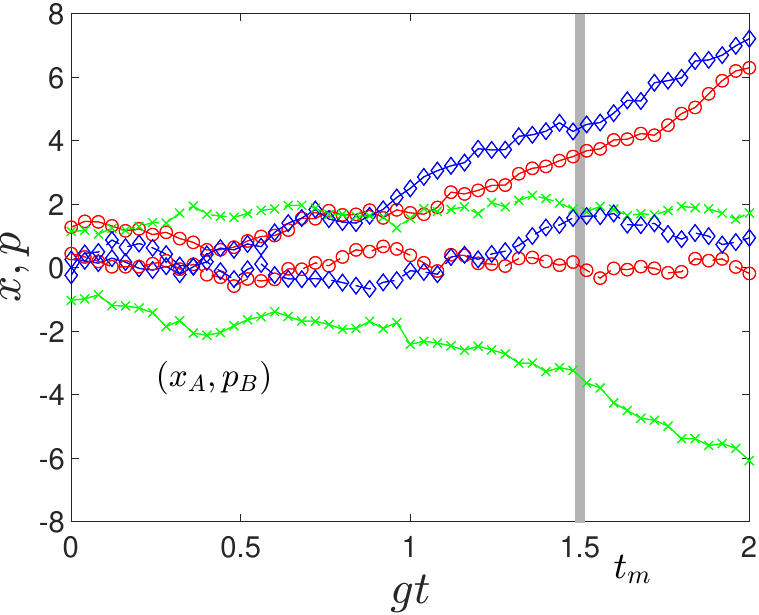}
\par\end{centering}
\caption{Simulation solutions for the EPR system. Here we present results for
the gedanken experiment considered by Schr\"{o}dinger, where one
measures $\hat{X}_{A}$ and $\hat{P}_{B}$. Example of trajectories
for $x_{A}$ (solid line) and $p_{B}$ (dashed line) are given. The
values of the same color are for the same run, hence depicting the
outcomes for joint measurement of $\hat{X}_{A}$ and $\hat{P}_{B}$.
Here, $r=2$ (top) and $r=0.5$ (lower).\label{fig:epr2}\textcolor{red}{}\textcolor{blue}{}}
\end{figure}

What if we measure $\hat{X}_{A}$ at site $A$ and $\hat{P}_{B}$
at the site $B$, as proposed by Schr\"{o}dinger \citep{s-cat-1,sch-epr-exp-atom}?
Plots of trajectory values are given in the Figure \ref{fig:epr2}.
In the model, the amplitudes $x(t_{m})$ and $p(t_{m})$ defined at
the time $t_{m}$ do \emph{give} the outcomes for $\hat{X}_{A}$ and
$\hat{P}_{B}$, in the run of the simulation. By correlation in the
model, from Fig. \ref{fig:epr1} (top), the value $p_{B}$ \emph{is
also the outcome} for $\hat{P}_{A}$, as Schr\"{o}dinger proposed.

However, in order for the prediction of $\hat{P}_{A}$ based on the
measurement $\hat{P}_{B}$ to be realized, it would be necessary to
change the setting at $A$ from $X$ to $P$. This requires interaction
according to $H_{A}$ where $g_{A}<0$, so that first there is reversal
of the amplification of $X_{A}$, followed by amplification of $P_{A}$
(as depicted in Fig. \ref{fig:schematic-meter}). The boundary condition
for the measurement is determined by the $Q$ function at the time
$t_{f}$ after the interaction, given by $Q_{epr}(p_{A},p_{B},t_{f})$.
Hence, \emph{assuming the setting at $B$ is unchanged}, the boundary
condition ensures that the outcome of the measurement $P_{A}$ is
correlated with the value $P_{B}$. In the model, the value $p_{B}(t_{m})$
at time $t_{m}$ \emph{does} predetermine the outcome for $P_{A}$.
In other words, the values $x_{A}$ and $p_{A}$ are simultaneously
determined at the time $t_{m}$.

Schr\"{o}dinger expressed the concern in his essay that if $\hat{P}_{B}$
is to be considered the measurement of $\hat{P}_{A}$, then ``the
quantum mechanician maintains that'' the system $A$ ``has a psi-function\emph{''
}in which $p$\emph{ ``is fully sharp'',} but $x$\emph{ ``fully
indeterminate}''. The simulation is consistent with quantum mechanics,
being derived via the $Q$ function, and allows us to address these
remarks, within this model. There is no conflict with the uncertainty
relation in quantum mechanics. This is because, although the simulation
shows in a given run the dynamics of individual amplitudes $x(t)$
that are part of a single band, the quantum state $|\psi\rangle$
at the time $t_{m}$ does not give this description. The quantum wavefunction
represents a superposition of states, giving the outcomes associated
with both bands. Also, at the time $t_{m}$, the system is in the
state 
\[
e^{-\frac{|g_{B}|t}{2}(\hat{b}^{\dagger2}-\hat{b}^{2})}e^{\frac{|g_{A}|t}{2}(\hat{a}^{\dagger2}-\hat{a}^{2})}|\psi_{epr}\rangle,
\]
prepared for the final stage of measurement of $\hat{X}_{A}$ and
$\hat{P}_{B}$. However, when the measurement setting has changed
to $\hat{P}_{A}$ at $A$, the system is in the different state, given
by $e^{-\frac{|g_{B}|t}{2}(\hat{b}^{\dagger2}-\hat{b}^{2})}e^{-\frac{|g_{A}|t}{2}(\hat{a}^{\dagger2}-\hat{a}^{2})}|\psi_{epr}\rangle$
(refer Fig. \ref{fig:schematic-meter}). 

As a final remark, the solutions of the simulation are consistent
with the premises given in Section \ref{sec:Incompleteness-criterion-based},
of weak macroscopic realism (wMR). This is because the values $x_{A}(t_{m})$
and $p_{B}(t_{m})$ give a determination of the outcome at time $t_{m}$
after amplification, as in the premise wMR(1). We see that the simulation
is consistent with the third premise wMR(3), because of the future
boundary condition that ensures the correlations between the outcomes
$P_{A}$ and $P_{B}$, once the setting for $P_{B}$ is fixed. The
second premise wMR(2) is that the value, $p_{B}(t_{m})$ say, for
the outcome of $\hat{P}_{B}$, is not changed by a change of setting
at the location $A$.

The weak macroscopic realism premises and the solutions of the simulation
are not inconsistent with the violations of Bell inequalities, as
shown in Refs. \citep{ghz-cat,wigner-friend-macro,weak-versus-det,q-measurement-wMR}.
The amplitudes $x_{A}(t_{m})$ and $p_{A}(t)=-p_{B}(t_{m})$ given
in the simulation (Figs. \ref{fig:epr1} and \ref{fig:epr2}) can
be distinguished from the ``elements of reality'' considered by
EPR \citep{epr-1}. The elements of reality considered by EPR were
defined for the system as it exists at time $t_{0}$, \emph{prior}
to the interactions (e.g. $H_{A}$ and $H_{B}$) which determine the
choice of measurement setting (i.e. whether to measure either $\hat{X}$
or $\hat{P}$). EPR's elements of reality can be negated in the experiments
proposed by Bell \citep{Bell-2}. While Schr\"{o}dinger did not fully
distinguish between these different types of ``elements of reality'',
Schr\"{o}dinger's analysis referred to the set-up where there is
a fixed choice of measurement setting, and hence differed from that
of EPR.

\section{Conclusion}

The main result of this paper is the derivation of a criterion for
the ``incompleteness'' of (standard) quantum mechanics. The criterion
is based on a \emph{subset} of local realistic premises that are not
falsified by Bell's work, and which can be applied to the version
of the EPR paradox put forward by Schr\"{o}dinger in his reply to
EPR in 1935. We refer to the less-restrictive premises as ``\emph{weak
macroscopic realism} \emph{(wMR)}'' (or ``\emph{weak local realism}'').
While the criterion is general, we have explained in detail how the
criterion can be applied to continuous-variable experiments in quantum
information. These experiments detect EPR correlations using quadrature-phase-amplitude
measurements on Gaussian systems that are based on the two-mode squeezed
state. Other realizations are possible e.g. for cat states.

Schr\"{o}dinger considered an EPR state, where two separated systems
$A$ and $B$ have correlated positions and anticorrelated momenta.
Schr\"{o}dinger examined the particular set-up where the measurement
settings are adjusted, so that one measures $X_{A}$ at $A$ and $P_{B}$
at $B$. The outcome of $P_{A}$ can be predicted precisely by making
a measurement of $P_{B}$ on the system $B$. This means that for
Schr\"{o}dinger's set-up, one is making a direct measurement of $X_{A}$
and an indirect measurement of $P_{A}$. The essence of Schr\"{o}dinger's
question is: \emph{At what time, if any, are the outcomes for $X_{A}$
and $P_{A}$ determined?}

In this paper, we consider Schr\"{o}dinger's EPR system at the time
$t_{m}$, \emph{after} the system has interacted with the devices
(e.g. a polarizing beam splitter) so that the measurement settings
have been fixed as $X_{A}$ and $P_{B}$. These interactions are reversible,
represented in quantum mechanics as a unitary operation $U_{\theta}$,
giving a rotation of the measurement basis. We also assume that there
has been a further amplification, in preparation for a final detection
or read-out. At the time $t_{m}$, the outcome of $P_{A}$ can be
predicted with certainty by a simple detection at $B$, which gives
the outcome of $P_{B}$. Hence, the premise of wMR implies that $P_{A}$
is determined at time $t_{m}$. The premises of wMR also imply that
$X_{A}$ is determined at time $t_{m}$, based on the assumption of
\emph{macroscopic realism}. In summary, the premises of wMR imply
the predetermination of both $X_{A}$ and $P_{A}$ at time $t_{m}$.

This brings us to the second main result of this paper. We provide
a phase-space forward-backward stochastic simulation of the measurement
on an EPR system, demonstrating how the predetermined values for $X_{A}$
and $P_{A}$ emerge. These values address queries raised by Schr\"{o}dinger.
The simulation is based on the $Q$ function for quantum fields, which
is always positive, and uniquely defines a quantum state. Trajectories
are given for the amplitudes $x_{A}$, $p_{A}$, $x_{B}$ and $p_{B}$
that are defined by the $Q$ function, $Q(x_{A},x_{B},p_{A},p_{B})$.
With sufficient amplification, the values for the outcomes of the
measurement are predetermined, \emph{if} that measurement proceeds
as a direct detection. (Recognizing that the amplification is reversible,
as is the setting dynamics given by $H_{\theta}$, we see that the
amplification and setting dynamics can be reversed and changed, in
which case the values would no longer apply). There is no conflict
with the uncertainty principle nor with Bell's theorem. On the other
hand, the simulation does reveal a predetermination of both outcomes
$X_{A}$ and $P_{A}$ at the time $t_{m}$, and hence presents a result
that is more complete than (i.e. not given by) standard quantum mechanics.

A second question implied by Schr\"{o}dinger's analysis (\emph{Does
the measurement of $\hat{P}_{B}$ at $B$ cause the outcome for $\hat{P}_{A}$
at $A$, or was that outcome determined prior?}) is difficult to address,
even within the simulation model. The Wigner function is positive
for this system, and provides a model in which all outcomes $\hat{X}_{A}$,
$\hat{P}_{A}$, $\hat{X}_{B}$ and $\hat{P}_{B}$ can be viewed as
determined at the initial time $t_{0}$. Such a model would not work
for Bohm's spin EPR paradox based on Bell states, and any resolution
must be applicable to the Bell-Bohm system. We see that in the simulation,
once the measurement setting is fixed at $B$, and at the time $t_{m}$
after amplification,  the value for the outcome of $\hat{P}_{A}$
is determined as $-p_{B}(t_{m})$. This is due to the future boundary
condition, specified by the $Q$ function $Q_{epr}(p_{-},p_{+},T)$
(refer to Eq. (\ref{eq:pmeasure})) of the amplified state, which
determines the correlation between $\hat{P}_{A}$ and $\hat{P}_{B}$.
However, the correlations specified by this function extrapolate from
those described by the initial $Q$ function at time $t_{0}$. Hence,
in the model, the value of $\hat{P}_{A}$ is determined once the amplification
takes place at $B$, but this arises in part due to the correlation
described by the initial $Q$ function. This initial correlation though
is weaker, being masked by a hidden non-amplifiable vacuum-noise level
(which arises from the retrocausal amplitudes). This effect is shown
by the solutions of Fig. \ref{fig:epr1}.

The question raised by Schr\"{o}dinger (\emph{when are the outcomes
of a measurement determined?}) is at the heart of understanding nonlocality
and measurement in quantum mechanics, problems attracting great interest
\citep{bohm-hv,auff-grangier-1,fr,pgrangier-2,brukner,wood,hossenfelder-toy-collapse,pegg,wharton-a-rmp,harrigan-spekkens,price,spekkens-toy,wharton-new-class,born,struyve-ward-pilot-wave,bell-against,price-time,bong-no-go},
but for which there appears to be as of yet no agreed resolution.
Here, we present a feasible proposal to demonstrate an incompleteness
of (standard) quantum mechanics, giving an argument that quantum mechanics
be completed by variables (``beables'' \citep{beables}) that are
not contradicted by Bell's theorem.
\begin{acknowledgments}
This research has been supported by the Templeton Foundation under
Project Grant ID 62843. We thank NTT Research for technical help and
motivation for this project.\textcolor{blue}{}
\end{acknowledgments}

\begin{widetext} 

\section*{Appendix A}

The Wigner function for the two-mode squeezed state is:\textcolor{red}{}\textcolor{black}{
\begin{equation}
{\color{red}{\color{red}{\color{black}{\color{red}}}}{\color{black}{\color{red}}W(\alpha,\beta)=\dfrac{4}{\pi^{2}}\exp\left(-2\cosh2r\left(\left|\alpha\right|^{2}+\left|\beta\right|^{2}\right)+2(\alpha\beta+\alpha^{*}\beta^{*})\sinh2r\right)}}
\end{equation}
}Here we define $\alpha=\left(x_{A}+ip_{A}\right)/\sqrt{2},\beta=\left(x_{B}+ip_{B}\right)/\sqrt{2}$,
upon a change of variables the igner function can be transformed into
distributions of the position and momentum coordinates for systems
$A$ and $B,$
\begin{equation}
W(\text{\textbf{x}},\text{\textbf{p}})=N\exp\left(-\cosh2r\left(\left(x_{A}^{2}+p_{A}^{2}\right)+\left(x_{B}^{2}+p_{B}^{2}\right)\right)+2(x_{A}x_{B}-p_{A}p_{B})\sinh2r\right)
\end{equation}
After solving for the normalization constant, here $N=1/\pi^{2}$
, the Wigner function for the TMSS can be recast as 
\begin{equation}
W(\text{\textbf{x}},\text{\textbf{p}})=\dfrac{1}{\pi^{2}}\exp\left(-\dfrac{e^{2r}}{2}\left(\left[x_{A}-x_{B}\right]^{2}+\left[p_{A}+p_{B}\right]^{2}\right)-\dfrac{e^{-2r}}{2}\left(\left[x_{A}+x_{B}\right]^{2}+\left[p_{A}-p_{B}\right]^{2}\right)\right)
\end{equation}
The marginal distribution $P(x_{A},p_{B})$ is obtained by integrating
over $p_{A}$ and $x_{B}$,

\begin{align}
P(x_{A},p_{B}) & =\dfrac{1}{\pi^{2}}\int\int dp_{A}dx_{B}\exp\left(-\dfrac{e^{2r}}{2}\left(\left[x_{A}-x_{B}\right]^{2}+\left[p_{A}+p_{B}\right]^{2}\right)-\dfrac{e^{-2r}}{2}\left(\left[x_{A}+x_{B}\right]^{2}+\left[p_{A}-p_{B}\right]^{2}\right)\right)\nonumber \\
 & =\dfrac{1}{\pi\cosh(2r)}\exp\left(-\dfrac{x_{A}^{2}}{\cosh(2r)}\right)\exp\left(-\dfrac{p_{B}^{2}}{\cosh(2r)}\right)
\end{align}
The distribution $P(p_{B})$ is given by integrating over $x_{A}$
for the marginal distribution $P\left(x_{A},p_{B}\right)$ which yields
a Gaussian
\begin{equation}
P(p_{B})=\int dx_{A}P(x_{A},p_{B})=\dfrac{1}{\sqrt{2\pi}\sigma_{p}}e^{-p_{B}^{2}/2\sigma_{p}^{2}}
\end{equation}
with zero mean and variance $\sigma_{p}^{2}=\dfrac{1}{2}\cosh2r$.

\section*{Appendix B}

 Following from Section \ref{sec:incompleteness-criterion:-quantu},
we consider the system $A$ which is either in state $\rho_{-0}$
or in state $\rho_{+0}$, given by $I=1$ and $I=2$ respectively,
with relative probabilities $P_{1}$ and $P_{2}$. We denote the variance
in $X_{A}$ for the system given by $I$ as $\sigma_{X_{A}|I}^{2}$.
The overall distribution $P(X_{A})$ for $\hat{X}_{A}$ can be measured,
and is predicted to be Gaussian, as in Figure \ref{fig:diagram-bins}.
We require to place an upper bound $U_{B}$ on (\ref{eq:real-inf-av-1})
\begin{equation}
\sigma_{real,X_{A}}^{2}=\sum_{I}P_{I}\sigma_{X_{A}|I}^{2}
\end{equation}
that can be deduced from the experimentally measurable quantities,
for a given value of $x_{1}$ (and hence $P_{0}$, refer to Figure
\ref{fig:diagram-bins}). \textcolor{black}{The value of $x_{1}$
determines the degree of ``macroscopic distinctness''. Hence, the
incompleteness criterion (Eq. (\ref{eq:real-inf-crit})) is}
\begin{equation}
\sigma_{real,X_{A}}^{2}\sigma_{inf,P_{A}}^{2}<1/4\label{eq:incompleteness-crit}
\end{equation}
which can be satisfied when
\begin{equation}
U_{B}\sigma_{inf,P_{A}}^{2}<1/4\label{eq:icc}
\end{equation}
where we note that the upper bounds are symmetric for $I=1$ and $2$.
Using that $\sigma_{inf,P_{A}}^{2}=1/(2\cosh2r)$, but which must
be measured in the experiment, we predict the incompleteness criterion
can be satisfied when $U_{B}<\frac{\cosh2r}{2}$. Below, we drop the
subscripts $A$ for convenience, defining $P(X)$ as $P(X_{A})$,
$X$ as $X_{A}$, and $\hat{X}$ as $\hat{X}_{A}$.

The derivation of an upper bound $U_{B}$ on the $\sigma_{X_{A}|I}^{2}$
is given in the Lemma below.\textcolor{green}{}
\begin{eqnarray}
\sigma_{X_{A}|I=2}^{2} & \leq & \frac{1}{P_{+}}\sum_{X>x_{1}}P(X)X^{2}-\frac{1}{P_{+}^{2}}\{\sum_{X>x_{1}}P(X)X\}^{2}+x_{1}^{2}P_{0}+\frac{2x_{1}P_{0}}{P_{+}}\sum_{X>x_{1}}P(X)X\label{eq:var+bound-1}
\end{eqnarray}
where $P(X)$ is the measurable full distribution for $\hat{X}$,
$P_{+}$ is the measurable probability of measuring $X>x_{1}$\textcolor{green}{}.
We see that for each given $x_{1}$, all of the quantities in (\ref{eq:var+bound-1})
can be determined experimentally from the measured distribution $P(X)$
(refer Figure \ref{fig:diagram-bins}).

We present the calculations for the predictions, based on the observation
that $P(X)=\frac{1}{\sqrt{2\pi}\sigma_{X}}e^{-\frac{X^{2}}{2\sigma_{X}^{2}}}$,
which is a Gaussian with variance $\sigma_{X}^{2}$ .Then\textcolor{black}{
\begin{align}
\sum_{X>x_{1}}P(X)X^{2} & =\frac{1}{\sqrt{2\pi}\sigma_{X}}\intop_{x_{1}}^{\infty}X^{2}e^{-\frac{X^{2}}{2\sigma_{X}^{2}}}\,dX\nonumber \\
 & =\frac{\sigma_{X}}{2\sqrt{2\pi}}\left[2x_{1}e^{-\frac{x_{1}^{2}}{2\sigma_{X}^{2}}}+\sqrt{2\pi}\sigma_{X}-\sqrt{2\pi}\sigma_{X}Erf\left(\frac{x_{1}}{\sqrt{2}\sigma_{X}}\right)\right]
\end{align}
where $\widetilde{x}_{1}=x_{1}/\sigma_{X}$ and where $Erf(z)$ is
the error function. Also, 
\begin{align}
\sum_{X>x_{1}}P(X)X & =\frac{1}{\sqrt{2\pi}\sigma_{X}}\intop_{x_{1}}^{\infty}Xe^{-\frac{X^{2}}{2\sigma_{X}^{2}}}\,dX=\frac{\sigma_{X}}{\sqrt{2\pi}}e^{-\frac{x_{1}^{2}}{2\sigma^{2}}}=\frac{\sigma_{X}}{\sqrt{2\pi}}e^{-\frac{\widetilde{x}_{1}^{2}}{2}}.
\end{align}
With these expressions, we can find the bound for $\sigma_{X_{A}|I=2}^{2}$.
Here, 
\begin{align}
P_{+} & =\frac{1}{\sqrt{2\pi}\sigma_{X}}\int_{x_{1}}^{\infty}e^{-\frac{X^{2}}{2\sigma_{X}^{2}}}=\frac{1}{2}\left[1-Erf\left(\frac{x_{1}}{\sqrt{2}\sigma_{X}}\right)\right]
\end{align}
and }\textcolor{red}{}\textcolor{black}{$P_{0}=1-2P_{+}.$ We show
the results of the calculation in Figure \ref{fig:bounds}.}

\begin{figure}[t]
\begin{centering}
\includegraphics[width=0.4\columnwidth]{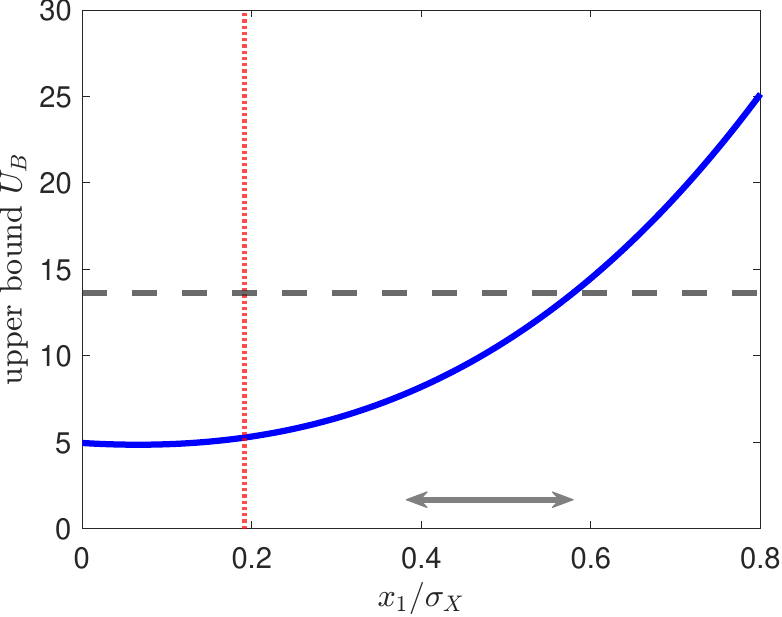}\,\,\,\,\,\,\,\,\,\,\,\,\includegraphics[width=0.4\columnwidth]{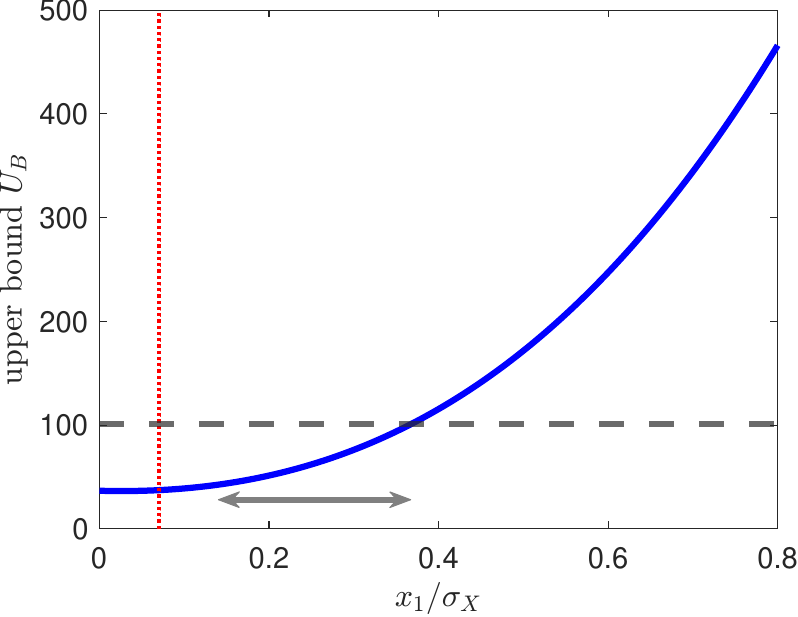}
\par\end{centering}
\caption{\textcolor{green}{}Here we show a regime where the incompleteness
criterion (Eqs. (\ref{eq:incompleteness-crit}) and (\ref{eq:real-inf-crit}))
can be satisfied for a macroscopic value of $x_{1}$. We plot the
upper bound $U_{B}$ on $\sigma_{X_{A}|I=2}^{2}$ (and hence $\sigma_{real,X_{A}}^{2}$)
as a function of $x_{1}/\sigma_{X}$, where $\sigma_{X}^{2}=\frac{1}{2}\cosh2r$.
The blue line corresponds to the bound $U_{B}$. The gray dashed,
horizontal line corresponds to the value of $\sigma_{X}^{2}=1/\sigma_{inf,p_{A}}^{2}=(\cosh2r)/2$.
The criterion is satisfied for values of $x_{1}$ when the blue solid
line $U_{B}$ is below the gray horizontal dashed line. The red dotted
vertical line corresponds to the value of $x_{1}=\frac{1}{\sqrt{2}}$.
We justify that $x_{1}$ becomes macroscopic as the value of $x_{1}$
exceeds that of the fluctuation in $x_{1}$ due to the quantum noise
level, given by $x_{1}=\frac{1}{\sqrt{2}}$, by a factor greater than
at least $2$. The regime of $x_{1}$ values of interest to us (where
the criterion is satisfied but $x_{1}$ is macroscopic in absolute
terms) is indicated by the gray double-arrow. We plot for $r=2$
(left) and $r=3$ (right).\textcolor{blue}{\label{fig:bounds}}\textcolor{red}{}\textcolor{blue}{}}
\end{figure}

To show the incompleteness criterion is predicted to be satisfied,
we note that the prediction for $P(X)$ is a Gaussian with variance
$\sigma_{X}^{2}=\frac{1}{2}\cosh2r$. We let $x_{1}/\sigma_{X}\rightarrow0$.
Then we consider the half-Gaussian $P_{1/2}(X)$ which is $P(X)$
defined over the range $X\geq0$. Since $P_{1/2}(X)=2P(X)$, the
Eq. (\ref{eq:var+bound-1}) becomes
\begin{eqnarray}
\sigma_{X_{A}|I=2}^{2} & \leq & \frac{1}{2P_{+}}\sum_{X>0}P_{1/2}(X)X^{2}-\frac{1}{4P_{+}^{2}}\{\sum_{X>0}P_{1/2}(X)X\}^{2}=Var_{1/2}\label{eq:var+step-1-1}
\end{eqnarray}
where $Var_{1/2}=\sigma_{X}^{2}(1-\frac{2}{\pi})$ is the variance
of the half-Gaussian and where we have used in the last line that
$P_{+}$ is predicted to be $1/2$. A similar result is given for
$\sigma_{X_{A}|I=1}^{2}$. We see that the estimate for $U_{B}$
is $\sim0.36\sigma_{X}^{2}$, so that
\begin{equation}
\sigma_{real,X_{A}}^{2}\leq\sigma_{X}^{2}(1-\frac{2}{\pi})=0.36\sigma_{X}^{2}.\label{eq:crit-real-1-1}
\end{equation}
We note that $1/\sigma_{inf,P_{A}}^{2}$ is predicted to be equal
to $\sigma_{X}^{2}$. Hence, as $x_{1}$ becomes small relative to
$\sigma_{X}$, the variance of the ``element of reality'' associated
with MR becomes less than $\sigma_{X}^{2}=\dfrac{1}{2}\cosh2r$\textcolor{blue}{,}
which is sufficient to satisfy the incompleteness criterion (\ref{eq:icc}).
This is seen in the plots, where as $x_{1}/\sigma_{X}\rightarrow0$,
the value of $U_{B}$ is estimated as $4.96$, and $36.65$ for $r=$
$2$ and $3$ respectively, giving consistency with the values ($4.92$
and $36.31$) of $0.36\sigma_{X}^{2}$ in each case.\textcolor{blue}{}

\textbf{Lemma:} First, we consider $I=2$, where the system in $\rho_{0+}$:
The distribution for $X$ given this state is denoted by $P_{I}(X)$.
The probability of obtaining an outcome $X$ in region $+$ is $P_{+|I=2}=\sum_{X>x_{1}}P_{I}(X)$
and the probability of obtaining $X$ in the region $0$ is $P_{0|I=2}$.
The mean is \textcolor{red}{\emph{}}
\begin{eqnarray}
\bar{X}= & \sum_{X}P_{I}(X)X= & \sum_{X>x_{1}}P_{I}(X)X+\sum_{-x_{1}\leq X\leq x_{1}}P_{I}(X)X\label{eq:mean}
\end{eqnarray}
Hence 
\begin{eqnarray}
\bar{X}^{2} & = & \{\sum_{X>x_{1}}P_{I}(X)X+\sum_{-x_{1}\leq X\leq x_{1}}P_{I}(X)X\}^{2}\nonumber \\
 & = & \{\sum_{X>x_{1}}P_{I}(X)X\}^{2}+\{\sum_{-x_{1}\leq X\leq x_{1}}P_{I}(X)X\}^{2}+2\{\sum_{X>x_{1}}P_{I}(X)X\}\{\sum_{-x_{1}\leq X\leq x_{1}}P_{I}(X)X\}\nonumber \\
 & \geq & \{\sum_{X>x_{1}}P_{I}(X)X\}^{2}-2x_{1}P_{0|I=2}\sum_{X>x_{1}}P_{I}(X)X
\end{eqnarray}
where we use that $\sum_{-x_{1}\leq X\leq x_{1}}P_{I}(X)X\geq-x_{1}P_{0|I=2}$,
on noting that for $I=2$, $X\geq-x_{1}$ and $\sum_{X<x_{1}}P_{I}(X)=P_{0|I=2}$.
Now,
\begin{eqnarray}
\langle X^{2}\rangle & = & \sum_{X}P_{I}(X)X^{2}=\sum_{X>x_{1}}P_{I}(X)X^{2}+\sum_{X\leq x_{1}}P_{I}(X)X^{2}\nonumber \\
 & \leq & \sum_{X>x_{1}}P_{I}(X)X^{2}+x_{1}^{2}P_{0|I=2}\label{eq:var}
\end{eqnarray}
where we use that for $I=2$, in the region $0$, $X^{2}\leq x_{1}^{2}$.
Hence, 
\begin{eqnarray}
\sigma_{X_{A}|I=2}^{2} & \leq & \sum_{X>x_{1}}P_{I}(X)X^{2}-\{\sum_{X>x_{1}}P_{I}(X)X\}^{2}+x_{1}^{2}P_{0|I=2}+2x_{1}P_{0|I=2}\sum_{X>x_{1}}P_{I}(X)X\label{eq:bound-1}
\end{eqnarray}
The first term requires knowledge of $P_{I}(X)$ where $X>x_{1}$.
For this domain, the outcome can only occur if the system is in $\rho_{+0}$.
Hence, over this domain, $P_{I}(X)$ is the same as $P(X)$, but renormalised.
Using conditional probabilities, $P(X)=P(X,I=2)=P(X|I=2)P_{2}$. The
distribution for $X$ \emph{conditioned on} $X>x_{1}$, given the
state $\rho_{+0}$, is hence 
\begin{equation}
P_{I}(X)=P(X)/P_{2}\label{eq:prob}
\end{equation}
The $P_{2}$ cannot be measured directly. Now the variance given
wMR given above is written
\begin{eqnarray}
\sigma_{X_{A}|2}^{2} & \leq & \frac{1}{P_{2}}\sum_{X>x_{1}}P(X)X^{2}-\frac{1}{P_{2}^{2}}\{\sum_{X>x_{1}}P(X)X\}^{2}+x_{1}^{2}P_{0|I=2}+\frac{2x_{1}P_{0|I=2}}{P_{2}}\sum_{X>x_{1}}P(X)X\nonumber \\
 & \leq & \frac{1}{P_{+}}\sum_{X>x_{1}}P(X)X^{2}-\frac{1}{P_{+}^{2}}\{\sum_{X>x_{1}}P(X)X\}^{2}+x_{1}^{2}P_{0}+\frac{2x_{1}P_{0}}{P_{+}}\sum_{X>x_{1}}P(X)X\label{eq:A-final}
\end{eqnarray}
\textcolor{green}{} where we use that $P_{2}\geq P_{+}$, $P_{0|I=2}\leq P_{0}$,
and note that the upper bound is measurable. The bound on $\sigma_{X_{A}|I=1}^{2}$
is identical, on replacing $P_{+}$ with $P_{-}$. $\square$

\section*{Appendix C}

\textcolor{red}{}The $Q$ function for the two-mode squeezed state
(\ref{eq:tmss}), in terms of the variables $\alpha$ and $\beta$,
is: 
\begin{equation}
Q_{epr}=\frac{1}{\pi^{2}}\langle\alpha|\langle\beta|\psi\rangle\langle\psi|\alpha\rangle|\beta\rangle=\dfrac{1}{\pi^{2}}(1-\eta^{2})e^{\left(\alpha^{*}\beta^{*}+\alpha\beta\right)\tanh r}e^{-(|\alpha|^{2}+|\beta|^{2})}.
\end{equation}
Here $\eta=\tanh r.$ The $Q$ function is rewritten as

\begin{align}
Q(\bm{\lambda},t_{0}) & =\dfrac{1}{4\pi^{2}}(1-\eta^{2})e^{-\frac{1}{4}(x_{A}-x_{B})^{2}(1+\eta)}e^{-\frac{1}{4}(p_{A}+p_{B})^{2}(1+\eta)}e^{-\frac{1}{4}(x_{A}+x_{B})^{2}(1-\eta)}e^{-\frac{1}{4}(p_{A}-p_{B})^{2}(1-\eta)}.\label{eq:92-1}
\end{align}
where $\alpha=(x_{A}+ip_{A})/\sqrt{2}$ and $\beta=(x_{B}+ip_{B})/\sqrt{2}$.
The normalization constant is $N=\dfrac{1}{4\pi^{2}},$ with $\bm{\lambda}=(x_{A},x_{B},p_{A},p_{B})$.
For large squeezing, as $r\rightarrow\infty,$ we find $\tanh r\rightarrow1,$
and in this instance the $Q$ function for the $|\psi_{epr}\rangle$
state becomes
\begin{equation}
Q(\bm{\lambda},t_{0})\rightarrow\dfrac{(1-\eta^{2})}{4\pi^{2}}e^{-\frac{1}{2}(x_{A}-x_{B})^{2}}e^{-\frac{1}{2}(p_{A}+p_{B})^{2}}\label{eq:epr-q-rlarge}
\end{equation}
which corresponds to an eigenstate of $\hat{X}_{A}-\hat{X}_{B}$ and
$\hat{P}_{A}+\hat{P}_{B}$. 

\end{widetext}


\begin{thebibliography}{100}
\bibitem{exp-variance} For any distribution $P(x_{i})$ of a variable
$x_{i}$ which is constrained to a range of values $a$ to $a+B$,
the largest variance $\sum_{i}P(x_{i})(x_{i}-\langle x_{i}\rangle)^{2}$
occurs when the values are equally placed at the boundaries i.e. when
all half the values are at $a$ and half at $a+B$. Hence, the variance
cannot exceed $B^{2}/4$.
\end{thebibliography}
\end{document}